\documentclass[prd,showpacs,superscriptaddress,nofootinbib]{revtex4}
\usepackage{graphicx}
\usepackage{epsfig}
\usepackage{symb}
\usepackage{epsf}
\usepackage{amsmath}
\usepackage{amssymb}
\usepackage{graphicx}
\usepackage{gensymb}
\usepackage{natbib}
\usepackage{color}
\usepackage{color}
\usepackage{mathrsfs}
\begin{document}
\def\h0{H_0}
\def\om{\Omega_m}
\def\ode{\Omega_{\mathrm{DE}}}
\def\ok{\Omega_k}
\def\p{\partial}
\def\w0{w_0}
\def\wa{w_a}
\def\({\left(}
\def\){\right)}
\def\pp{\p^2}
\def\da{d_A(z)}
\def\intzinf{\int_z^\infty}
\def\intoz{\int_0^z}

\def\arg{\sqrt{\Omega_k}\chi(z)}

\def\ia{SNIa}
\def\dqdom#1{\frac{\p \mathscr{E}(#1) }{\p \om}}
\def\dqdok#1{\frac{\p \mathscr{E}(#1) }{\p \ok}}
\def\dqdwa#1{\frac{\p \mathscr{E}(#1) }{\p \wa}}
\def\dqdw0#1{\frac{\p \mathscr{E}(#1) }{\p \w0}}

\def\dk{(\mathbf{d-t})}
\def\dkt{(\mathbf{d-t})^\mathrm{T}}
\def\cov{\mathbf{C^{-1}}}
\def\dcov#1{\mathbf{C_{,#1}^{-1}}}
\def\vchi{\mathbf{\chi}^2}
\def\bt{\mathbf{t}}
\def\bd{\mathbf{d}}
\def\tran{^{\mathrm{T}}}

\def\om{\Omega_m}
\def\ode{\Omega_{\mathrm{DE}}}
\def\ok{\Omega_k}
\def\p{\partial}
\def\({\left(}
\def\){\right)}
\def\intzinf{\int_z^\infty}
\def\integ0a{\int_0^a}

\def\be{\begin{equation}}
\def\ee{\end{equation}}
\def\bea{\begin{eqnarray}}
\def\eea{\end{eqnarray}}

\def\name{Fisher4Cast}
\def\versionnum{Version 2.2}
\def\fig_dir#1{#1}
\def\da_0/do_k{\(-\frac{1}{2} \frac{a_0}{\ok}\)}
\def\de/dE/do_m{\frac{1}{2} \frac{1}{E} \left [ \left ( \frac{a}{a_0} \right ) ^{-3} - f\right ]}

\def\df/da_0{\left( \( \frac{3(1+w_0+w_a)}{a}  \(\frac{a_0}{a}\)^{3(1+w_0+w_a)-1} \times \exp \left \{3w_a\left(\frac{a_0}{a}-1 \right) \right \}\)  +  \(\frac{a_0}{a}\)^{3(1+w_0+w_a)} \times 3\frac{w_a}{a} \exp \left \{3w_a\left (\frac{a_0}{a}-1 \right ) \right \}\right)}

\def\df/do_k{\(\df/da_0\) \(\da_0/\do_k\)}

\def\dedom#1{\frac{1}{2E(#1)} \left\{(1+#1)^3 -f(#1) \right\}}
\def\dedok#1{\frac{1}{2E(#1)}\left\{(1+(#1))^2 -f(#1)\right\}}
\def\dedf#1{\frac{1}{2E(#1)}\left[ 1 - \om - \ok \right]}
\def\dfdw0#1{f(#1) 3 \ln(1+#1)}
\def\dfdwa#1{3 f(#1) \left( \ln (1+#1) - \frac{#1}{1+#1} \right)}
\def\dedok#1{\frac{1}{2E(#1)}\left\{ (1+#1)^2 -f(#1) \right\}}
\def\prederA#1{\frac{-3(1+#1)}{E(#1)^4}}

\def\preA#1{\frac{(1+#1)}{E(#1)^3}}
\title{Fisher Matrix Preloaded -- \name{}}

\author{Bruce A. Bassett}
\affiliation{South African Astronomical Observatory, Observatory, Cape Town, South Africa} \affiliation{Department of Mathematics and Applied Mathematics, University of Cape Town, Rondebosch, 7700, Cape Town, South Africa} \affiliation{Astrophysics and Cosmology Group, African Institute for Mathematical Sciences, 6-8 Melrose Road, Muizenberg, 7945, South Africa} \email{fisher4cast@gmail.com}
\author{Yabebal Fantaye}
\affiliation{South African Astronomical Observatory, Observatory, Cape Town, South Africa}\affiliation{Department of Mathematics and Applied Mathematics, University of Cape Town, Rondebosch, 7700, Cape Town, South Africa}
\affiliation{Department of Astrophysics, SISSA, Via Beirut 4, 34014 Trieste, Italy}
\author{Ren\'{e}e Hlozek}
\affiliation{South African Astronomical Observatory, Observatory, Cape Town, South Africa} \affiliation{Department of Mathematics and Applied Mathematics, University of Cape Town, Rondebosch, 7700, Cape Town, South Africa} \affiliation{Department of Astrophysics, Oxford University, Denys Wilkinson Building, Keble Road, OX1 3RH, United Kingdom}
\author{Jacques Kotze}
\affiliation{Department of Mathematics and Applied Mathematics, University of Cape Town, Rondebosch, 7700, Cape Town, South Africa}
\affiliation{Department of Astrophysics, Oxford University, Denys Wilkinson Building, Keble Road, OX1 3RH, United Kingdom}
\affiliation{Institute of Cosmology and Gravitation (ICG), University of Portsmouth, Dennis Sciama Building, Burnaby Road, Portsmouth PO1 3FX, UK}

\begin{abstract}
The Fisher Matrix is the backbone of modern cosmological forecasting. We describe the \name{}~software: a general-purpose, easy-to-use, Fisher Matrix framework. It is open source, rigorously designed and tested and includes a Graphical User Interface (GUI) with automated \LaTeX{}  file creation capability and {\em point-and-click} Fisher ellipse generation. \name{}~was designed for ease of extension and, although written in Matlab, is easily portable to
open-source alternatives such as Octave and Scilab. Here we use \name{}~to present new 3-D and 4-D visualisations
of the forecasting landscape and to investigate the effects of growth and curvature on future cosmological surveys.
Early releases have been available at http://www.mathworks.com/matlabcentral/fileexchange/20008 since mid 2008. The current release of the code is Version 2.2 which is described in here. For ease of reference a Quick Start guide and the code used to produce the figures in this paper are included, in the hope that it will be useful to the cosmology and wider scientific communities.
\end{abstract}

\maketitle

\section{Introduction}

The need for rapid forecasts of constraints from proposed new
surveys has played an important role in the approach of the cosmology
community to the mysteries of dark energy. Faced with an almost total
lack of understanding of the physics underlying dark energy, focus has shifted
towards designing surveys that are optimal in some information theory sense.
A general appreciation for the full parameter space is required before one can make definitive statements about its specific components; the Fisher Matrix formalism allows one to do this in a computationally inexpensive way. Despite its limitations, the Fisher Matrix formalism has become the {\em de facto} standard
in cosmology for comparing surveys and for forecasting parameter constraints\footnote{For a
list of over 100 recent astronomy/cosmology papers using the Fisher Matrix formalism see\\
http://lanl.arxiv.org/find/astro-ph/1/abs:+AND+Fisher+matrix/0/1/0/all/0/1}. It
can also be easily adapted to different scenarios and has thus informed the design and
funding of essentially all modern cosmological surveys.

Nevertheless, while the Fisher Matrix formalism is relatively simple
in principle, students can find it challenging at first, implementations are at times
buggy and usually limited -- often assuming a flat universe, for example -- and there
is unnecessary repetition of code, without a clear development
path accessible to the entire community to build on. \name{}~was initially developed in
2007 and 2008 to address these issues with the aim of providing a free, graphical
framework that would make it easy for students to learn the formalism while
also simultaneously providing a rigorous, robust and general code-base for researchers to build on and extend.
Version 2.2, released with this paper, continues with the same aims\footnote{An IDL codeset,
iCosmo, was released subsequent to \name{}~and also provides powerful routines
for cosmological Fisher Matrix analysis with a convenient web interface and
extensive online documentation as part of the {\em Initiative
for Cosmology}. While the two codes use different prescriptions, the output of iCosmo \cite{icosmo} has been compared to \name{}~and are consistent.}.

We discuss the Fisher formalism in Section~\ref{formalism} and introduce the cosmology used in \name{}~in Section~\ref{example}. The effect of including cosmic curvature as a parameter is discussed in Section~\ref{curv}~and the growth of structure as an observable in Section~\ref{growth_section}. The applications of \name{}~are outlined in Section~\ref{fisherapp} while new features in Versions 1.2 through to Version 2.2 are described in Section~\ref{newfeat}. The appendices give the general and explicit derivatives for the Fisher Matrices and the Quick Start guide for \name{}, as well as the samples of the Matlab code used to produce the figures in this paper.

\section{Forecasting and the Fisher Matrix \label{formalism}}

Forecasting of survey constraints for a proposed survey can be achieved either through full Monte Carlo simulations of the survey which are typically time consuming and computationally intensive, or by using the much simpler and quicker Fisher Matrix technique, which we now discuss in detail and which forms the basis of \name{}.  The Fisher Matrix translates errors on observed quantities measured directly in the survey into constraints on parameters of interest in the underlying model. Put more directly, it is the elegant way of doing propagation of errors in the case of multiple, correlated, measurements and many parameters \cite{tegmark}. As an example, consider an arbitrary function $y = f(z,\theta)$ of some parameter $\theta$ and an independent variable $z$. Assuming a perfect measurement of $z$, the error $\delta \theta$, for a given measured $\delta y$ is, by simple calculus, $\delta \theta = (\frac{\partial f}{\partial \theta})^{-1} \delta y$, or equivalently  $(\delta \theta)^{-2} = (\frac{\partial f}{\partial \theta})^2 (\delta y)^{-2}$. This is perhaps the simplest example of a Fisher Matrix; with a single element, cf. Eq~(\ref{fishereq}). Here $\theta$ represents the parameter we want to measure and $f$ the observable quantity\footnote{Technically, $H, d_A$ and $G$ are derived quantities, since we are only able to observe photons with telescopes, but from here onwards we treat them as `observables' as we quantify the relationship between these derived observational quantities and the underlying cosmology, which is the aim of Fisher Matrix analysis.} (e.g. $H(z)$).

In more generality, the Fisher Matrix formalism predicts the constraints on a vector of parameters ${\boldsymbol \theta} = (\theta_{\mathrm{1}}, \theta_{\mathrm{2}},...,\theta_\mathrm{A},.. )$ - such as the coefficients in a parameterisation of dark energy - resulting from measurements of one or more observables ${\bf X^\alpha}= {\bf X}^\alpha({\boldsymbol \theta,\bf z})$ (such as $H(\boldsymbol \theta,\bf z)$ or $d_A(\boldsymbol \theta,\bf z)$), each at a range of redshifts, ${\bf z} = (z_1,z_2,...,z_i,..)$ e.g. in a Baryon Acoustic Oscillation (BAO) survey one might measure  $H(\boldsymbol \theta,\bf z)$ and $d_A(\boldsymbol \theta,\bf z)$ at a single redshift, while a Type Ia supernova (SNIa) survey may measure $d_L(\boldsymbol \theta,\bf z)$ at hundreds of redshifts. There are therefore three indices in general to keep track of, ($\mathrm{A},\alpha,i$) corresponding to parameter, observable and redshift. The number of observables is arbitrary and combining results from independent observables is essentially trivial, hence we will often suppress observable index, ${\bf X^\alpha} = {\bf X}$ for simplicity. Boldface indicates the entire vector, either of parameters ${\boldsymbol \theta},$ observables ${\bf X}$ or redshifts ${\bf z}$.

The Fisher Matrix estimates not only the individual errors on the parameters, ${\boldsymbol \theta}$, evaluated at a given input/base/fiducial model ${\boldsymbol \theta = \boldsymbol \theta^*}$, but also the correlations between them, leading to the characteristic Fisher error ellipsoids (ellipses if one considers only pairs of $\theta_A$). To make this clear, consider the likelihood, $\mathscr{L} = P({\bf d}|{\boldsymbol \theta}),$ for a general survey, which gives the conditional probability of observing the data ${\bf d} = (d_1,d_2,...,d_i,..)$ assuming the cosmological model ${\boldsymbol \theta}$ is correct. We can expand the likelihood around the fiducial model:
\bea
\ln\mathscr{L}({\boldsymbol \theta^*}+ \delta{\boldsymbol \theta}) &=& \ln\mathscr{L}({\boldsymbol \theta^*}) + \sum_{A}\left.\frac{\partial\ln\mathscr{L}({\boldsymbol \theta})}{\partial_{\mathrm{A}}}\right|_{ {\boldsymbol \theta} = {\boldsymbol \theta^*}} \delta\theta_\mathrm{A} \nonumber \\
&&~~~~~~~~~~~~~~+ \frac{1}{2}\sum_{\mathrm{AB}} \left.\frac{\partial^2\ln \mathscr{L}({\boldsymbol \theta})}{\partial_{\mathrm{A}}\partial_{\mathrm{B}}}\right|_{{\boldsymbol \theta}={\boldsymbol \theta^*}} \delta\theta_\mathrm{A}\delta\theta_\mathrm{B} \nonumber \\
&&~~~~~~~~~~~~~~~~+\frac{1}{6}\sum_{\mathrm{ABD}}\left.\frac{\partial^3\ln\mathscr{L}({\boldsymbol \theta})}{{\partial_{\mathrm{A}}\partial_{\mathrm{B}}\partial_{\mathrm{D}}}}\right|_{\boldsymbol \theta=\boldsymbol\theta^*}\delta\theta_\mathrm{A}\delta\theta_\mathrm{B}\delta\theta_\mathrm{D} \nonumber \\ &&~~~~~~~~~~~~~~~~~~~~~+\frac{1}{12}\sum_{\mathrm{ABDE}}\left.\frac{\partial^4\ln\mathscr{L}({\boldsymbol \theta})}{{\partial_{\mathrm{A}}\partial_{\mathrm{B}}\partial_{\mathrm{D}}\partial_{\mathrm{E}}}}\right|_{\boldsymbol\theta=\boldsymbol\theta^*}\delta\theta_\mathrm{A}\delta\theta_\mathrm{B}\delta\theta_\mathrm{D}\delta\theta_\mathrm{E} \, + ... \, ,
\label{expansion}
\eea
where $\partial_{A}\equiv \partial \theta_\mathrm{A}$ represents the partial derivatives with respect to the parameter $\theta_\mathrm{A}$. The first term in the expansion is a constant depending on the fiducial model. The fiducial model is expected (after averaging over many data realisations) to be the point of maximum likelihood, hence the first derivative of the likelihood vanishes. The third term is the curvature matrix/Hessian of the likelihood, and is the term used in the Fisher Matrix which is formally defined as the expectation value of the derivatives of the log of the likelihood with respect to the parameters ${\boldsymbol \theta}$, or
\begin{equation}
F_{\mathrm{AB}} = -\left\langle \frac{\partial^2 \ln \mathscr{L}}{\partial \theta_{\mathrm{A}}\partial \theta_{\mathrm{B}}}\right\rangle\,.
\label{fab_lnl}
\end{equation}
Here the angle brackets indicate the expectation value which, for an arbitrary function $g(x)$, is defined to be $\langle g(X)\rangle \equiv \int_{-\infty}^\infty g(x) f_X(x)dx$ with $f_X(x)$ the probability distribution function of the random variable $x$, which here is the noise on the data, assumed to be Gaussian of mean zero.

\begin{figure}[htbp!]
\begin{center}
$\begin{array}{@{\hspace{-0.15in}}l@{\hspace{-0.3in}}c}
\epsfxsize=3.5in
	\epsffile{\fig_dir{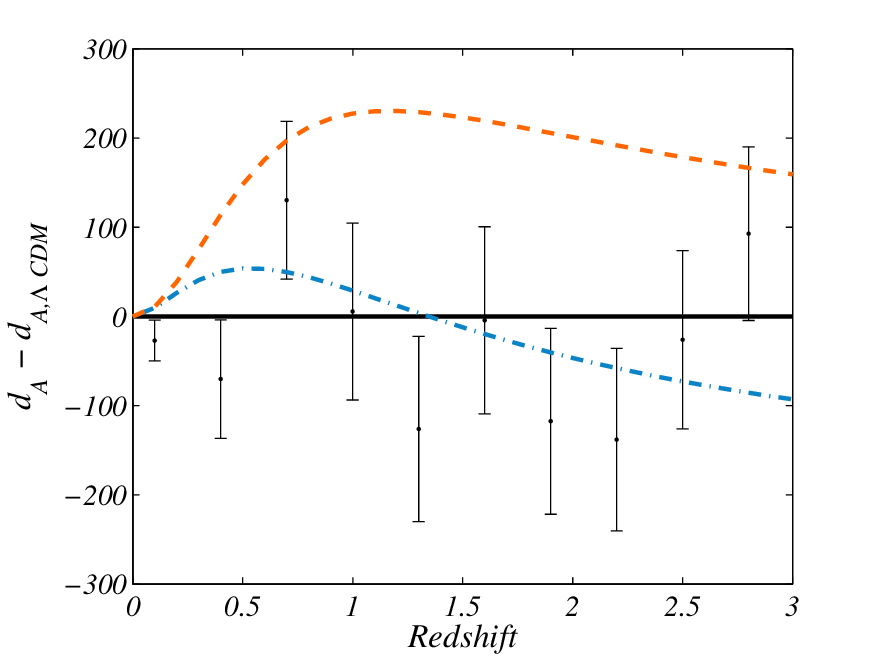}} &
\epsfxsize=3.5in
	\epsffile{\fig_dir{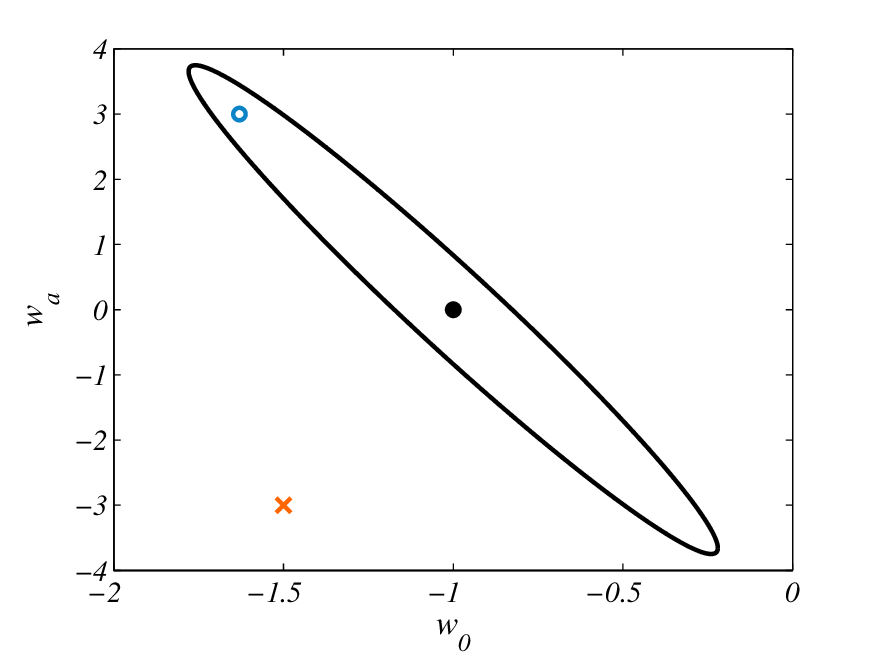}} \\ [0.0cm]
 \end{array}$
\caption{{\bf The link between model curves, data and error ellipses.} Parameter values inside the ellipse are a good fit to the data, such as the blue dot-dashed curve (left) corresponding to the open blue circle (right). The horizontal line $d_A - d_{A,\Lambda\mathrm{CDM}} = 0$ corresponds to the fiducial model on the right $(w_0=1,w_a=0)$. A parameter value outside the ellipse is not a good fit to the data however, as illustrated by the orange dashed line (left) and orange cross (right). The ellipse and simulated error bars in the two panels both correspond to a nominal survey measuring the angular diameter distance, $d_A(z)$, with $6\%$ error in ten bins between $0.1<z<3$, with $\Lambda$CDM assumed to be the correct model in both cases. \label{intuitive_fm}}
\end{center}
\end{figure}
The likelihood for a given observable ${\bf X}$ is expressed in terms of the theoretical value of the observable $X_i$ evaluated at the redshifts $z_i$ and data for that specific observable $d_i$ as $\mathscr{L} \propto \exp(-{\bf \Delta}^T {\bf C}^{-1} {\bf \Delta}/2)$ where ${\bf \Delta} \equiv {\bf X - d}$, generalising the usual chi-squared statistic relating the theory by allowing for a general data covariance matrix ${\bf C}$. Substituting the above expression into Eq.~(\ref{fab_lnl}) converts the equation from derivatives of the likelihood itself into a sum over derivatives of the {\em observable} ${\bf X}$ with respect to the parameter $\theta_\mathrm{A}$:
\bea
F_{\mathrm{AB}} &=& \frac{\partial {\bf X}}{\partial \theta_\mathrm{A}}^{\mathrm{T}} {\bf C}^{-1} \frac{\partial {\bf X}}{\partial \theta_\mathrm{B}} + \frac{1}{2}\mathrm{Tr} \left( {\bf C}^{-1} \frac{\partial {\bf C}}{\partial \theta_\mathrm{A}}{\bf C}^{-1} \frac{\partial {\bf C}}{\partial \theta_\mathrm{B}}\right) \label{fishereq}\\
&=& \sum_{i} \frac{1}{\sigma^2_i}\frac{\partial {X}}{\partial \theta_\mathrm{A}}(z_i)\frac{\partial {X}}{\partial \theta_\mathrm{B}}(z_i)
\nonumber
\eea
where ${\partial {\bf C}}/{\partial \theta_\mathrm{A}}$ is the derivative of the data covariance matrix with respect to the parameter $\theta_\mathrm{A}$ which is assumed to vanish in the second equality implying that the data errors are independent of cosmological parameters. This is often the case, e.g. the errors on measurements of Type Ia supernova (SNIa) flux are independent of the dark energy parameters $w_0, w_a$ (of the Chevallier-Polarski-Linder (CPL) parameterisation \cite{cp,linder_w}, described in Section~\ref{example}) to good accuracy\footnote{In cases where the mean is zero (such as CMB analyses), however, the dependence of the covariance on the parameters can no longer be ignored, see for example \cite{tegmark}.}. The second equality also requires that the data are uncorrelated, in which case $\mathrm{{\bf C}}$ is diagonal with entries $\sigma_i^2$, with the $\sigma_i$ the $1-\sigma$ error on the $i$-th data point.

In the case where we have multiple independent measurements of different observables ${\bf X}^\alpha$ (e.g. $H(z)$ and $d_A(z)$), the total Fisher Matrix is just the sum of the individual Fisher matrices indexed by $\alpha$. Similarly, if we have independent prior information, this is encoded in a prior matrix between the cosmological parameters. In this paper we will refer to the prior on a single parameter $\theta_\mathrm{A}$ as $\mathrm{Prior}(\theta_\mathrm{A}) = (\Delta \theta_\mathrm{A})^{-2}$, where $\Delta \theta_\mathrm{A}$ is the uncertainty on the parameter as measured from prior surveys, see e.g. Figure~(\ref{curvature_plot}). In the case where the different measurements are not independent, they must be combined with the suitable data covariance matrix. The inverse of the Fisher Matrix, $F^{-1}_{\mathrm{AB}}$, provides an estimate of the {\em error covariance matrix} for the parameters $\theta_\mathrm{A},$ as we now expand upon.

\begin{figure}[htbp!]
\centering
\includegraphics[width = 4in]{\fig_dir{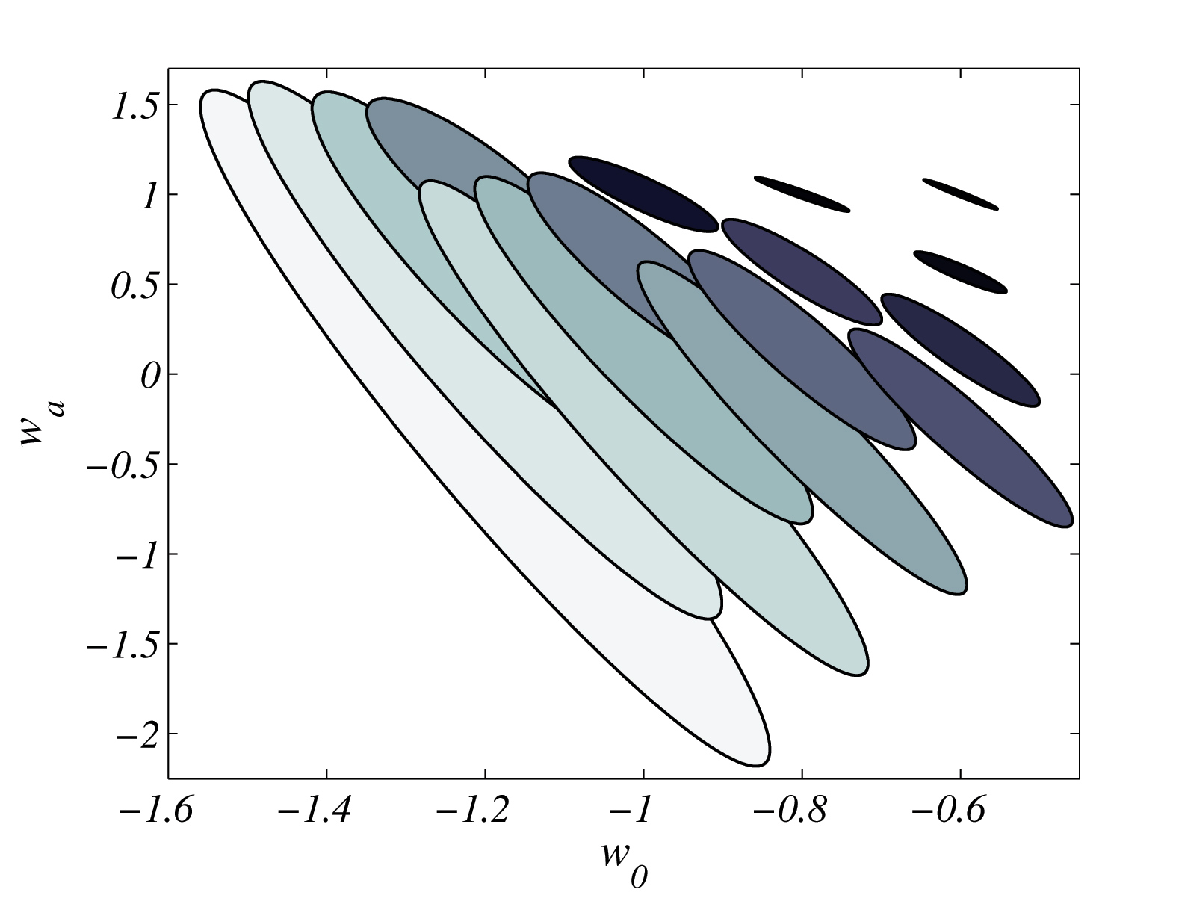}}
\caption{{\bf Varying $w_0$ and $w_a$ of the fiducial model --} the generated ellipses for a measurements of the Hubble parameter $H(z)$ and angular diameter distance $d_A(z)$ survey characterised in Table~\ref{table_seo}. The values for the coefficients in the CPL parameterisation for $w(z)$, $w_0$ and $w_a$, have been varied on a grid over $-1.3 < w_0 < -0.6, -0.7 < w_a < 1 $. As $w_0,w_a$ change, not only does the ellipse centre shift, but the size of the ellipse changes, as well as the slope of the degeneracy direction between the two parameters. Since the Dark Energy Task Force Figure of Merit is linked to the inverse of the area of the ellipse the value of the FoM increases for with increasing $w_0$ and $w_a$. \label{fig:wa_w0_vary_ellipse}}
\end{figure}
In general one often considers a $p+n$-dimensional likelihood which includes not only the $p$ parameters of interest but an additional $n$ nuisance parameters that form a natural part of the problem but are of no direct interest (such as $H_0$ in studies of dark energy dynamics). As a result we are usually only interested in the posterior $\mathscr{P}({\boldsymbol \theta}) = \mathscr{L}({\boldsymbol \theta}) \Pi({\boldsymbol \theta})$ as a function of the $p$ key parameters, which is obtained by marginalising over the nuisance parameters, {\em viz.}
\be
\mathscr{P}(\theta_1,...\theta_p) = \int_{-\infty}^{\infty} \mathscr{L}(\theta_1,..,\theta_p,..,\theta_{p+n}) \Pi(\theta_1,...\theta_{p+n}) d\theta_{p+1}...d\theta_{p+n}. \label{marg},
\ee

where $\Pi({\boldsymbol \theta})$ is the prior on the parameter vector ${\boldsymbol \theta}$.
For a likelihood of arbitrary shape marginalisation must be performed numerically and is well-suited to Markov Chain Monte Carlo (MCMC) methods but in the special case where it is a multivariate Gaussian, the marginalisation can be performed analytically and simply in terms of the Fisher Matrix. Let us write the full Fisher Matrix in terms of sub-matrices, as
\be
F=\begin{pmatrix}
\Theta&{\cal O}\\
{\cal O}^{\mathrm{T}}&{\cal N}
\end{pmatrix},\ee
where $\Theta$ is the $p\times p$ sub-matrix corresponding to the parameters of interest, ${\cal O}$ is an $p\times n$ matrix describing the correlation between the nuisance parameters and the parameters of interest\footnote{If ${\cal O} = 0$ then the nuisance parameters have no impact on $\Theta$.}
and ${\cal N}$ is the $n \times n$ matrix representing the nuisance parameters we wish to marginalise over. The marginalised Fisher Matrix for the parameters of interest is then given by \cite{matsubara}:
\be
\widetilde{F} = \Theta - {\cal O} {\cal N}^{-1} {\cal O}^{\mathrm{T}}, \label{marg_mat}
\ee
where the first term is the matrix of the reduced parameter space of interest, the second term encodes the effects of the marginalisation over the other nuisance parameters and ${}^T$ represents matrix transpose.

As we mentioned before, the inverse of the Fisher Matrix provides an estimate of the parameter covariance matrix. For an unbiased estimator (that is one whose expected value of ${\boldsymbol \theta}$ is equal to the fiducial model ${\boldsymbol \theta^*}$ assumed to be correct), and in the case where one does not marginalise over any other parameters (i.e. we consider all other parameters perfectly known), the expected error on any parameter $\theta_\mathrm{A}$ satisfies the Cram\'{e}r-Rao bound\footnote{For a proof see e.g. http://en.wikipedia.org/wiki/CRLB or e.g. p. 426 in \cite{leon}}
\be
\Delta \theta_\mathrm{A} \geq \frac{1}{\sqrt{F_{\mathrm{AA}}}}\,,
\ee
while in the more realistic case that one wants to marginalise over all the other parameters in the problem, the bound becomes
\be
\Delta \theta_{\mathrm{A}} \geq {\sqrt{({\bf F}^{-1})_{\mathrm{AA}}}}
\ee
i.e. one first inverts the Fisher Matrix, then takes the $\mathrm{AA}$ component of the resulting matrix. One can show that the latter is always greater than or equal to the former, i.e. marginalisation cannot decrease the error on a parameter, and only has no effect if all other parameters are completely uncorrelated from the parameter of interest.  Note that in the case where the likelihood is exactly Gaussian in the parameters, the Cram\'{e}r-Rao bound becomes an equality and not just a lower bound.

Since Fisher Matrix analysis assumes the likelihood is a multivariate Gaussian (an approximation that can be made arbitrarily good by considering better and better surveys), contours of constant probability are ellipsoids within the Fisher formalism. These ellipsoids (ellipses for two parameters) are given by solving the equation
\be
{{\bf \Delta \boldsymbol \theta}}^\mathrm{T} \widetilde{F} {\bf \Delta \boldsymbol\theta} = \beta
\label{ellipseeq}
\ee
where ${\bf \Delta \boldsymbol\theta} = {\boldsymbol \theta - \boldsymbol\theta^*}$ is the parameter vector around the fiducial model, ${\boldsymbol \theta^*}$, and $\beta$ is a constant determined by the desired confidence level and the number of parameters. For two parameters, the $1$ and $2-\sigma$ contour levels correspond to $\beta = 2.31$ and $6.17$ respectively \cite{numerical_recipes, matsubara}. The \name{}~GUI allows plotting of both $1$- and $2$-dimensional contours and hence always marginalises the full $p+n$ -dimensional Fisher Matrix to achieve this. Marginalisation over some or all of the other parameters can be effectively switched off by making the corresponding diagonal elements of the prior matrix very large.
The Fisher Matrix and the corresponding ellipses provide the Gaussian estimate for how well the parameters of the model will be constrained by a given experiment assuming the true model is that at which the Fisher Matrix was evaluated (e.g. $\Lambda$CDM). This is illustrated in Figure~(\ref{intuitive_fm}), which shows the $1-\sigma$ error ellipse around the fiducial $\Lambda$CDM model with the coefficients in the Chevallier-Polarski-Linder (CPL) \cite{cp, linder_w} (see Eq.~(\ref{wparam})) parameterisation $(w_0, w_a) = (-1,0)$, for a survey consisting of measurements of the angular diameter distance between redshifts of 0.1 and 3. Values of $w_0, w_a$ inside this ellipse will have expected likelihoods that differ from the fiducial model by less than $1-\sigma$. The Fisher Matrix allows us to estimate which sets of parameter values we will be able to rule out at a given significance level if the fiducial cosmological model is correct. We will see in Section~\ref{fisherapp} (see Figure~(\ref{fig:wa_w0_vary_ellipse})) that changing the assumed fiducial/base model has a big effect on the ellipses for the same survey.

While the ellipses provide significant insight they do not allow immediate comparison between different surveys, a feature required if one wants to optimise or compare surveys head-to-head \cite{huterer, bruceipso, correct_detf, parkinson06, trotta07, design,parkinson,parkinson09}. One common way to perform such comparison is to formulate a Figure of Merit which ascribes a single real number to each survey, the simplest of which is to use the volume of the ellipsoid or of the marginalised ellipse. The volume of the $n$-dimensional error ellipsoid (corresponding to an $n$-dimensional Fisher Matrix) is:
\be
\mathrm{Vol_n} = \mathrm{V_{S^n}}\times \left(\frac{\beta}{\mathrm{det}({\bf F})}\right)^{1/2}\,,
\label{vol}
\ee
where $\beta$ is defined in Eq~(\ref{ellipseeq}), $\mathrm{V_{S^n}} = \pi^{n/2}/\Gamma(\frac{n}{2}+1)$ is the volume of the $n$-dimensional unit sphere and $\Gamma(u)$ is the Gamma function. For the interesting case $n = 2$, $\mathrm{V_{S^2}} = \pi$ is, of course, the area of the unit circle. Note that it is common in the literature to ignore the $\mathrm{V_{S^n}}$ and $\beta$ factors and to incorrectly refer to the determinant factor alone as the area or volume of the ellipse/ellipsoid.

In fact, since the Fisher Matrix is a metric, the square root of the determinant is a natural volume element providing the Jacobian for the action of the linear mapping induced by the Fisher Matrix, i.e. one should think of the Fisher Matrix as inducing a linear mapping rather than `being' an ellipse itself.

\name{}~includes the standard FoMs as well as some new ones available through the GUI and command line. Those using the volume are based on Eq~(\ref{vol}) with $n=2$. Although some of the Figures of Merit are only defined for the error ellipse in the $w_0-w_a$ plane, where $w_0, w_a$ are the coefficients in the CPL \cite{cp, linder_w} parameterisation of the equation of state of dark energy (see for e.g. \cite{correct_detf}), the FoMs in \name{}~are calculated by the code for any pair of cosmological parameters being considered rather than the full 5-D matrix. We briefly outline the FoMs used in \name{}:
\begin{itemize}
\item{DETF}\\
This Figure of Merit in the Report of the Dark Energy Task Force \cite{correct_detf} is defined as $\mathrm{det}(F^{1/2})$, which is the inverse of the $1-\sigma$ ellipse in the $w_0-w_a$ plane of the CPL dark energy parameterisation \cite{cp, linder_w}, in units of the area of the unit circle. {Given that one requires the ellipse to be as small as possible, this FoM increases monotonically for survey configurations that best constrain the dark energy parameters $w_0, w_a$.}
\item{Area$^{-1}_{1-\sigma}$}\\
This Figure of Merit is the reciprocal of the $1-\sigma$ error ellipse area in the parameter plane currently plotted, i.e. $\mathrm{det}(F^{1/2})/(\pi\sqrt{2.31}).$ While this FoM has a similar definition as the DETF FoM (and differs by a constant factor for the $w_0-w_a$ example) it is however defined for all parameter combinations, while the DETF FoM is defined specifically for the dark energy parameters.
\item{Area$_{1-\sigma}$}\\
Simply the inverse of the previous FoM and hence is smaller for surveys which constrain the specific 2-parameter combination more tightly.
\item{Tr${\mathbf{C}}$}\\
This FoM is defined as the trace of the covariance matrix of the parameters, $\mathbf{C} = \mathbf{F^{-1}}$, estimated as the inverse of the marginalised Fisher Matrix. This FoM is simply the sum of the squares of the marginalised errors on each parameter. {As the errors decrease on the parameters (for more optimal surveys), the FoM decreases in magnitude.}
\item{$\Sigma_{AB} C_{AB}^2$} \\
This FoM is defined as the sum of the squares of the entries of the whole covariance matrix, $\mathbf{C} = \mathbf{F^{-1}}$. Unlike the previous definition this FoM is sensitive to the off-diagonal components of the covariance matrix as well as the diagonal components. {This FoM also decreases in magnitude for surveys that reduce the error on (and the correlation between) the parameters}
\end{itemize}
Designing a FoM which is both robust and easy to use, while not favouring a particular model space or paradigm has been the topic of recent work \cite{albrecht_fomswg}. Other suggestions have been proposed, such as using principal component analysis \cite{huterer_pcs}, as the basis of an effective FoM to compare survey efficiency We include a module which plots the principal components from standard survey configurations in the latest release - this module is discussed in Section \ref{newfeat}.

Finally we note that in addition to the Fisher Matrix, one can include the non-Gaussian terms in the expansion of the log likelihood (Eq.~(\ref{expansion})). Study of these {\em flex} corrections is left to future work. 

\section{The Cosmology of Hubble, Distance and Growth \label{example}}
Although \name{}~is a completely general Fisher Matrix framework at the command-line level, the GUI is coded as a cosmology interface, since this is its primary application. In the context of modern cosmological surveys, the primary observables are the expansion rate of the Universe, measured through the Hubble rate $H(z),$ cosmological distances such as the angular diameter distance, $d_A(z),$ and the growing mode of dark matter density perturbations, $\delta({\bf x},z) \propto G(z)$. $H(z)$ and $d_A(z)$ are provided by for e.g. BAO surveys while growth can be measured using lensing or number count surveys and potentially also the BAO (see \cite{rassat_bao} for a review of methods to obtain the BAO) if the bias is measured independently (e.g. through redshift distortions - see \cite{zhang, percival_white}). {We do not assume that these measurements come {\it a priori} from BAO surveys, and there is nothing specific do BAO in the formulation of \name{}. The additional modules provided within \name{}~provide assistance in calculating some of the required quantities, such as $H(z), d_A(z)$ from input survey characteristics, but the code does not require these modules to run, and can be used for analysing any Fisher Matrix data.}

The \name{}~GUI uses the observables $H, d_A$ and $G$ in a general Friedmann-Lema\^{i}tre-Robertson-Walker (FLRW) universe. The cosmic parameters assumed for the GUI are $(H_0, \Omega_m, \Omega_k, w_0, w_a),$ where $H_0$ is the value of the Hubble constant in kms$^{-1}$Mpc$^{-1}$, $\Omega_m$ is the energy density of matter today in units of the critical density, $\Omega_k$ is the curvature energy density ($\Omega_\mathrm{DE} = 1 - \Omega_m - \Omega_k$)\footnote{Unless explicitly indicated elsewhere, references in this paper to $\Omega_i$ will mean the current value of the density parameter and not its value as a function of time.} and $w_0, w_a$ are the coefficients in the CPL expansion of the dark energy equation of state \cite{cp, linder_w}:
\begin{equation}
w(z) = w_0 + w_a\frac{z}{1+z} = w_0 + w_a\left(1-\frac{a}{a_0}\right)\label{wparam}\,,
\end{equation}
where $a_0 = c/(H_0\sqrt{|\Omega_k|})$ is the curvature radius of the cosmos \cite{peebles1993}. Other dark energy expansions can be easily accommodated in \name{}~by changing the appropriate input functions.
{In the context of a FLRW universe, these parameters are sufficient to explore the space of available models. We choose to parameterise the Hubble constant separately as this reduces degeneracies in the parameter space, however if one was considering measurements of the Cosmic Microwave Background (for example), other cosmological parameters would be included.}
\begin{figure*}[htbp!]
$\begin{array}{@{\hspace{-0.2in}}l@{\hspace{-0.15in}}l@{\hspace{-0.15in}}l}
	\epsfxsize=2.4in
	\epsffile{\fig_dir{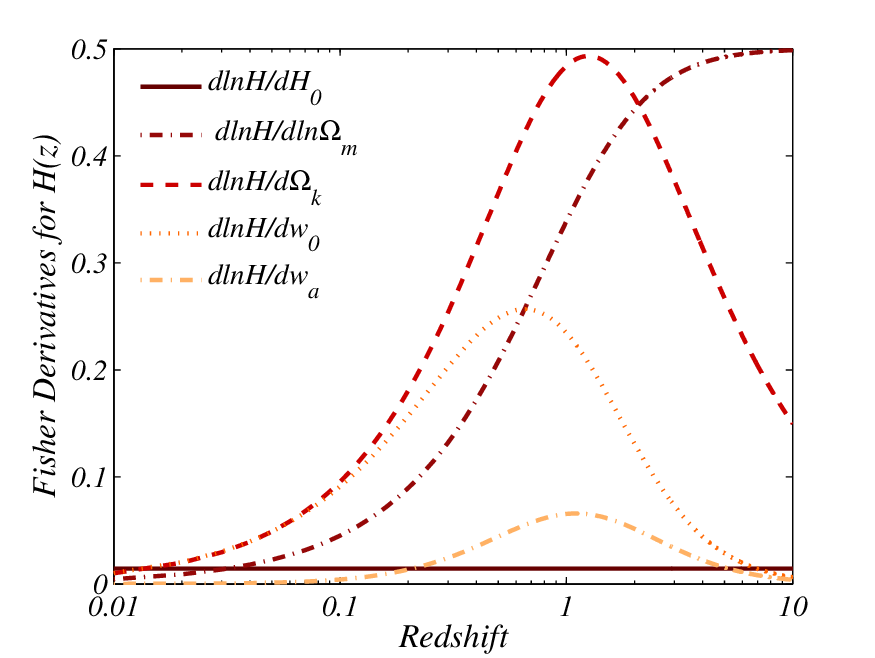}} &
	\epsfxsize=2.4in
	\epsffile{\fig_dir{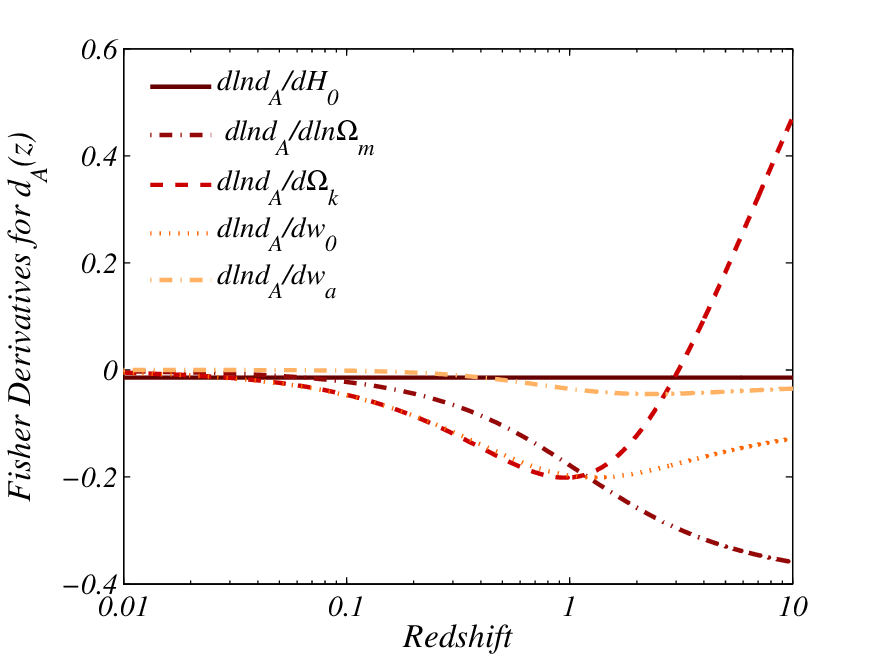}}  &
    \epsfxsize=2.4in
    \epsffile{\fig_dir{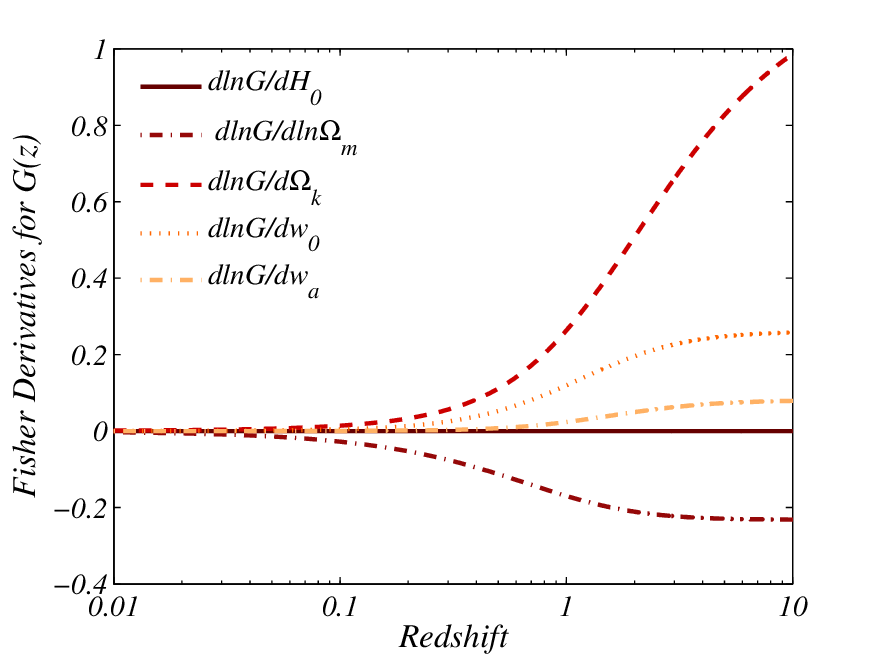}} \\ [0.0cm]
 \end{array}$
\caption{{\bf Derivatives} for the Hubble parameter (left panel), angular diameter distance (middle panel) and growth function (right panel) are shown for the parameters considered in our cosmological example: $H_0$ ({solid dark red line}), $\ln \Omega_m$ ({dot-dashed red line}), $\Omega_k$ ({dashed dark orange line}), $w_0$ ({dotted orange line}) and $w_a$ ({dot-dashed peach line}). The full set of analytical derivatives of the $H(z)$ and $d_A(z)$ are found in Appendix~\ref{app_deriv}. In the case of the growth function $G(z)$ the derivatives of the solution to Eq.~(\ref{newg}) are taken numerically, using  double-sided central derivatives. The full procedure is outlined in the Users Manual for \name{}.\label{derivs}}
\end{figure*}
The expansion history of a FLRW universe is described by the Hubble parameter:
\begin{equation}
H^2(z) = H_0^2 E^2(z) \equiv H_0^2 \left(\om (1+z)^{3} + \ok(1+z)^{2} +(1-\om-\ok) f(z, w_0, w_a)\right),
\label{eeq}
\end{equation}
with the evolution of the dark energy density, $\rho_\mathrm{DE}(z) \propto f(z)$ determined by
\begin{equation}
f(z) =\exp \left(3 \int_0^z \frac{1+w(z')}{1+z'}dz' \right) = (1+z)^{3(1+w_0+w_a)} \exp \left\{-3w_a \frac{z}{1+z}\right\}\, \label{feq}
\end{equation}
where the last equality is specifically assuming the CPL parameterisation.

The angular diameter distance, $d_A(z)$ relates the angular size of an object to its known length, providing a measure of the distance to the object, and is given by:
\begin{equation}
d_A(z) = \frac{1}{1+z} \frac{c}{H_0 \sqrt{\ok}}\sinh \( \sqrt{\ok} \chi(z) \),
\label{daeq}
\end{equation}
where
$ \chi(z) \equiv \int_0^z \frac{dz'}{E(z')}, \label{chiz} $
and $E(z)$ is as defined in Eq.~(\ref{eeq}). These forms are valid for all values of $\Omega_k$ via continuity and the trigonometric identity $\sinh(ix) = i\sin(x)$. The often-used equation for the angular diameter distance contains three equations, depending on the sign and magnitude of $\Omega_k$, however this is redundant, at least conceptually. In numerical analysis we use the Taylor series expansion for very small $\Omega_k$, Eq.~(\ref{taylorok}),
\bea
 \left.\frac{\p d_A(z)}{\p \ok}\right|_{\ok \rightarrow 0} &=& \frac{c}{H_0}\frac{1}{1+z}\left\{\frac{1}{6}\chi^3(z,0) + \frac{\p \chi(z,0)}{\p \ok}\right\}
 \eea
where $ X(z,0)\equiv \left.X(z)\right|_{\ok\rightarrow 0}$ are the functions (for example $E(z),~\chi(z)$) assuming flatness.

Finally we discuss the governing equation for the growth of structure, a potentially powerful probe of dark energy \cite{correct_detf, amendola, amendola07, loeb_growth, wang_growth,linder_growth05, linder_growth09,Zhan_08_growthDE, Lee_09_growthDE,Wei_08_growthDE,Bertschinger_08_growthDE}. In general one needs to solve the differential equation for the perturbations in the matter density $\delta$ (assuming the pressure and pressure perturbations of the matter are zero - $p=\delta p=0$) \cite{peebles1993,wang_steinhardt,linder_jenkins}:
\begin{equation}
\ddot{\delta} + 2H\dot{\delta} = 4\pi G\rho_m \delta \,.
\label{deltaeq}
\end{equation}
We discuss this equation in the context of curved universes with dynamical dark energy in Section~\ref{growth_section}. \name{}~takes as input constraints on the growth $G(z)$ which provides the temporal evolution of density perturbations, i.e. $\delta({\bf x},z) \propto G(z)$.
\begin{figure}[h]
\begin{flushleft}
\begin{tabular*}{\textwidth}{l l l l l }
 \includegraphics[width=1.35in]{\fig_dir{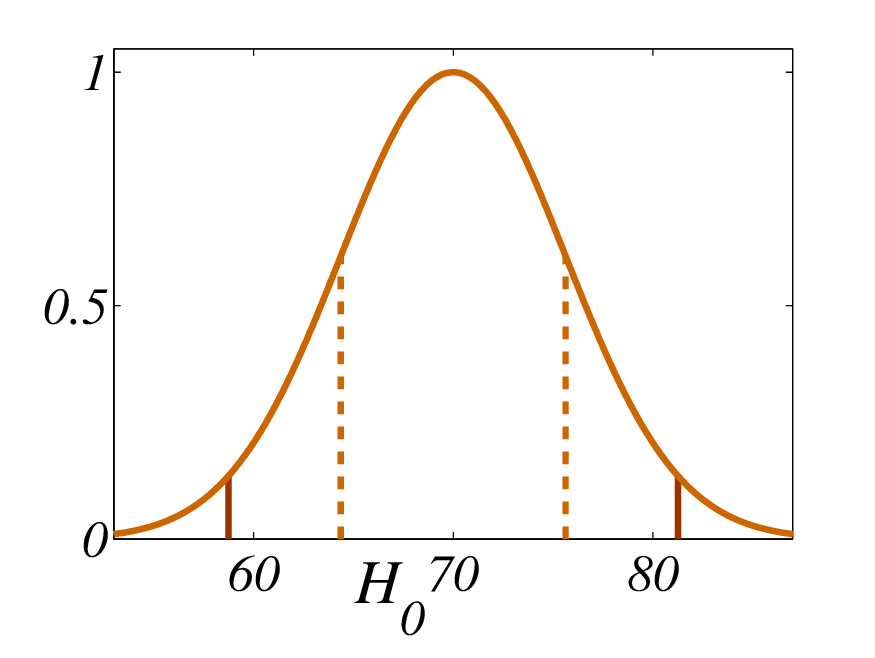}}& & & & \\
 \includegraphics[width=1.35in]{\fig_dir{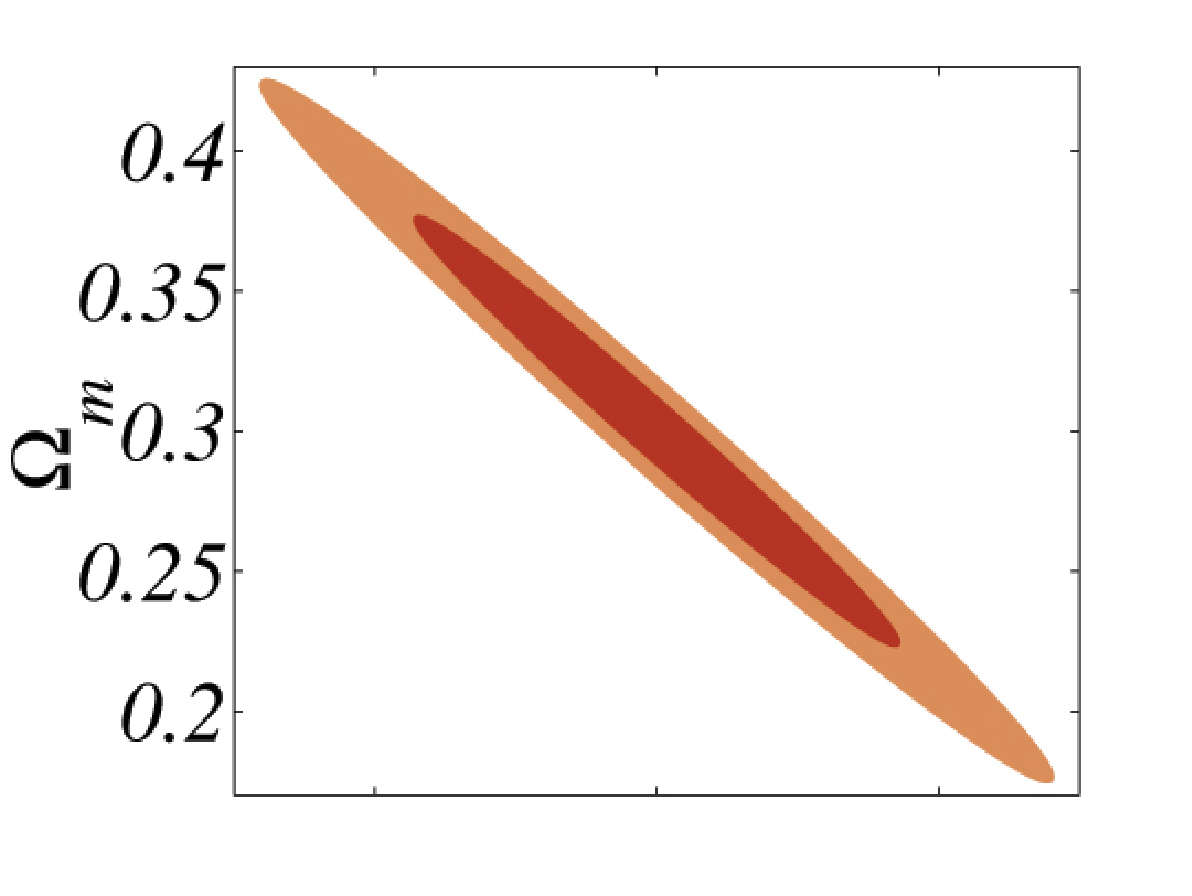}}&
  \includegraphics[width=1.35in]{\fig_dir{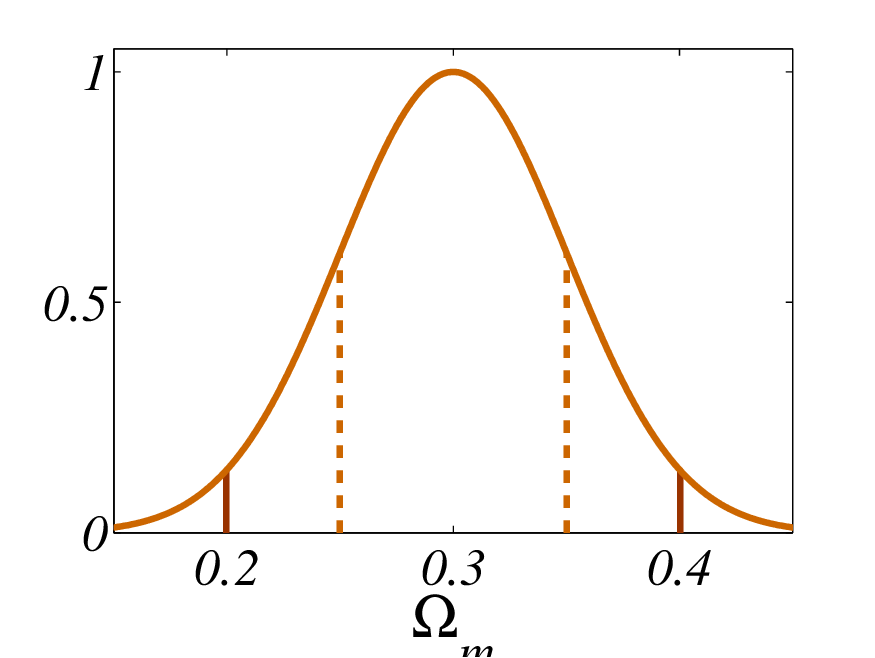}}& &&\\
  \includegraphics[width=1.35in]{\fig_dir{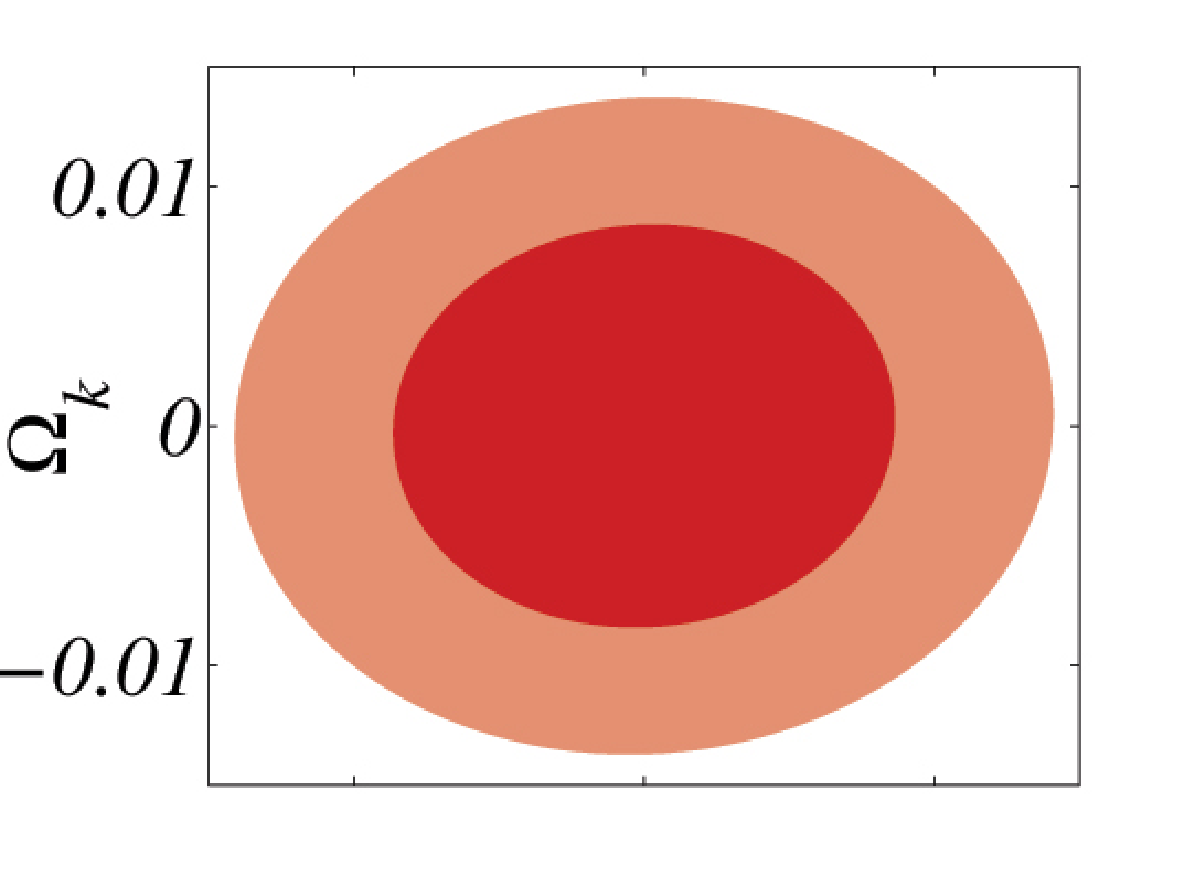}}&
   \includegraphics[width=1.35in]{\fig_dir{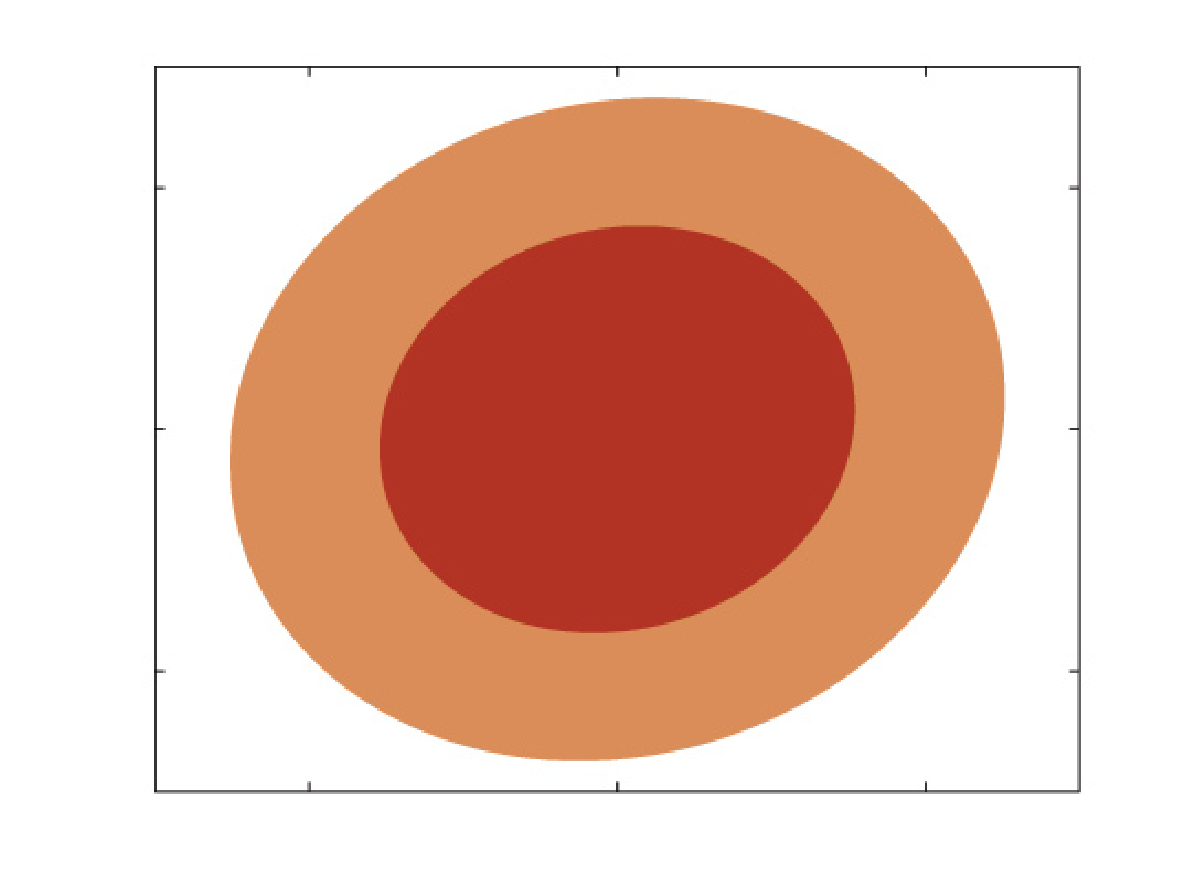}}&
    \includegraphics[width=1.35in]{\fig_dir{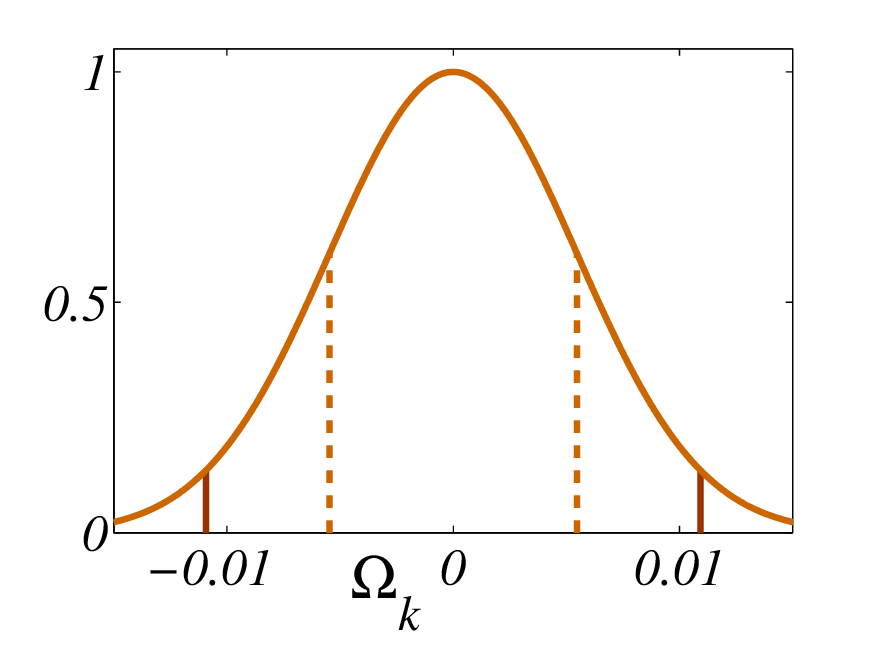}}&&\\
\includegraphics[width=1.35in]{\fig_dir{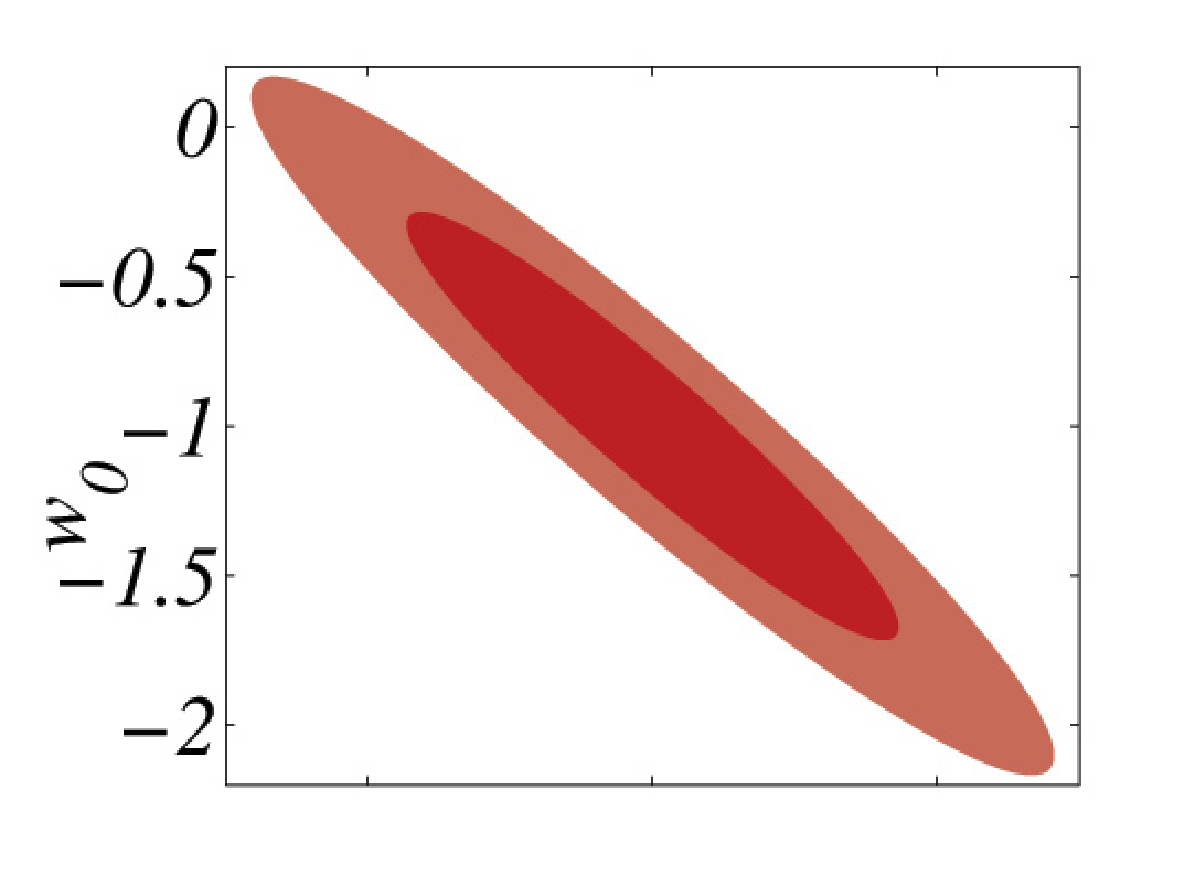}}&
\includegraphics[width=1.35in]{\fig_dir{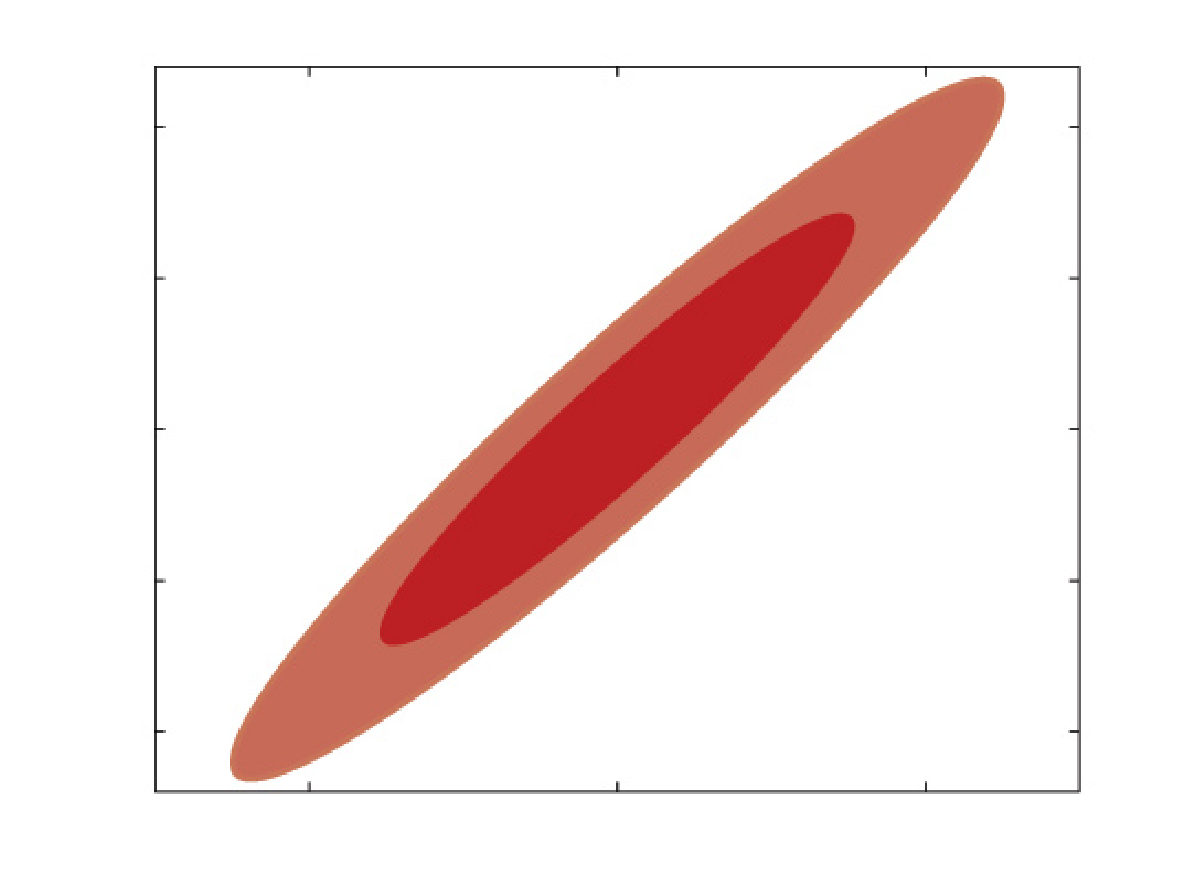}}&
\includegraphics[width=1.35in]{\fig_dir{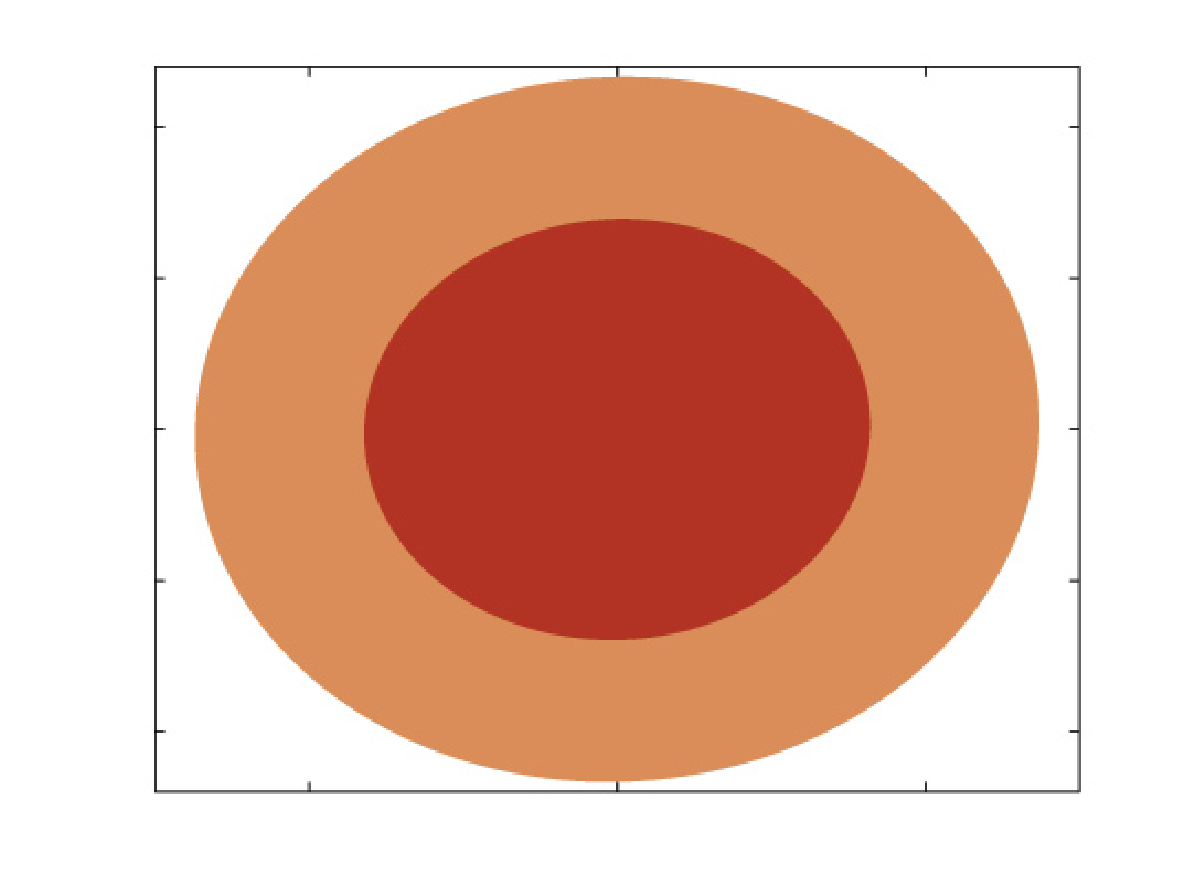}}&
 \includegraphics[width=1.35in]{\fig_dir{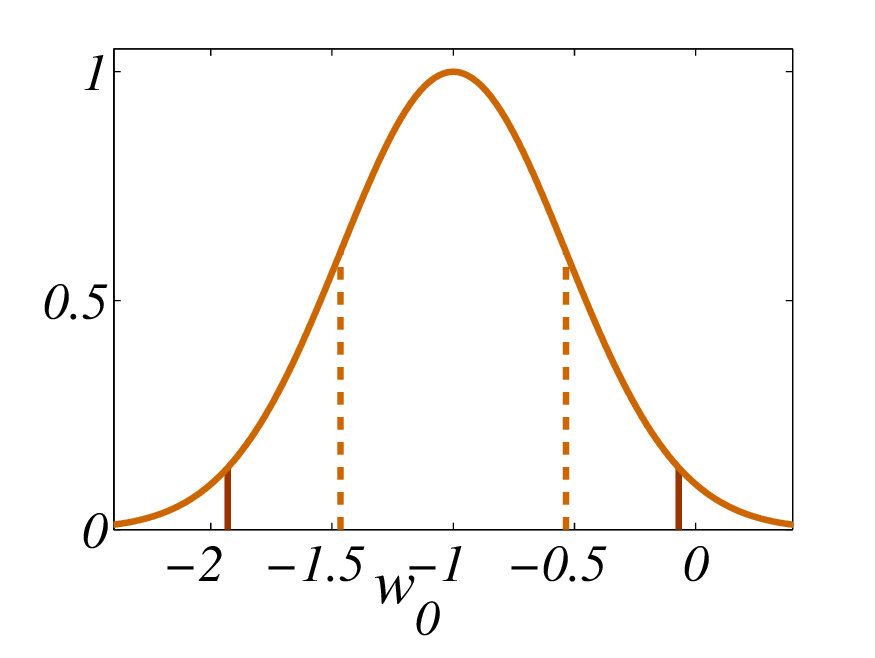}}& \\
 \includegraphics[width=1.35in]{\fig_dir{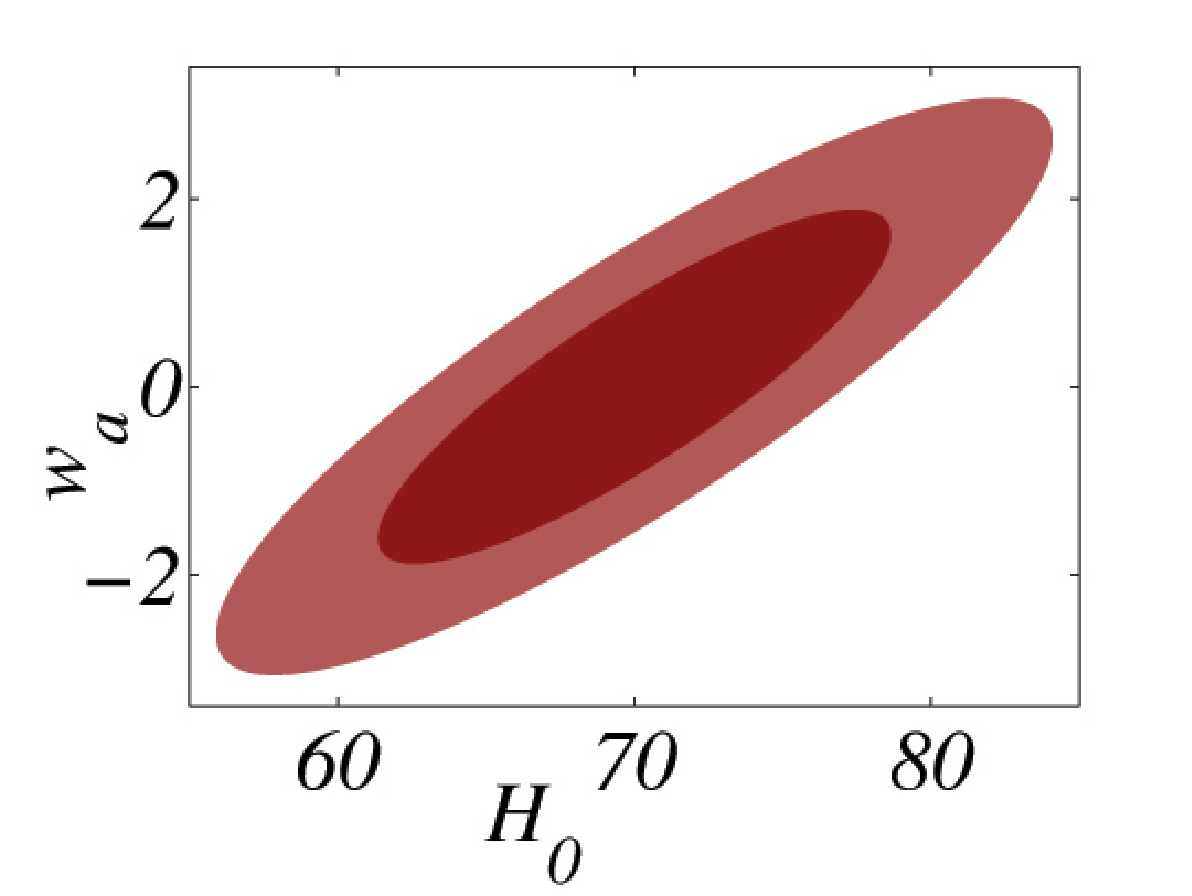}} &
 \includegraphics[width=1.35in]{\fig_dir{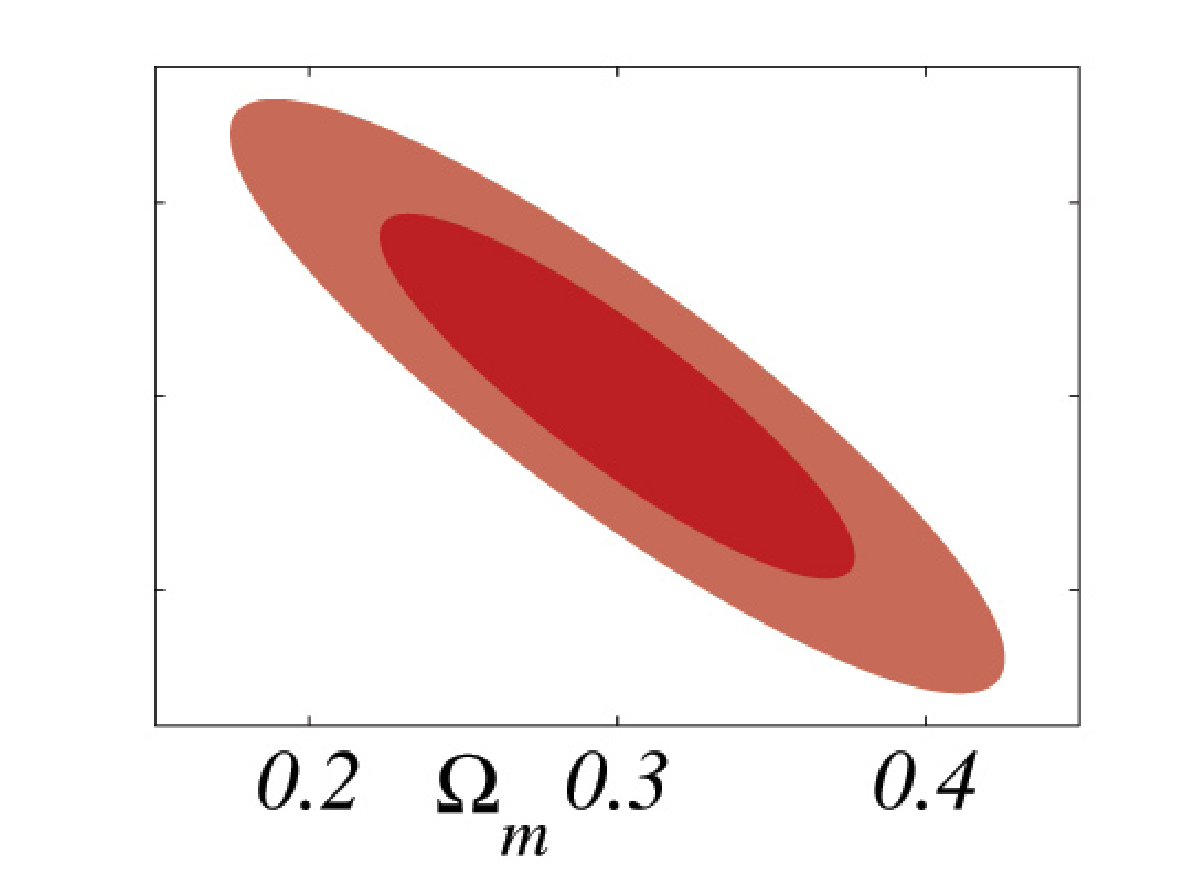}}&
 \includegraphics[width=1.35in]{\fig_dir{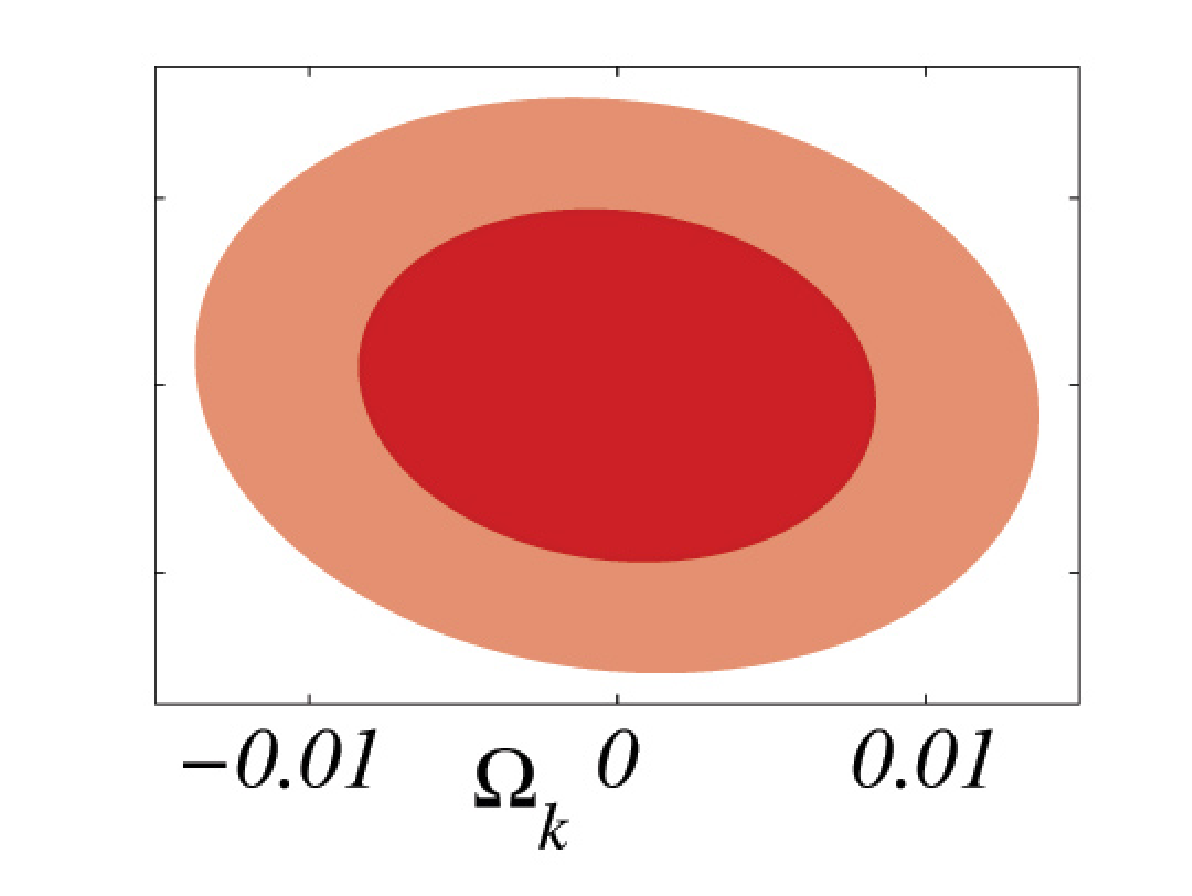}}&
 \includegraphics[width=1.35in]{\fig_dir{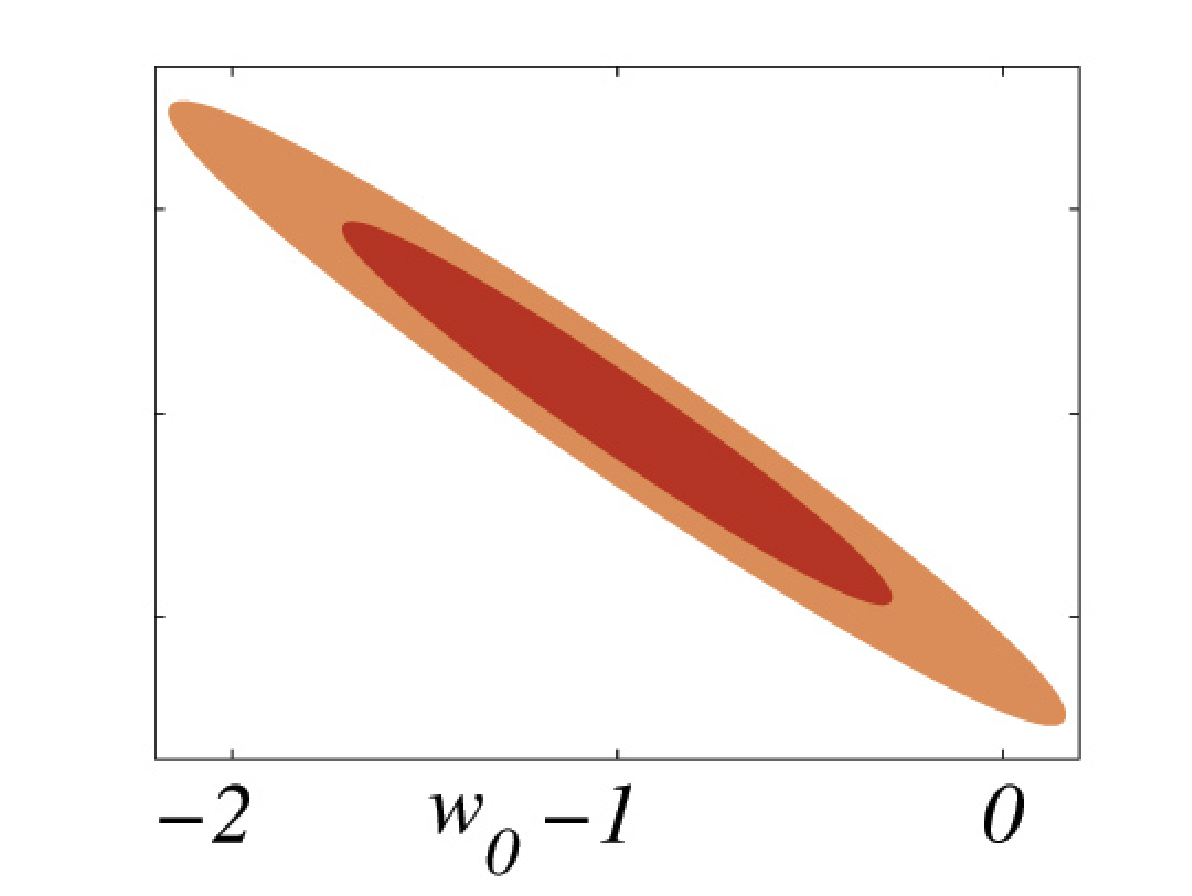}}&
  \includegraphics[width=1.35in]{\fig_dir{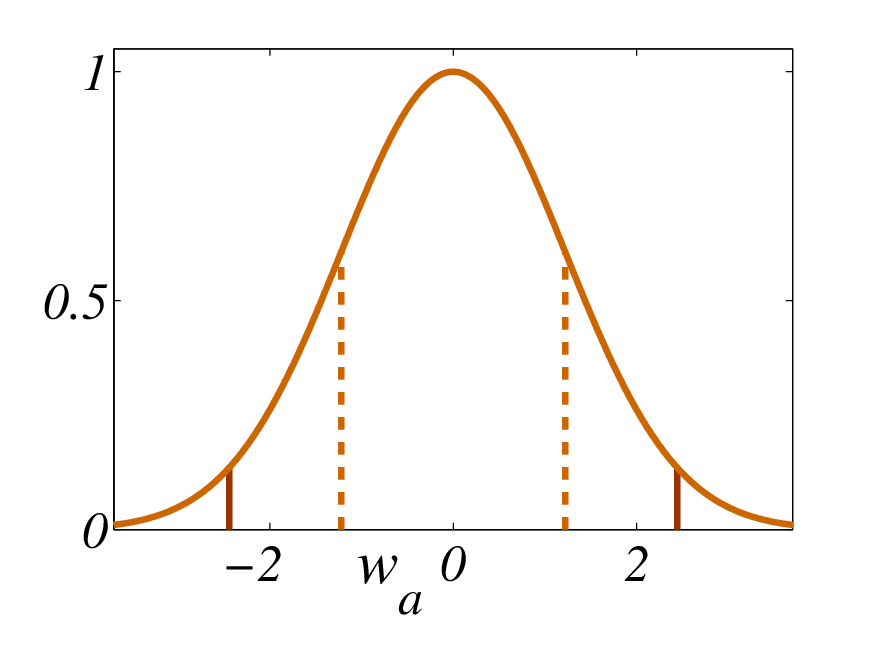}}
   \end{tabular*}
   \end{flushleft}
   \begin{center}
\caption{{\bf Marginalised Fisher ellipses for all parameters.} Fisher error ellipses for the full range of cosmological parameters, namely $H_0, \om, \ok, w_0, w_a,$ (for a survey characterised by Table \ref{table_seo} with the addition of growth measurements at the same redshifts and with the same precision as the Hubble parameter) are produced by calling \name~in a simple loop over all parameters. The dark and light filled ellipses indicate the $1-$ and $2-\sigma$ contours respectively marginalised over all other parameters. The diagonal panels show the fully marginalised one-dimensional likelihood for each parameter; $1$ and $2-\sigma$ limits shown as solid and dashed vertical lines respectively. The code used to produce these plots is included in the latest release of \name. \label{general_constraints_fig}}
\end{center}
\end{figure}

Error ellipses for all the cosmic parameters in our example are shown in Figure~(\ref{general_constraints_fig}), where they have been computed around the flat-$\Lambda$CDM fiducial model - $(H_0, \Omega_m, \Omega_k, w_0, w_a)= (70, 0.3, 0, -1,0)$ - and the full set of analytical derivatives for $H(z), d_A(z)$ are given in the Appendix~\ref{app_deriv}. In the case of the derivative of the angular diameter distance in terms of the curvature parameter $\partial d_A/\partial \Omega_k$, which will be zero in the flat-$\Lambda$CDM model, a Taylor expansion of the derivative (Eq.~\ref{ddadok})) around $\Omega_k = 0$ is also provided in Section~\ref{app_deriv}. The derivatives for growth are computed numerically since no general analytical solution for the growth exists; see Section~\ref{growth_section}. {The errors on the parameters $H(z), d_A(z), G(z)$ are given for sample survey configurations in Table~\ref{table_seo}; this configuration is used in many of the figures. The errors in the table are calculated using the prescription presented in \cite{seo2003}  for moving of moving from the radial and transverse oscillation scales $r_{||}, r_{\perp}$ to the observables $H(z), d_A(z)$, using $r_{||} = c\Delta z/H(z), r_{\perp} = (1+z)d_A(z)\Delta \theta.$ The method from moving from input survey characteristics to errors on the radial and transverse oscillation scales is described in full in the \name{}~ manual for the prescriptions of Blake {\em et al.} \cite{blake_etal_ff} and Seo and Eisenstein \cite{SE2007}.}
\section{Fisher4Cast and its Applications}
\name{}~is written in Matlab\footnote{See http://www.mathworks.com.} following an object-oriented model and using general software engineering principles for implementation and testing. The code is not specific to cosmology; Fisher matrices can be generated given any ${\bf X}({\boldsymbol \theta}).$ The derivatives $\partial {\bf X}/\partial \theta_\mathrm{A}$ are computed analytically (if they are known) or numerically, as in the case of growth, allowing the code to handle complex cases without analytical formulae for ${\bf X}$. The numerical derivative method used is the complex-step algorithm \cite{complex_step} (as of the latest release), although a ordinary two-sided derivative is also coded as an option in the software. The code suite includes a Graphical User Interface (GUI), which is specific to the cosmological example we discussed above in Section~\ref{example}. \name{}~includes an automated \LaTeX{}~report generating feature described in detail in the Users' Manual\footnote{The Users' Manual is also bundled with the \name~code software available at \cite{cosmo_org}.} \cite{usersmanual}, which automatically creates a summary of relevant data, matrices and figures.

\name{}~facilitates novel research and education in two different ways. Apart from being well-tested against existing Fisher Matrix results, it is general and modular. Because of this modular nature, a natural application of \name{}~is to visualisation. The code can easily be called repeatedly in large loops, enabling one to study large-scale properties of the Fisher Matrix for a wide range of surveys and cosmologies. We give examples of such studies in the following subsection. Secondly, \name{}~is coded for a general FLRW universe (the curved case has rarely been studied in the literature). We study the issues of curvature and growth in dark energy Fisher analysis in the last two subsections.

\begin{figure}[htbp]
\begin{center}
$\begin{array}{@{\hspace{-0.15in}}c@{\hspace{-0.15in}}c}
\epsfxsize=3.5in
	\epsffile{\fig_dir{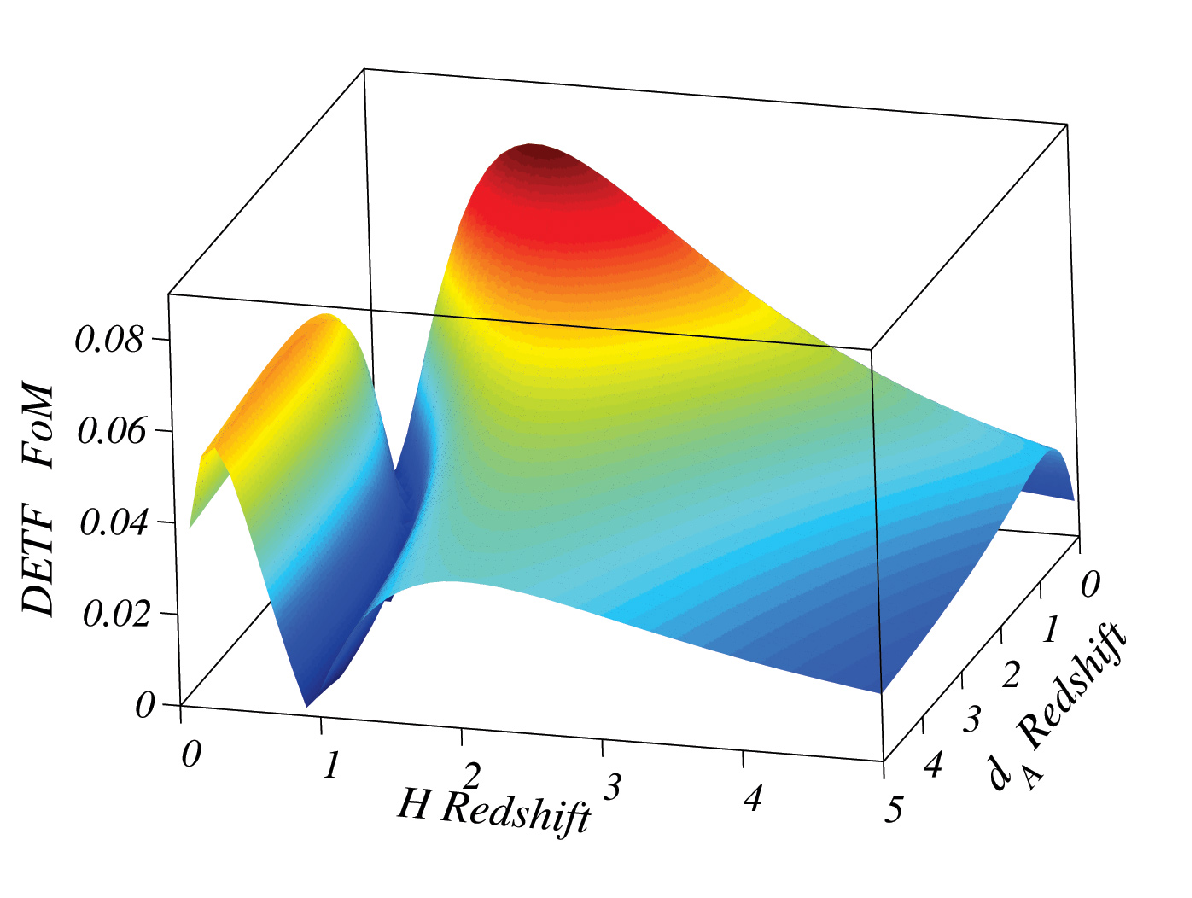}} &
   \epsfxsize=3.5in
    \epsffile{\fig_dir{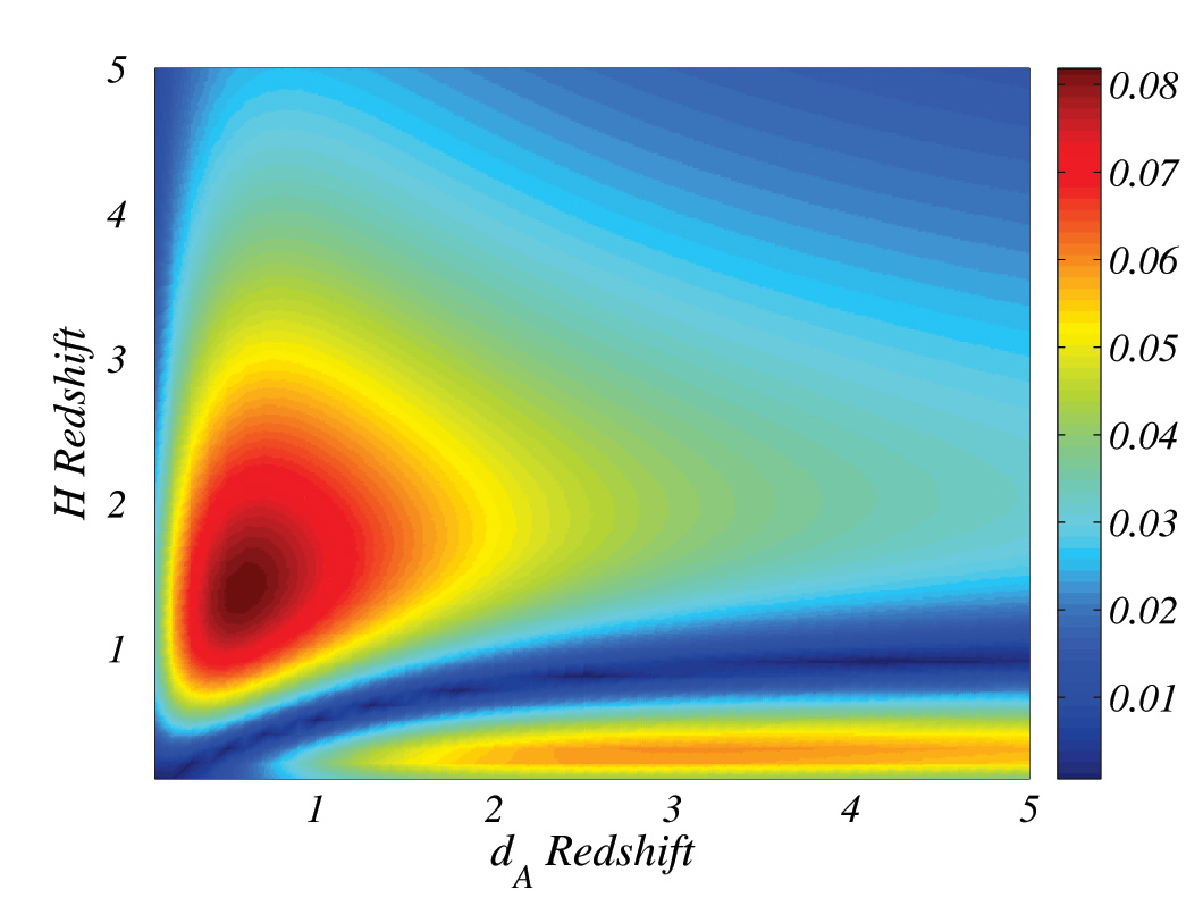}} \\ [0.0cm]
    \end{array}$
\caption{{\bf Figure of Merit plane -- } the Dark Energy Task Force Figure of Merit (FoM) for a survey consisting of one bin each of the Hubble parameter and angular diameter distance, with the fractional errors $\sigma_H/H = \sigma_{d_A}/d_A = 0.1$. The redshifts of the $H,d_A$ measurements are varied separately from $z=0.1$ to $z=5$, and the resulting FoM for each survey configuration is plotted as a 3-dimensional landscape {\bf (left panel)}, or a flat 2-dimensional plane {\bf (right panel)}. The colourmap in both panels reflects the value of the FoM, from low values of FoM $\sim 0$(blue) to higher values (FoM $\sim 0.08$) \label{fig:vary_H_da_grid_topview}.}
\end{center}
\end{figure}
\subsection{Visualisations \label{fisherapp}}
One of the key design principles behind \name{}~is ease of use - the GUI was specifically created to provide users without an in-depth knowledge of Fisher Matrix theory access to the power of the formalism. \name{}~can also be called from the command-line, as shown in the code examples in Appendix~\ref{fig_code}. In this subsection we investigate applications of \name{}~to probe and visualise the Fisher Matrix. 
\begin{figure}[h!]
\centering
\begin{tabular*}{\textwidth}{l l l}
 \includegraphics[width=0.31\textwidth]{\fig_dir{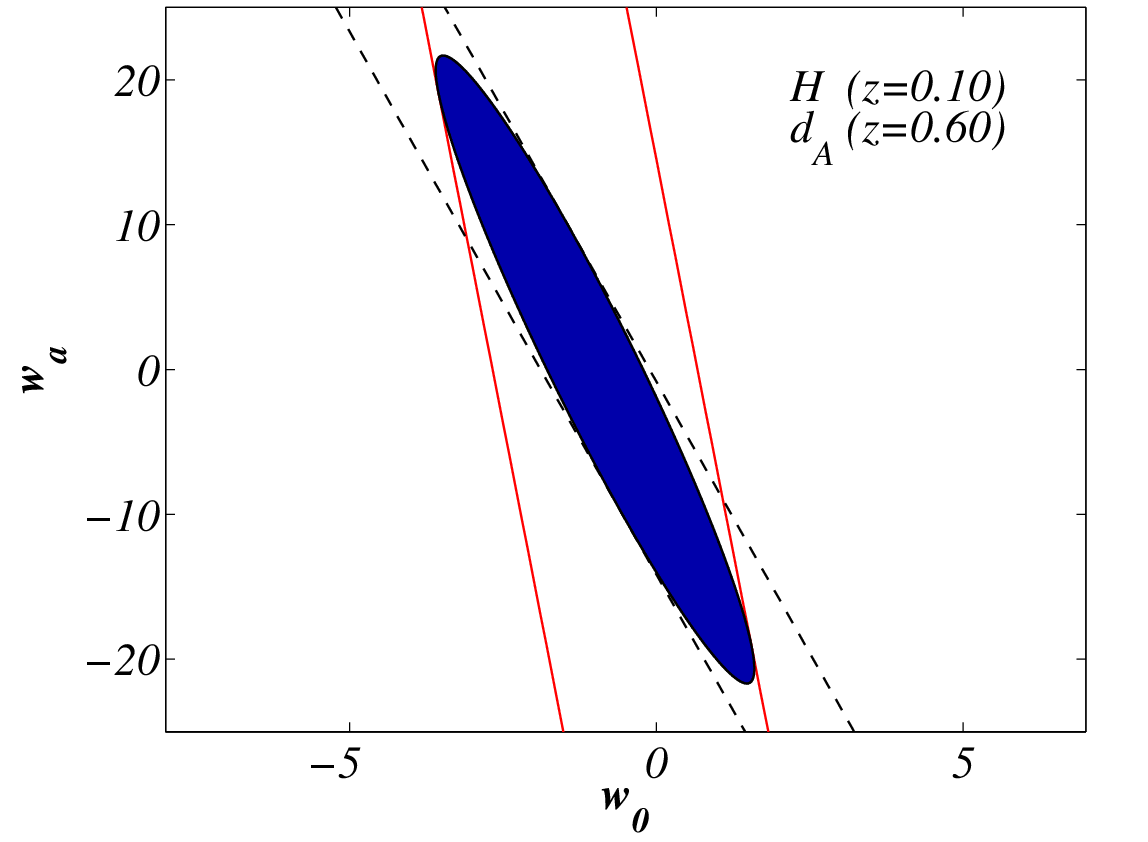}}&
 \includegraphics[width=0.31\textwidth]{\fig_dir{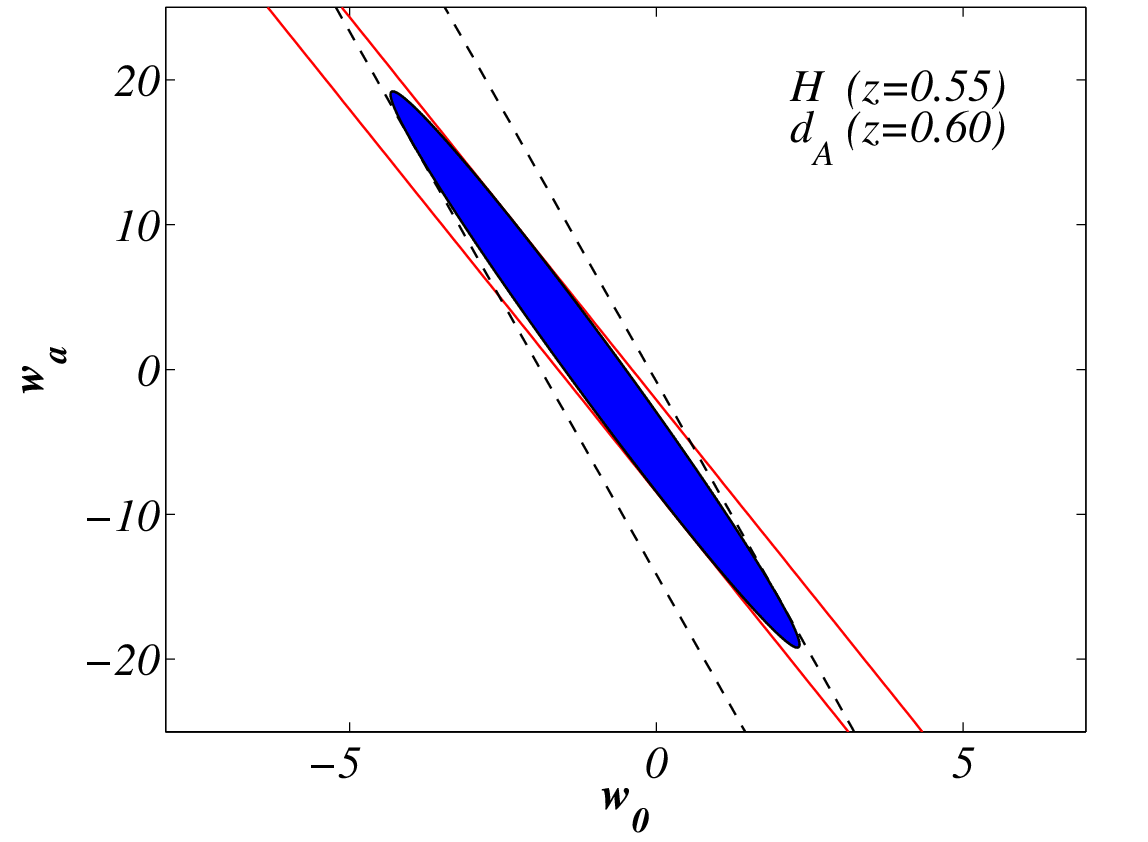}}&
 \includegraphics[width=0.31\textwidth]{\fig_dir{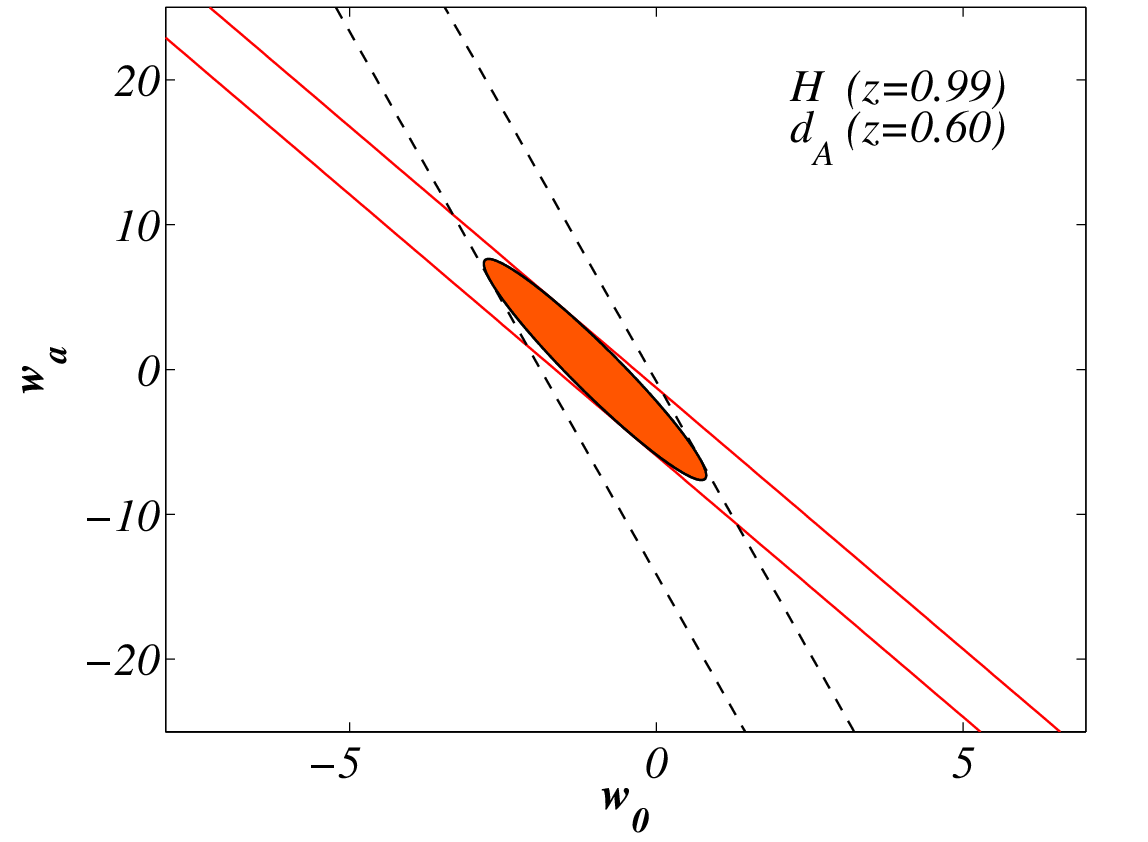}}\\
 \includegraphics[width=0.31\textwidth]{\fig_dir{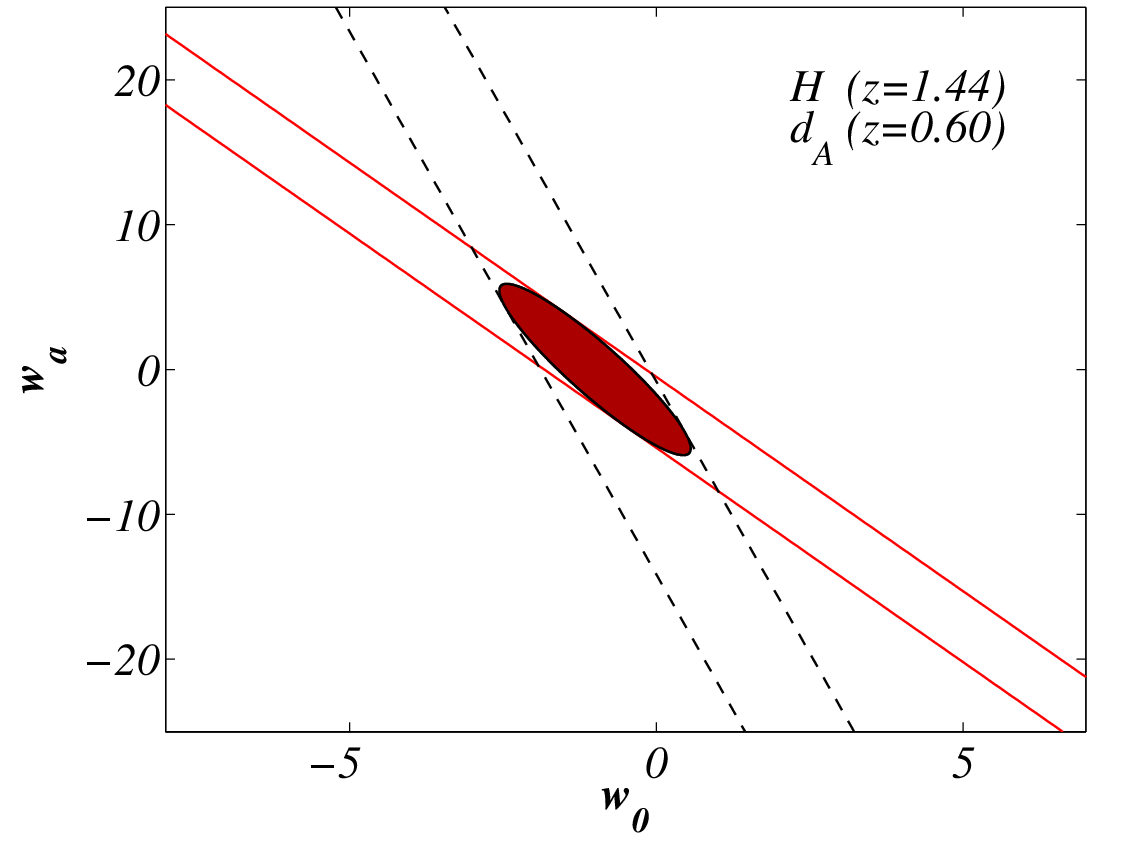}}&
 \includegraphics[width=0.31\textwidth]{\fig_dir{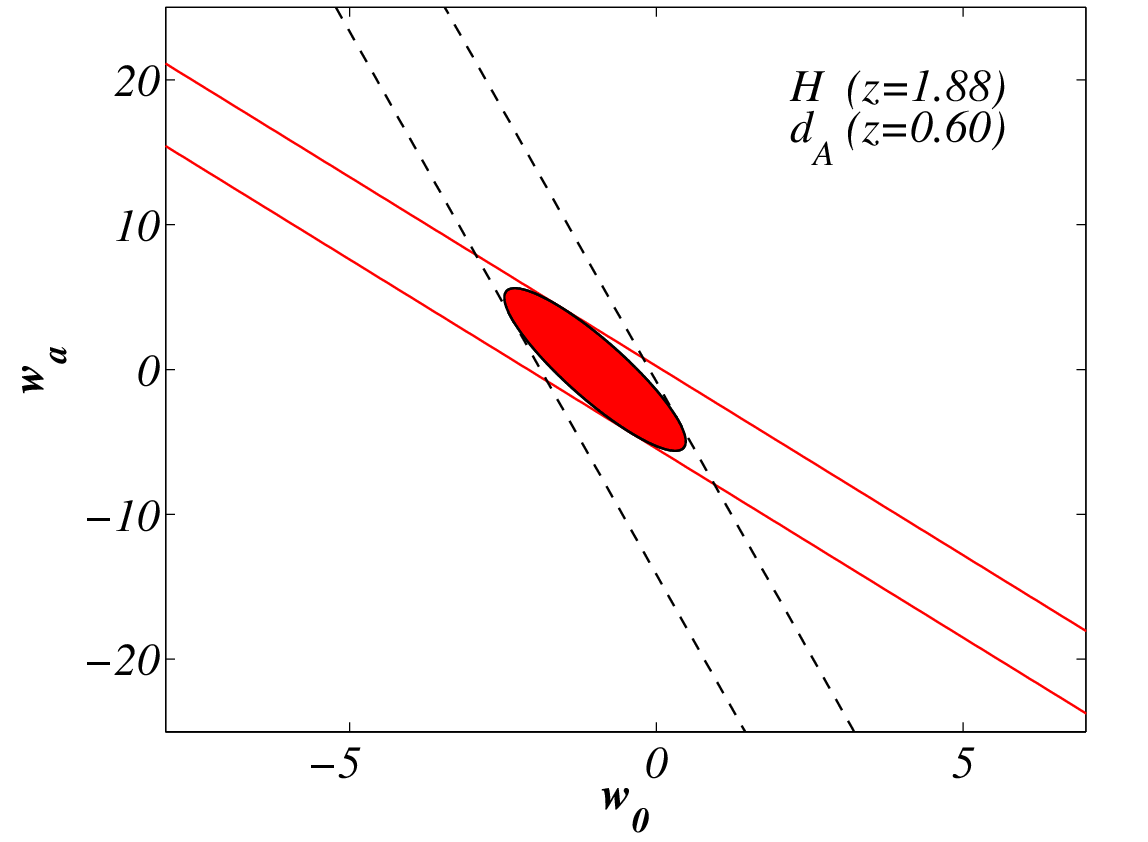}}&
 \includegraphics[width=0.31\textwidth]{\fig_dir{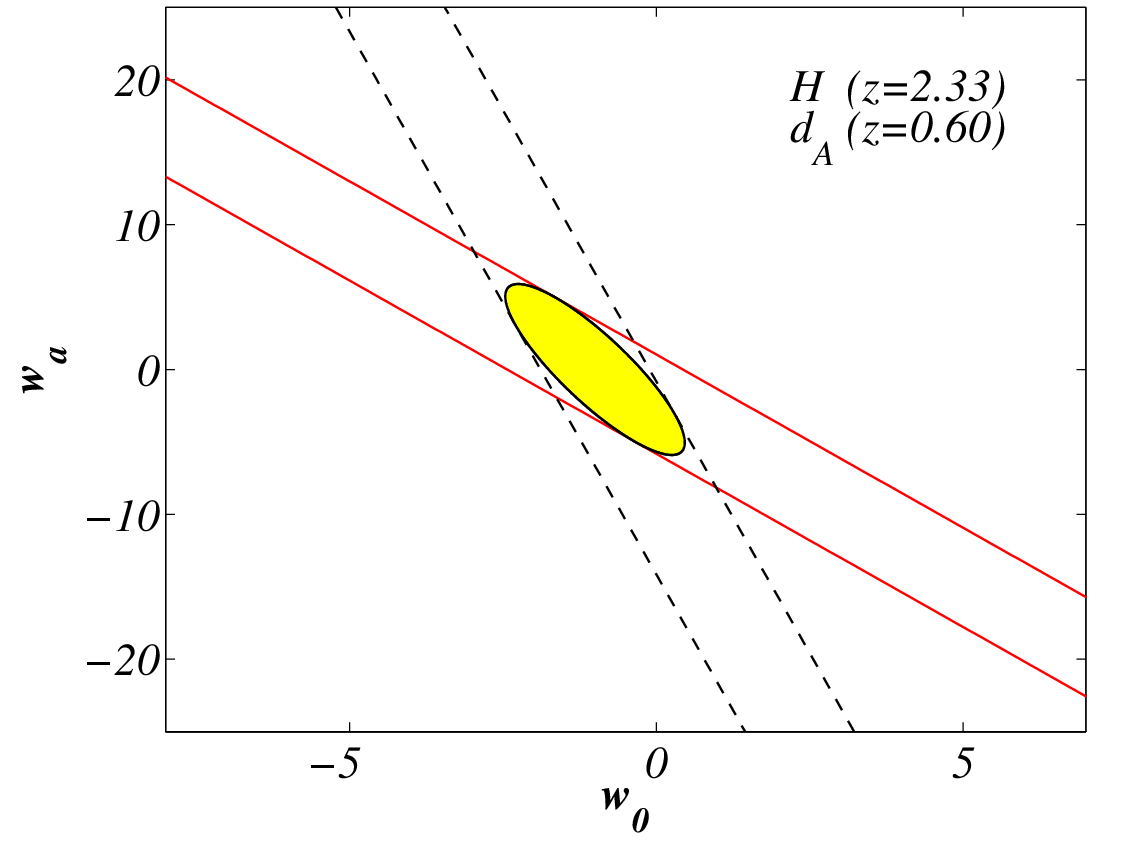}}\\
 \includegraphics[width=0.31\textwidth]{\fig_dir{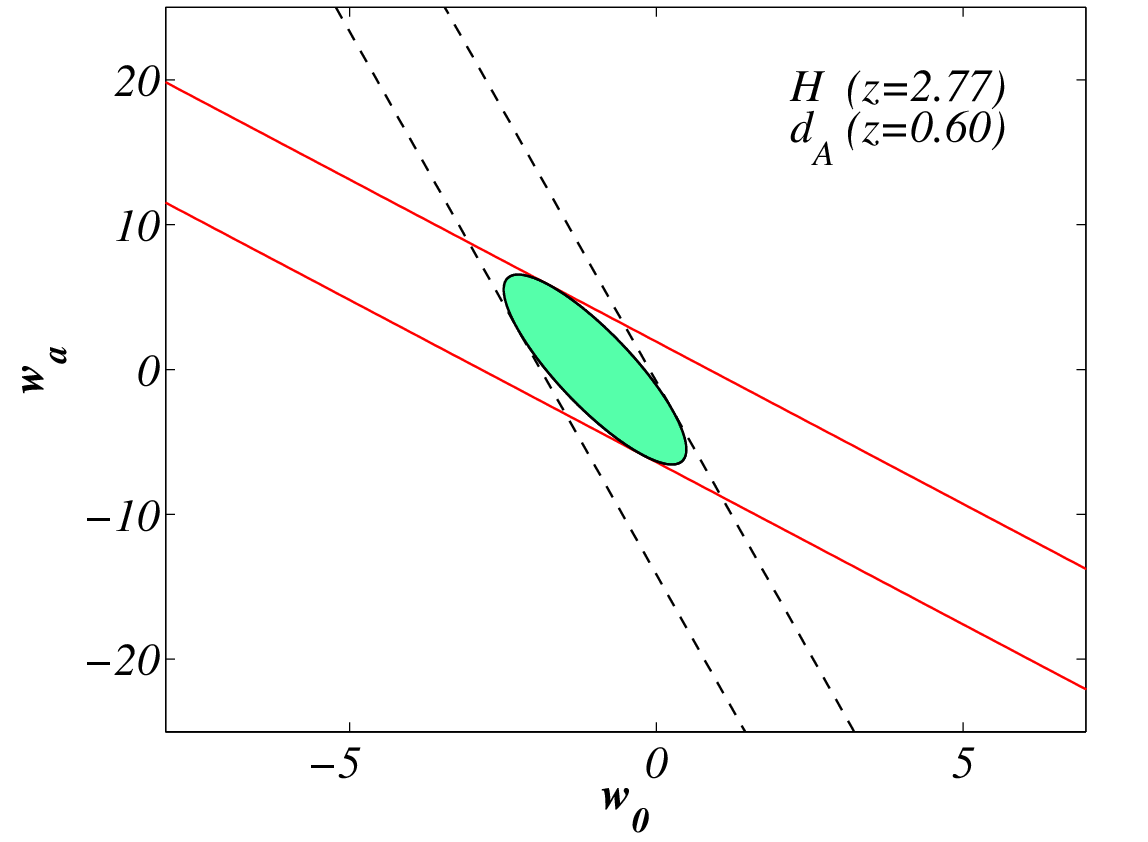}}&
 \includegraphics[width=0.31\textwidth]{\fig_dir{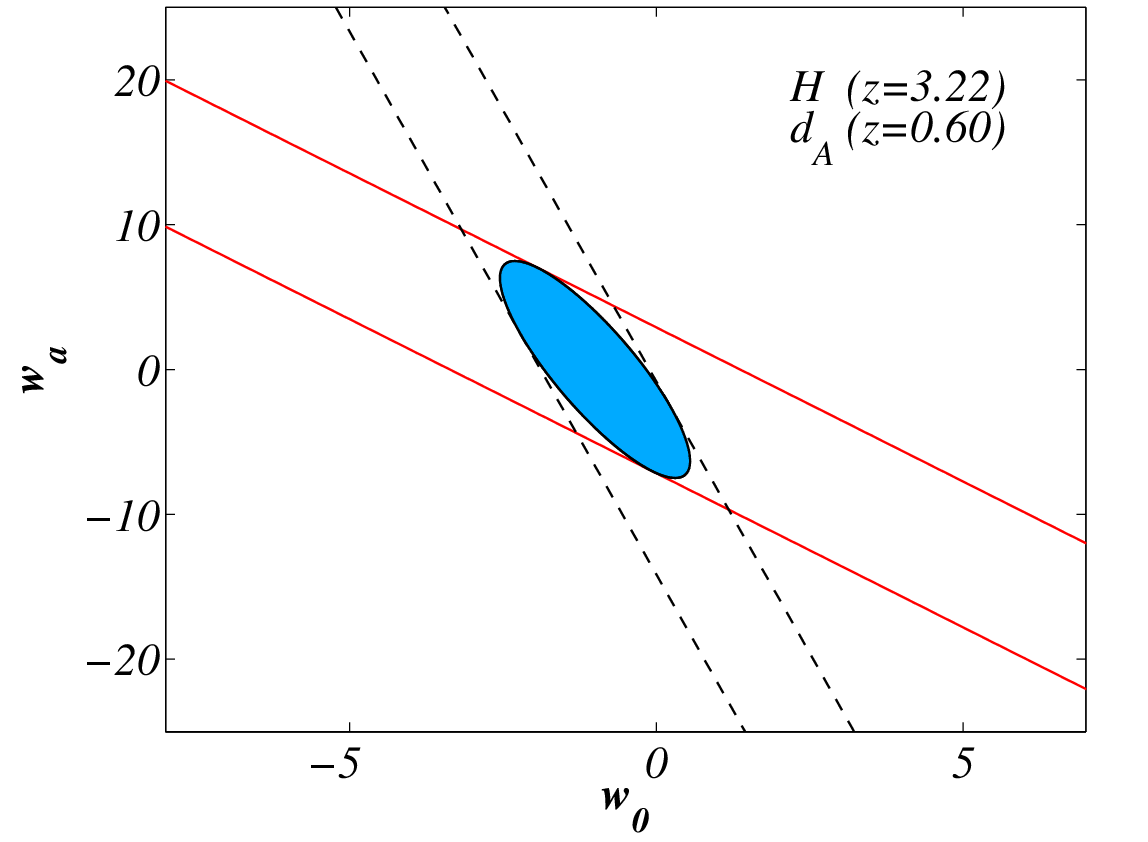}}&
 \includegraphics[width=0.31\textwidth]{\fig_dir{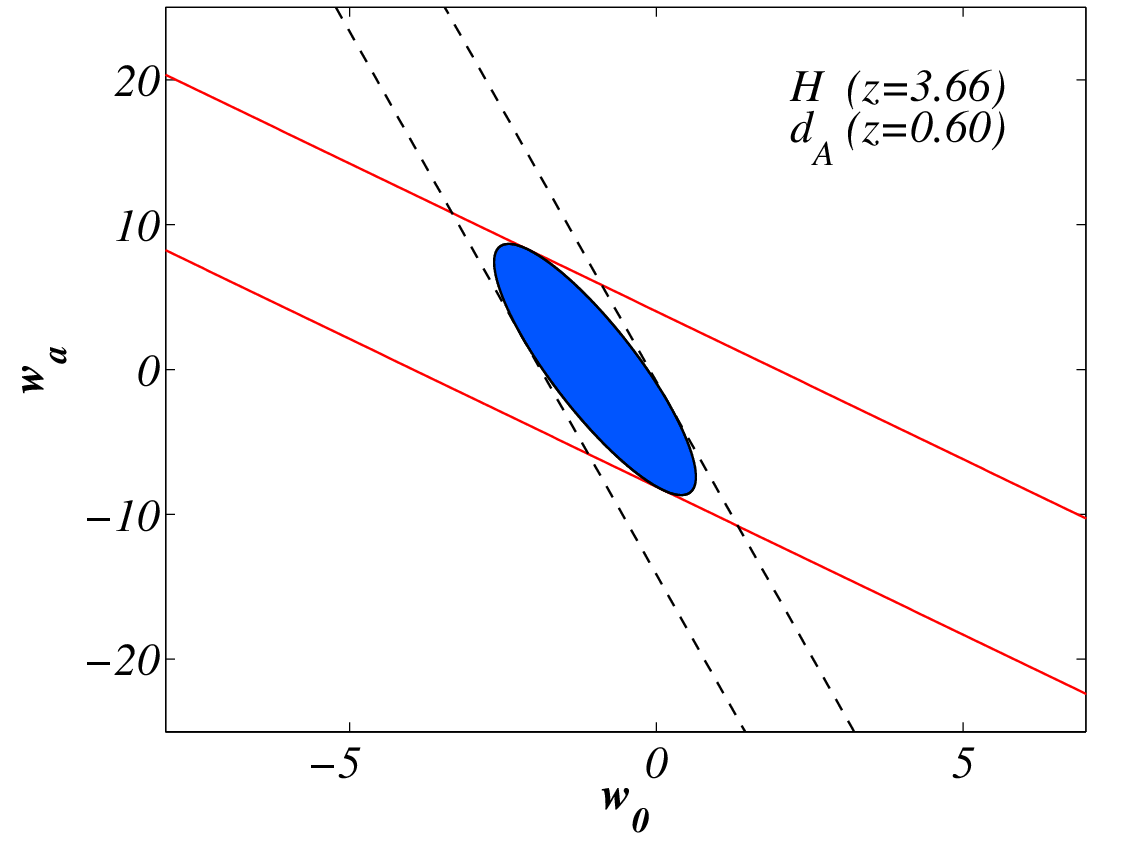}}\\
 \includegraphics[width=0.31\textwidth]{\fig_dir{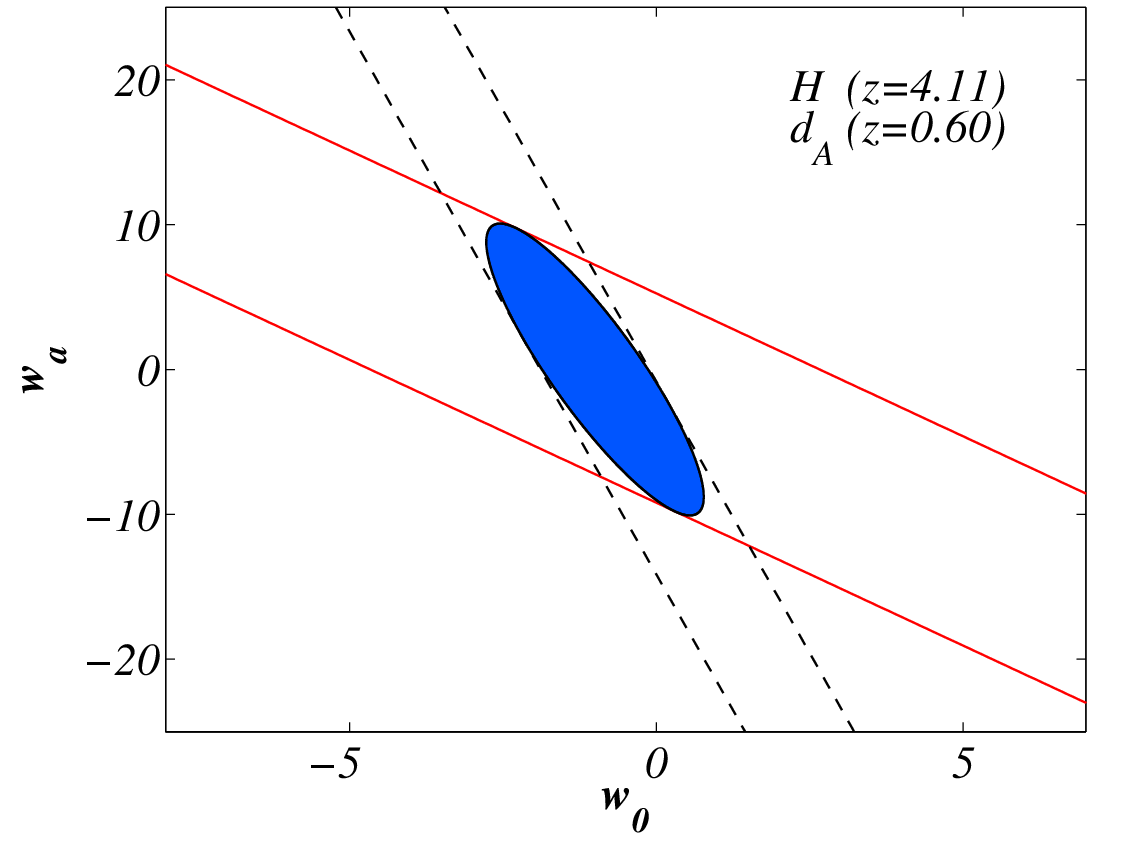}}&
 \includegraphics[width=0.31\textwidth]{\fig_dir{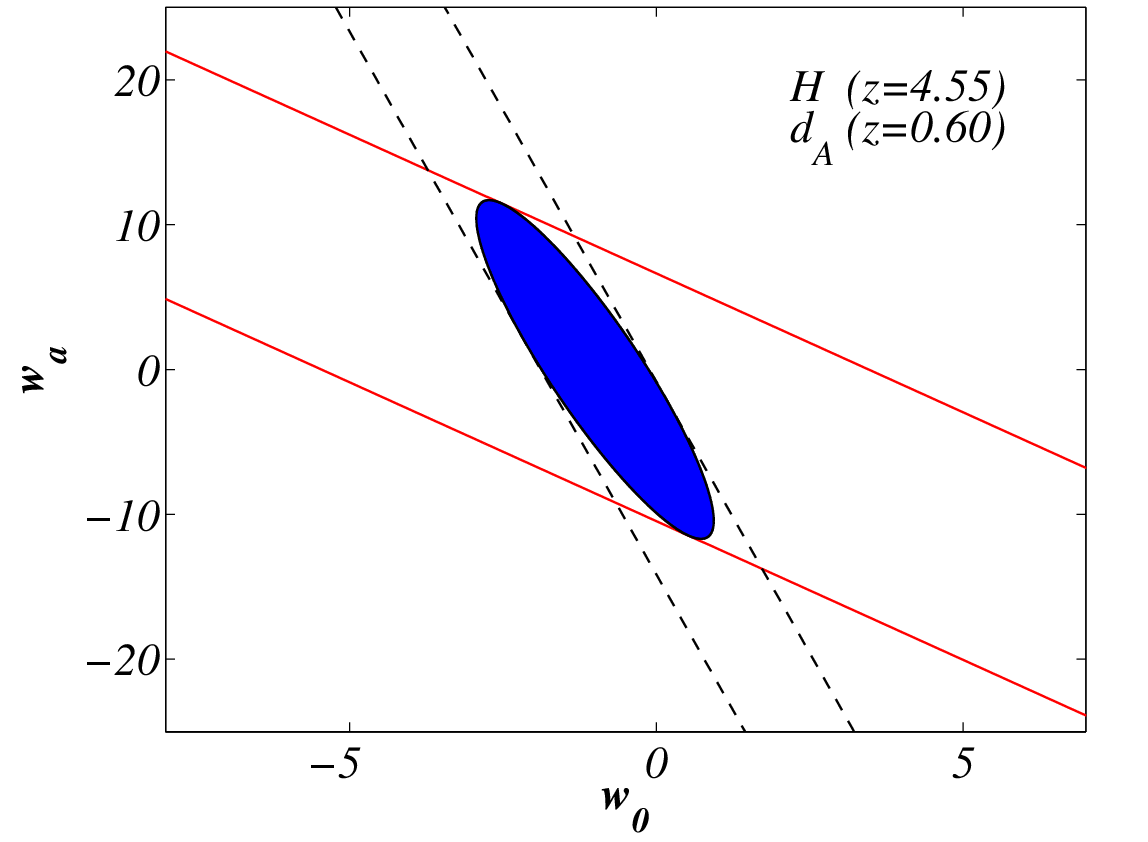}}&
 \includegraphics[width=0.31\textwidth]{\fig_dir{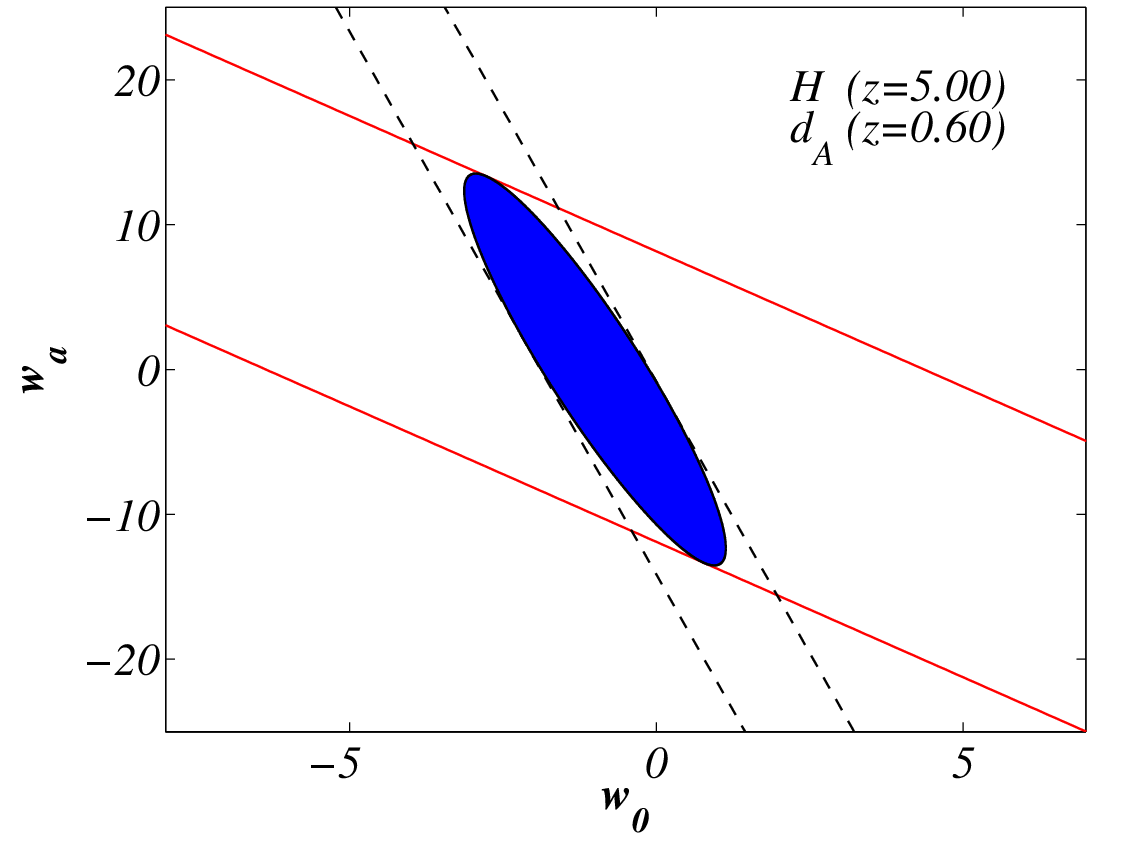}}\\
\end{tabular*}
\caption{{\bf Varying redshift of $H$ -- } examples of ellipses corresponding to particular values in the FoM landscape of Figure~(\ref{fig:vary_H_da_grid_topview}), in the particular case where the redshift bins for $H$ are varied while keeping the bin for $d_A$ fixed at $z=0.6$. The degeneracy direction for $d_A$ (black dashed line) and $H$ (red solid line) are included to help see how their relative orientations contribute to the orientation and size of the combined ellipse. The degeneracy direction of $H$ rotates anti-clockwise with the most rapid rotation experienced initially for the low redshift range of $H$. After $z=1$ the rotation of $H$ slows down and no longer is significant. As the $H$ and $d_A$ degeneracy directions become more orthogonal so the resulting constraints improve, yielding a higher FoM. The ellipses have colours which corresponding to their FoM, where red indicates the largest (around 0.08), and blue the smallest FoM. \label{fig:vary_H}}
\end{figure}
Figure~(\ref{fig:wa_w0_vary_ellipse}) illustrates the resulting ellipses when \name{}~is called in a simple two-dimensional loop that varies the values of $w_0-w_a$ in the cosmological model assumed to be true. The strong dependence of both the orientation and size of the resulting ellipse on the assumed model is clearly evident, as the ellipses rotate and shrink for larger values of $w_0$ and $w_a$. This is understood by recalling that $w_a$ controls the steepness of the $w(z)$ function; the steeper the slope of the function (especially at low redshift), the easier it is to discern the `true' cosmological model from other models, given the same survey characteristics.
Figure~(\ref{fig:vary_H_da_grid_topview}) illustrates the complementary case when the assumed cosmological model is kept fixed, but where instead the survey parameters are varied. In all cases the survey consisted of one measurement of $d_A(z)$ and $H(z)$ in a single redshift bin (with fractional errors of $10\%$ on either observable), while the redshifts of these two bins were varied independently over the range $0.1<z<5$. The DETF Figure of Merit (FoM) of the survey ($\propto 1/\mathrm{Area}$ of the $w_0-w_a$ ellipse) is plotted as a FoM ``landscape", where the colourmap is related to the value of the FoM - the higher values of the FoM are depicted in red, and the lower values in blue. The FoM ranges between $10^{-5}$ and $\sim 0.08$. The right-hand panel of Figure~(\ref{fig:vary_H_da_grid_topview}) shows this landscape projected into two dimensions, with the same FoM colourmap as in the three dimensional case.

Two interesting features are immediately apparent. First, the peak in the FoM landscape, or ``FoM hotspot'' occurs for a survey with measurements of $H(z)$ at $z=1.5$ and $d_A(z)$ at $z=0.6$, while a ridge of moderately good values of the FoM (relative to the average) emerges along the line corresponding to $H(z=0.2)$ for measurements of $d_A$ and redshifts larger than 1.6. Secondly these two regions of higher Figure of Merit are separated by a ``cold valley'' of lower FoM values. This is of particular interest given the optimisation of current and future Baryon Acoustic Oscillation (BAO) surveys \cite{bruceipso, design,correct_detf, huterer, rassat_bao}. Future BAO surveys will measure the radial and tangential oscillation scale (see \cite{BAOChapter} for a recent review of BAO in cosmology), producing measurements of $d_A$ and $H$ at the same redshifts, i.e. along the diagonal line in this plot. For $z$ roughly between $0.5$ and $3$, this line intersects regions of relatively high FoM. This landscape is explained by the geometric interplay between the size and orientation of the degeneracy directions between the $H(z)$ and $d_A(z)$ ellipses. This interplay is clearly evident in Figure~(\ref{fig:vary_H}) where we show the two degenerate ellipses (red solid lines for $H$ and black dashed lines for $d_A$) and the combined ellipse for a series where we keep the $d_A$ bin fixed at $z=0.6$ while varying $H$ in the redshift range $0.1<z<5$. This moves the perspective of the figure along a line in the landscape that intersects the peak at an $H$ bin of $z=1.5$. The combined ellipses have colours that correspond to the FoM in Figure~(\ref{fig:vary_H_da_grid_topview}), the highest value of the DETF FoM ($\sim 0.08$) coloured in red, and the lowest ($\sim0$) coloured in dark blue. One can see a distinct rotation of the degeneracy direction for $H$ as the redshift changes. This anti-clockwise direction rotation is most pronounced at low redshift from $z=0.1$ to $z=1.0$ after which it becomes more subtle and stops rotating completely at higher redshifts. It is this rapid initial rotation of the $H$ degeneracy direction and the narrowing of the width of the $d_A$ degeneracy direction, at $H(z=1.44)$,  which accounts for the peak of the FoM forming and being observed.

\begin{figure}[htbp!]
\centering
$\begin{array}{@{\hspace{-0.45in}}l@{\hspace{-0.62in}}l}
\includegraphics[width = 4in,  height = 3.1in]{\fig_dir{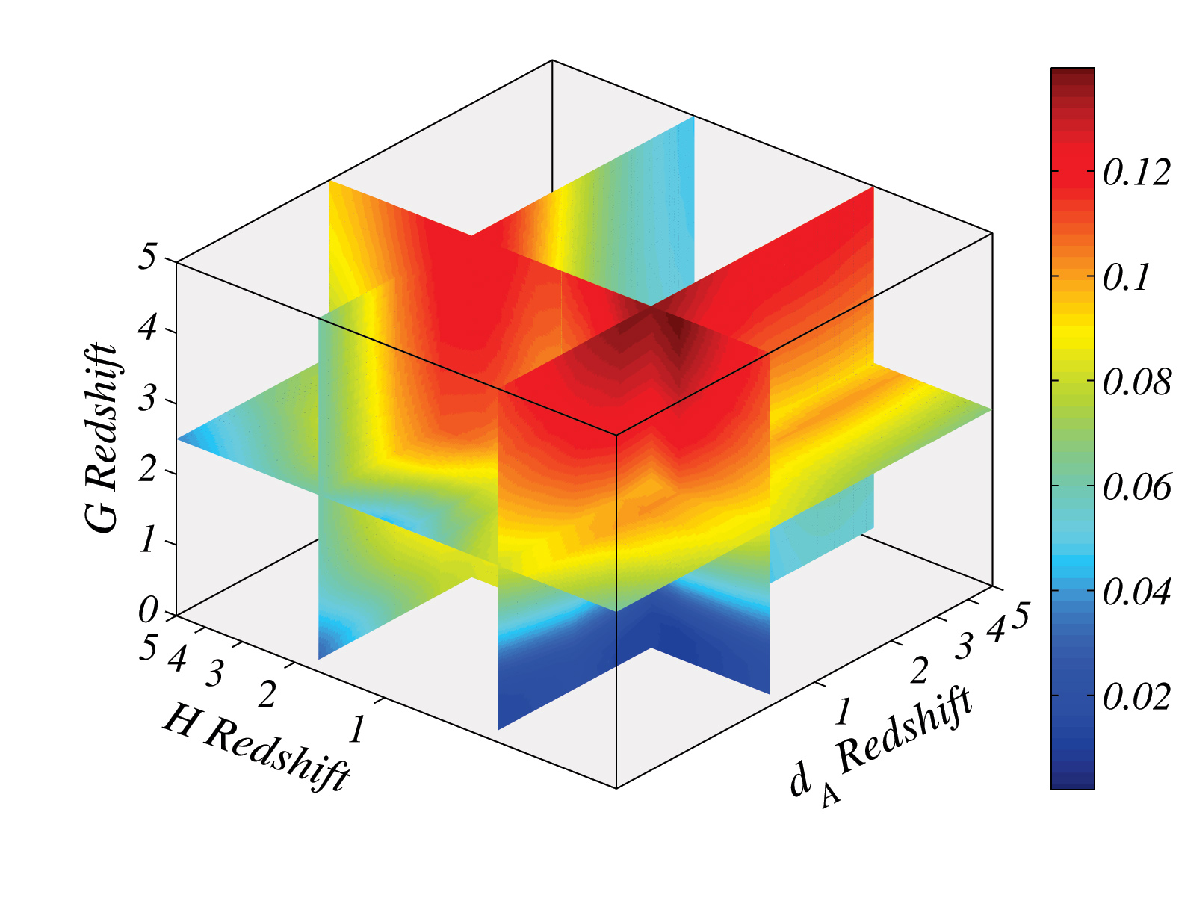}} &
\includegraphics[width = 4.1in, height = 2.9in]{\fig_dir{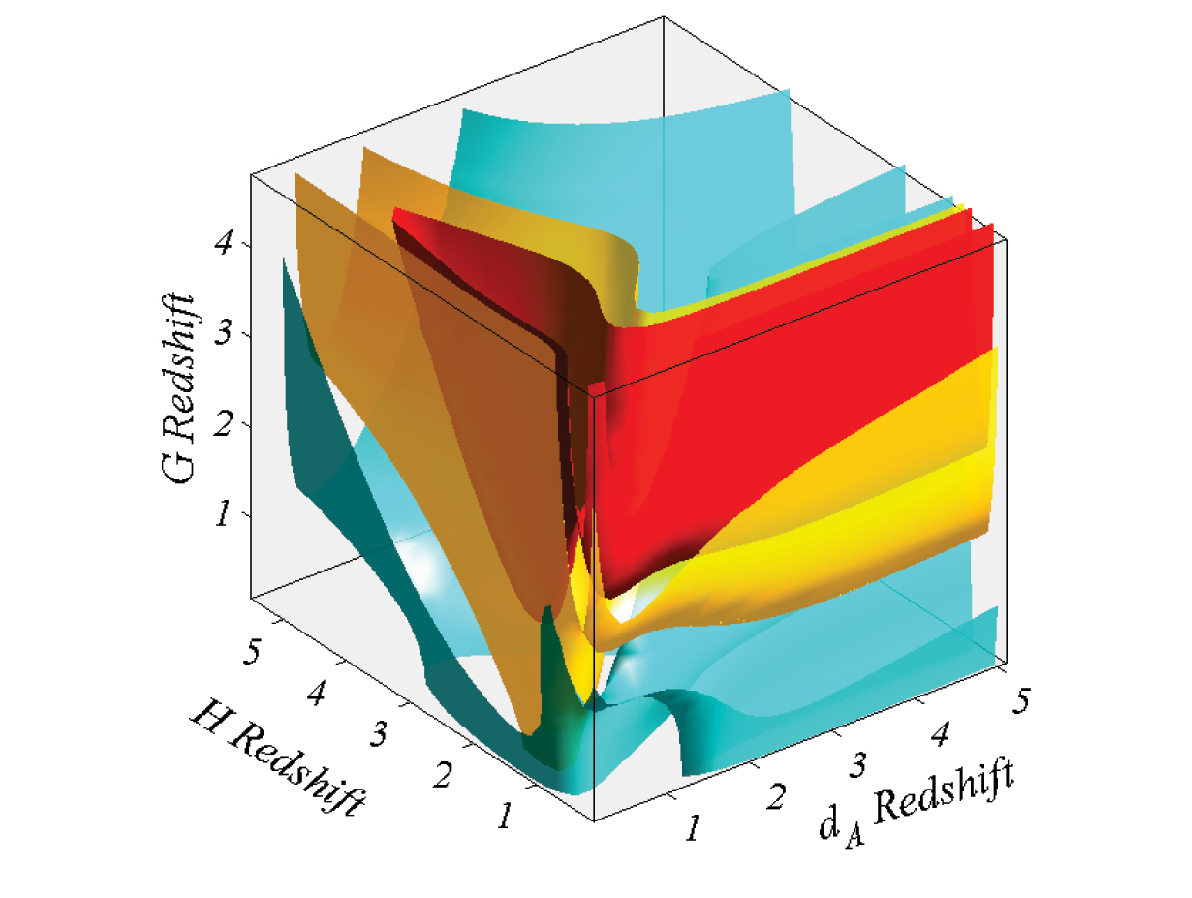}} \\ [0.0cm]
   \end{array}$
\caption{{\bf 4D Fisher Hypersurface Visualisations --} DETF FoM for the combination of a single measurement each of $H$, $d_A$ and $G$ (where the growth is normalised to unity at $z = 0$). The redshift of each measurement {(which can be thought of as the mean redshift in a redshift bin)} is allowed to vary, generalising Figure~(\ref{fig:vary_H_da_grid_topview}) by including growth. Slices through this hypersurface {\bf (left-hand panel)} at $z_H =  1.67$, {at a value of} $z_{d_A} = 0.67$ and $z_G = 2.50$ show the relationship between the redshifts at which the measurement of the observables are made and the FoM. The colourmap goes from a FoM of $9.5\times10^{-4}$ (dark blue) to 0.14 (dark red). Adding a high-redshift measurement of the growth function tightens the constraints on the dark energy parameters $w_0$ and $w_a$, shown by the red ridge of high FoM values. This hypersurface can be illustrated in a complementary way: surfaces of constant FoM are shown in the {\bf right-hand panel} - ranging from 0.03 (transparent light blue outer surface) to 0.15 (dark red opaque centre surface). Comparing the left and right panels, one notes that the intersection ``hotspot'' region in the left-hand panel at $(z_H = 0.4, z_{d_A} = 0.67, z_G = 5)$ is contained within in the red iso-surface with the high value of the FoM. This iso-surface extends right down to redshift $z_G \sim 1,$ re-iterating the improvement on dark energy constraints when including growth. \label{fig:slice}}
\end{figure}
Finally we extend the $H-d_A$ landscape to include a loop over the redshift of a single growth measurement in the range $0.1 < z < 5$. In all cases we normalise the growth today; $G(z=0) = 1$. The full four dimensional surface cannot be plotted in general, we show slices through the hypersurface in Figure~(\ref{fig:slice}), illustrating how a measurement of the growth at high-redshift measurement leads to much higher values of the FoM overall, and opens up interesting new ``hotspots''. This is can be seen in the dark red region in the top left-hand panel of Figure~(\ref{fig:slice}) which contains the high-redshift growth measurement, compared to the dark blue region at the bottom of the same figure. This illustrates the power of \name: not only can one general perform survey optimisations including different observables and parameters, but the ease of use of the visualisation modules capabilities means that one can trace out the FoM surface for a given survey configuration.
\subsection{Dark Energy Constraints from Growth of Structure \label{growth_section}}
The growth function $G(z)$, defined as the solution to Eq.~(\ref{deltaeq}) is sensitive to dark energy; the Hubble parameter acts as a friction term in the differential equation, increasing or suppressing the growth of structure. While in general there is no analytical solution to Eq.~(\ref{deltaeq}), under the assumption of a {\em flat universe} and a cosmological constant (or pure curvature) the growing mode satisfies the following integral form \cite{growth_eis, heath77}:
\begin{equation}
G(z) =\frac{5 \om E(z)}{2} \intzinf \frac{(1+z')dz'}{E(z')^3},
\label{growthz}
\end{equation}
where the $5/2$ coefficient is chosen to ensure that $G(z) \rightarrow 1/(1+z)$ during matter domination.
\begin{figure}[htbp!]
\begin{center}
$\begin{array}{@{\hspace{-0.15in}}c@{\hspace{-0.25in}}c}
\includegraphics[width = 3.5in, height = 2.5in]{\fig_dir{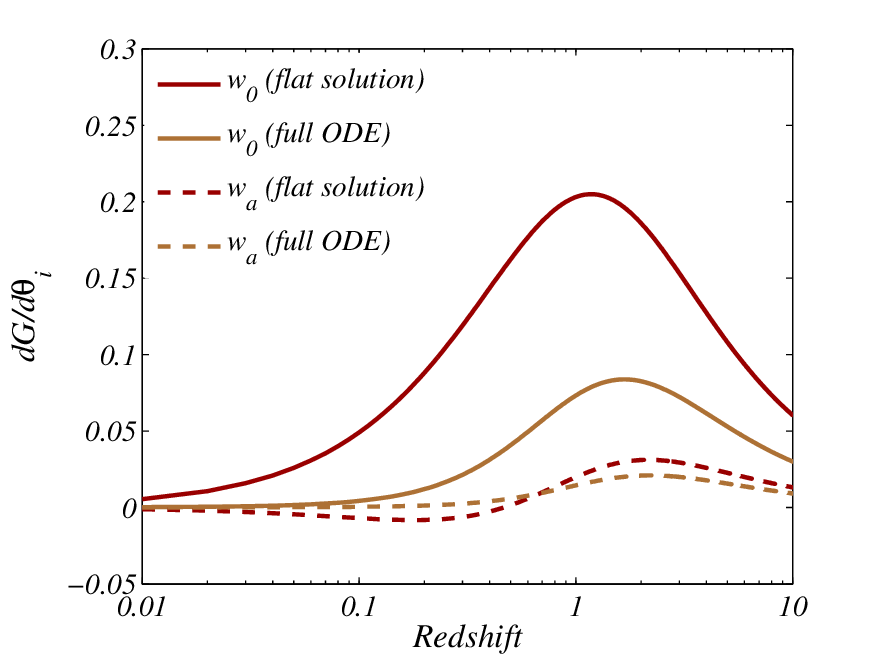}}  &
\includegraphics[width = 3.3in, height = 2.5in]{\fig_dir{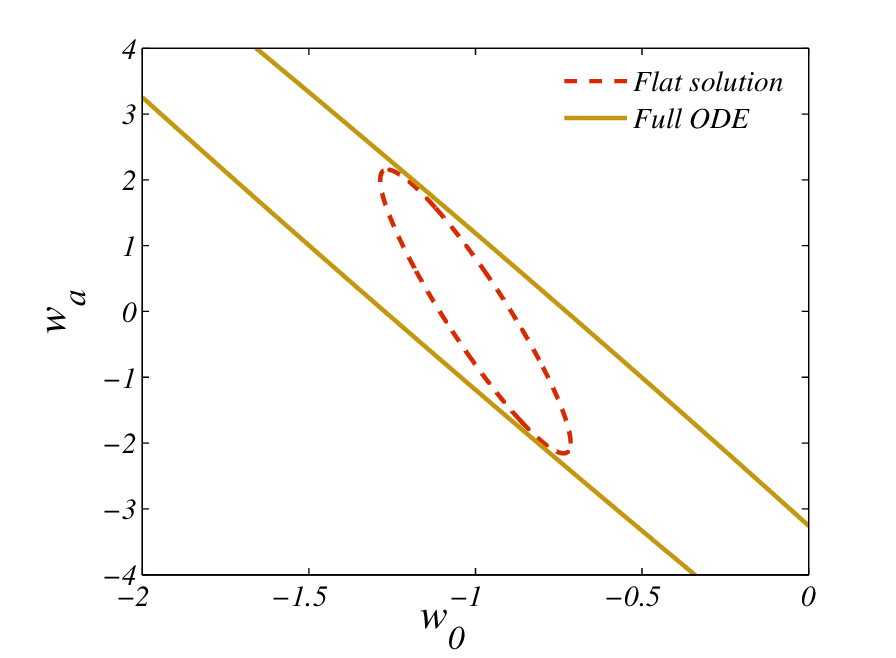}} \\ [0.0cm]
    \end{array}$
\caption{{\bf Comparing analytical derivatives to the full solution of the ordinary differential equation}. The left-hand panel in this plot shows the derivatives of the growth function $G(z),$ where analytical derivatives are taken of Eq.~(\ref{growthz}) {dashed red lines}, or where derivatives have been taken numerically of the solution to Eq.~(\ref{newg}) {solid brown lines}. The parameters considered are $w_0$ (light red for the analytical derivatives and dark brown for the numerical derivatives) and $w_a$ (dark red and light brown for analytical and numerical derivatives respectively). The right-hand panel shows the ellipses that result from a survey of the growth measured in 20 redshift bins from 0.1 to 2, with $10\%$ error on the growth function (normalised to unity at $z=0$), and priors of $10^4$ on the Hubble parameter, matter and curvature densities respectively. The ellipse that results from the analytical derivatives {red dashed line} suggests (incorrectly) much tighter constraints on the dark energy parameters, whereas the ellipse from the numerical solution ({solid brown line}) is much more degenerate in the dark energy parameters. \label{gderivs}}
\end{center}
\end{figure}
\begin{table}[htbp!]
\begin{center}
\begin{tabular}{|c|c|}\hline \hline
{\bf Parameter}& {\bf Value} \\
\hline
\hline
Redshifts of& $H:z = [0.3, 0.6, 0.8, 1.0, 1.2, 3]$\\
measurement& $d_A:z = [0.3, 0.6, 0.8, 1.0, 1.2, 3, 1000]$\\
\hline
{\bf Percentage error}& {\bf Value $[\%]$}\\
\hline
$H(z)$& $\sigma_{H}/H = $[5.80, 5.19, 3.59, 2.84, 2.53, 1.48]\\
\hline
$d_A(z)$& $\sigma_{d_A}/d_A = [5.19, 4.30, 3.22, 2.3, 2.03, 1.19, 0.22]$\\
\hline
{\bf Cosmological model}& {\bf Value}\\
\hline
($H_0, \om, \ok, w_0, w_a$)& $(70\mathrm{kms}^{-1}\mathrm{Mpc}^{-1}, 0.3, 0, -1, 0)$\\
\hline
Priors on model& $(10^4, 10^4, 10^4, 0, 0)$\\
\hline \hline
\end{tabular}
\caption{{\bf Survey data from the Seo \& Eisenstein survey configuration \cite{seo2003} -- } used in Figures~(\ref{growth_error}), (\ref{cooray_ok_ellipse}), (\ref{fig:wa_w0_vary_ellipse}) and (\ref{general_constraints_fig}). In some cases measurements of the growth function were added, taken at the same redshifts as the Hubble parameter; in others the prior information on various parameters was changed. See the captions of the relevant figures for the specific details. {The priors are added as a diagonal matrix with trace given in the final row of this table, and so can be thought of an additional Fisher Matrix from prior surveys and knowledge (for example priors on the Hubble parameter from \cite{riess/etal:2011}). A prior entry of zero indicates no knowledge of the value of that parameter.}  \label{table_seo} }
\end{center}
\end{table}
This expression should not be used however, to compute the derivatives $\partial G/\partial \Omega_k, \partial G/\partial w_0$ or $\partial G/\partial w_a$ since all of these derivatives violate the validity of the equation. Instead, the growth derivatives should be computed numerically from the solution of the full differential equation for $\delta(x)$. Rewriting the Raychaudhuri equation in terms of the Friedmann equation and the curvature density allows one to find an equation explicitly showing the curvature and dynamical dark energy contributions to the friction term: \be
G'' + \frac{3}{2}\left(1 +\frac{\Omega_k(x)}{3} - w(x)\Omega_{\mathrm{DE}}(x)\right)\frac{G'}{x} -\frac{3}{2} \Omega_m(x)\frac{G}{x^2} = 0, \label{newg}
\ee
where the new independent variable is $x \equiv a/a_0 = 1/(1+z)$, $a_0$ is the radius of curvature and $\Omega_k(x) = -k/(a_0^2x^2H(x)^2) ; \Omega_{\mathrm{DE}}(x) = \rho_{\mathrm{DE}}(x)/\rho_{crit}(x)$ are the fractions of the critical density in curvature and dark energy respectively. Alternatively, this can be written as a differential equation in terms of $\ln(x)$:
\be
\frac{d^2 G}{d\ln^2 x} + \frac{3}{2}\left(\frac{1}{3} +\frac{\Omega_k(x)}{2} - w(x)\Omega_{\mathrm{DE}}(x)\right)\frac{d G}{d\ln x} -\frac{3}{2}\Omega_m(x)G = 0, \label{newgln}
\ee
which is the equation actually solved in \name{}~since it is typically more stable numerically. Appropriate initial conditions for this differential equation are set deep in the matter dominated era: \\
$G(z_i)=1, dG/d\ln x(z_i) = G(z_i)$ for $z_i \geq 100.$\footnote{One can compare with the growth code at http://gyudon.as.utexas.edu/$\sim$komatsu/CRL/.} Note that as a result, the growth solutions will be unreliable if $w(z \rightarrow \infty) = w_0 + w_a \geq 0$ (or even if it is close to zero from below) since then there will be significant or even dominant early dark energy. \name{}~allows the user to choose the redshift where the growth is normalised to unity. The derivatives all satisfy $\partial G/\partial \theta_i = 0$ at the normalisation redshift.

While the growth functions Eq.~(\ref{growthz}) and Eq.~(\ref{newgln}) agree for $\Lambda$CDM, analytical derivatives taken of Eq.~(\ref{growthz}) will {\em not} agree with the numerical derivatives of Eq.~(\ref{newgln}), and will thus produce very different error ellipses when used incorrectly. The left-hand panel of Figure~(\ref{gderivs}) shows the $w_0,w_a$ derivatives of both solutions for $\Lambda$CDM. The derivatives are overestimated for both parameters, leading to (spurious) tight constraints on the parameters in the Fisher ellipse, as is illustrated in the right-hand panel of Figure~(\ref{gderivs}), for a survey characterised by 20 measurements (at $10\%$ accuracy) of the growth function between $z=0.1$ and $z=2.$ This illustrates the danger of using derivatives based on the analytical form of $G,$ even when evaluated at the model for which the analytical form is itself valid.

\begin{figure}[t]
\begin{center}
$\begin{array}{l@{\hspace{-0.1in}}c@{\hspace{-0.1in}}c}
	\epsfxsize=3.2in
	\epsffile{\fig_dir{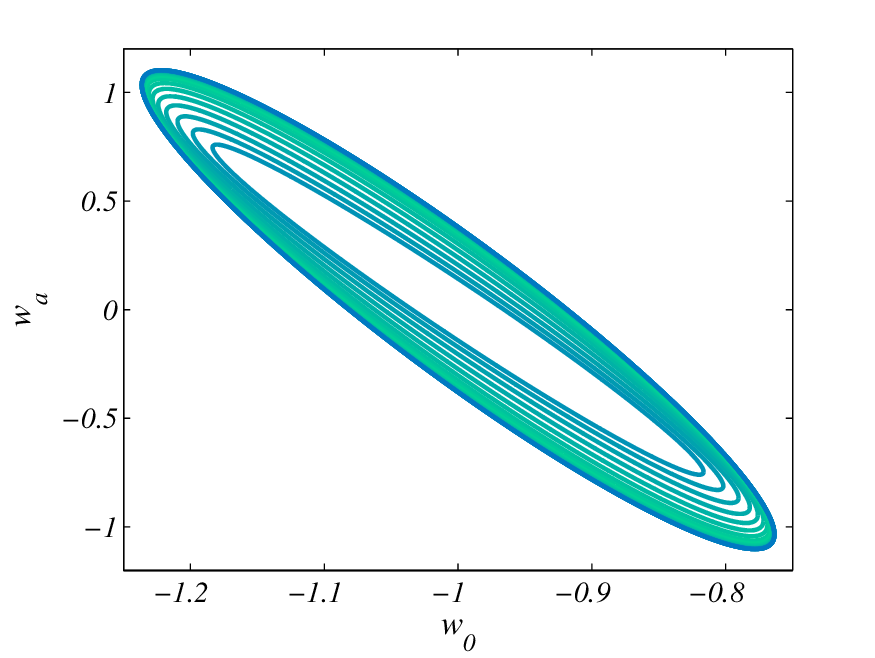}} &
	\epsfxsize=3.2in
	\epsffile{\fig_dir{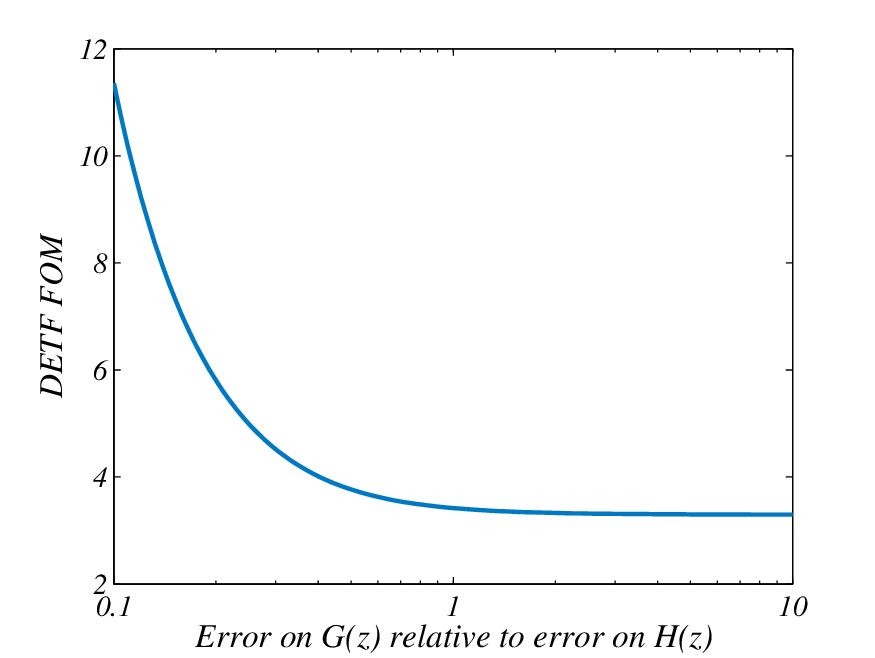}} \\ [0.0cm]
 \end{array}$
  \caption{{\bf Precision measurements of the Growth tighten dark energy constraints --} we show how changing the error on the growth relative to the Hubble parameter, $\delta G/G = \alpha \delta H/H,$ from $\alpha = 0.1$ {\bf (innermost ellipses)} to $\alpha = 10$ {\bf (outermost ellipses)} affects the power of the survey. We consider the survey summarised in Table~\ref{table_seo}, but with the addition of growth function measurements with $10\%$ error evaluated at the same redshifts as the Hubble parameter, for simplicity. The left-hand panel shows the Fisher ellipses in the $w_0-w_a$ plane, while the right-hand panel shows the DETF FoM as a function of $\alpha$. As the error on growth is decreased towards very small values ($\sigma G/G \rightarrow 0.1-\sigma_H/H$), the FoM increases as expected. For large values of $\alpha$, however, the FoM flattens out as there is essentially no pertinent information from the growth function. \label{growth_error}}
  \end{center}
 \end{figure}
As the precision of growth measurements increases, it is natural to ask how this impacts constraints on dark energy. This is easily investigated in \name{}. Consider Figure~(\ref{growth_error}), which illustrates constraints on the CPL parameters around $\Lambda$CDM for a survey characterised by Table~\ref{table_seo}, with the addition of growth measurements at the same redshifts as the Hubble parameter. As the error on growth relative to the error on $H$ decreases, the Figure of Merit of the survey increases dramatically, highlighting the merit in making high-precision measurements of the growth function, as is the focus of recent interest \cite{rhook, amendola07, ishak,linder_growth05, linder_growth09, abdalla_ska09, xia_ede, Zhan_08_growthDE, Lee_09_growthDE,Wei_08_growthDE,Bertschinger_08_growthDE}.

\begin{figure}[htbp!]
\begin{center}
$\begin{array}{l@{\hspace{-0.1in}}c@{\hspace{-0.1in}}c}
	\epsfxsize=3.4in
	\epsffile{\fig_dir{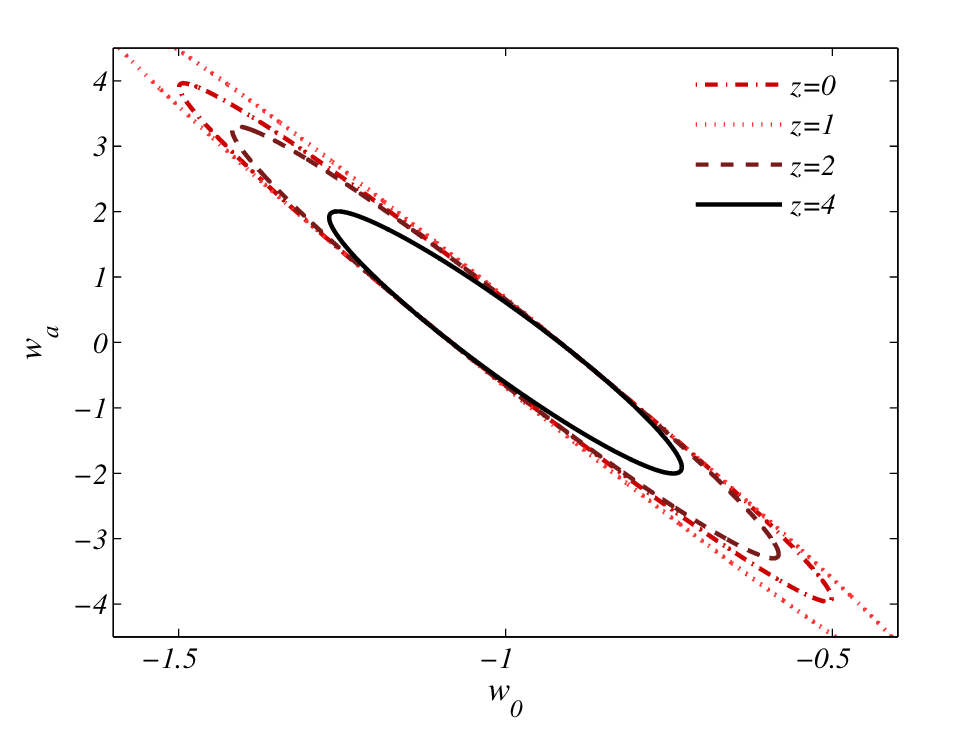}} \\ [0.0cm]
 \end{array}$
  \caption{{\bf Normalisation of the Growth function changes constraints --} the $w_0-w_a$ error ellipse for measurements of the growth combined with one measurement of $H$ at low redshift. The normalisation redshift of the growth function is varied from $z=0$ {(dot-dashed line)} to $z=4$ {(solid central black line)} to illustrate the change in the degeneracy direction of the dark energy parameters with growth normalisation. Note that the constraints do not scale monotonically with redshift: the ellipse is most degenerate for normalisation of the growth at a redshift of $z=1$. \label{norm_g}}
  \end{center}
 \end{figure}
As a final remark, we show how the redshift at which one normalises the growth function (normalisation redshifts $z=0,1,2,4$ are shown) influences dark energy constraints in Figure~(\ref{norm_g}). The constraints on the dark energy parameters are plotted for 5 measurements of the growth function at $z = (0.3, 0.6, 0.8, 1.2, 3)$ each with a $10\%$ accuracy, and one measurement of the Hubble parameter at $z=0.3$ with $\sigma_H/H = 5.19\%.$ An interesting point is that the resulting dark energy constraints do not scale monotonically with the normalisation redshift.
\subsection{The Effect of Cosmic Curvature On Dark Energy Constraints\label{curv}}
\begin{figure}[htbp!]
\begin{flushleft}
$\begin{array}{@{\hspace{-0.35in}}c@{\hspace{-0.0in}}c}
\includegraphics[width = 3.6in, height = 2.95in]{\fig_dir{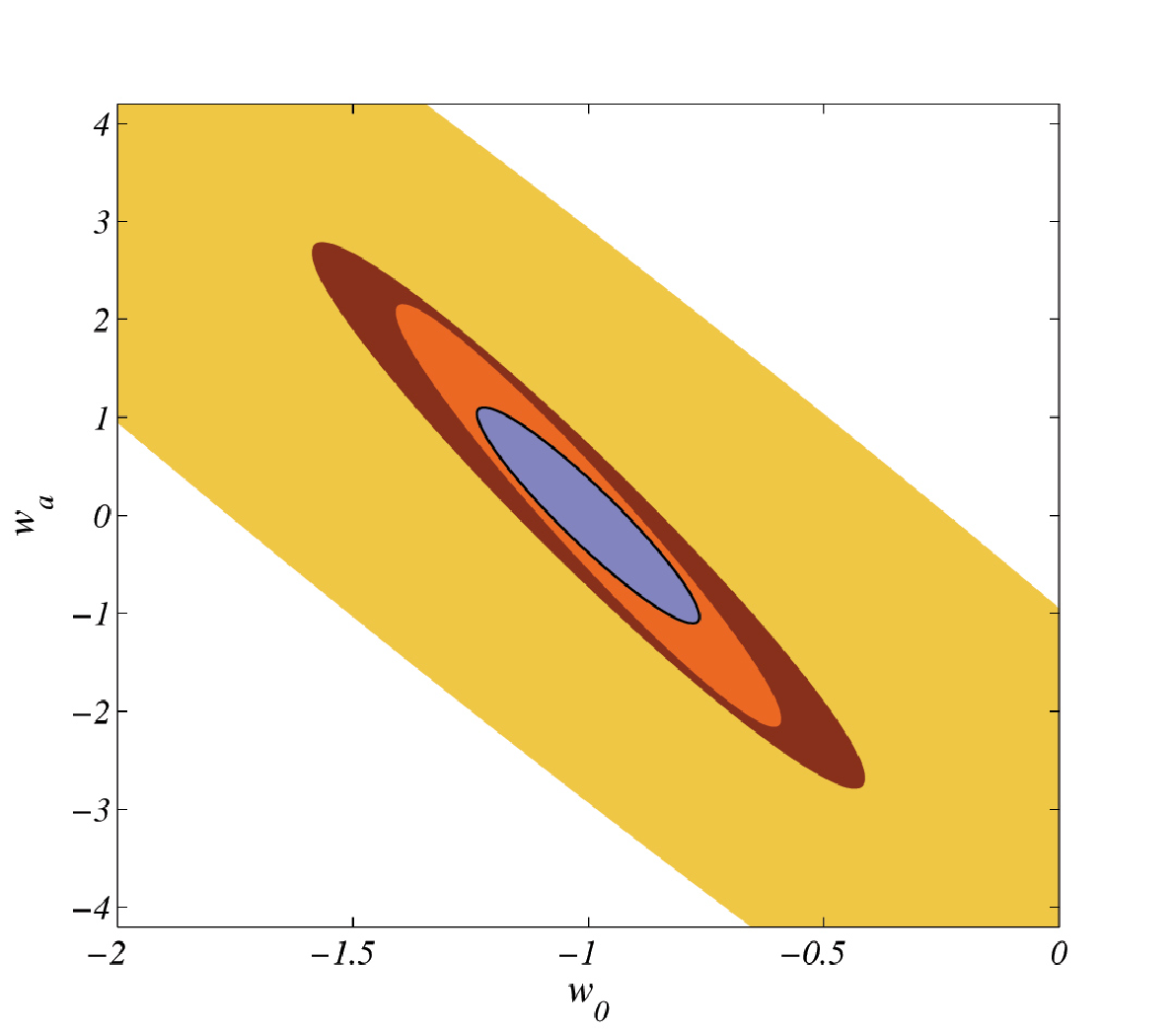}}&
\includegraphics[width = 3.2in, height = 2.7in]{\fig_dir{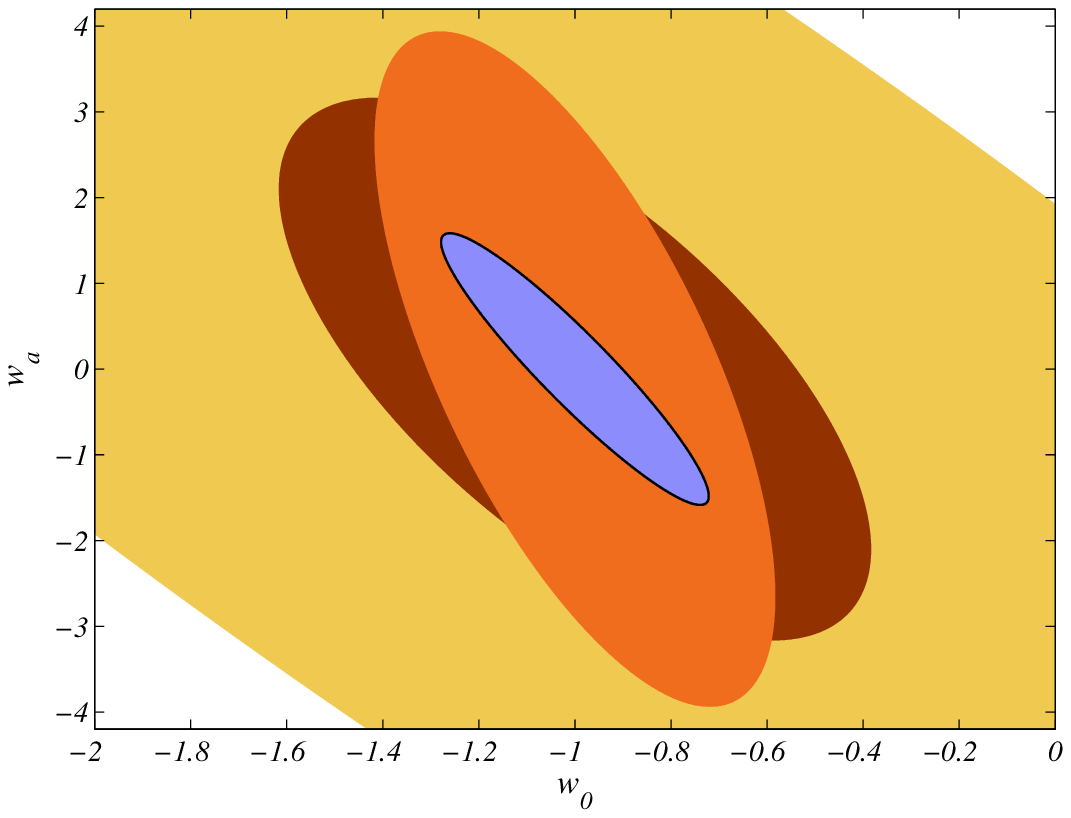}}   \\ [0.0cm]
    \end{array}$
\caption{{\bf Curvature marginalisation and dark energy constraints} - \name{}~ellipses on the $w_0-w_a$ plane where the prior on the curvature density is changed from very strong, with Prior$(\Omega_k)=10^6$ (left-hand panel), to very weak with Prior$(\Omega_k=10)$ (right-hand panel).
As the prior is decreased (i.e. we are less confident in the value of $\Omega_k$), the ellipse in the $w_0-w_a$ plane enlarges, in turn reducing the Dark Energy Task Force Figure of Merit (for the inner blue ellipse) from 3.298 to 1.890. The fiducial model is flat and the ellipses correspond to measurement of (moving from outwards in): $G(z)$ (light brown outer band), $H(z)$ (dark brown filled ellipse), $d_A(z)$ (orange filled ellipse) and the combination of all three (blue filled inner ellipse, outlined in black). The surveys details are given in Table~\ref{table_seo}, with the addition of $10\%$ measurements of the growth at the same redshifts as those of $H(z)$. The prior on the matter density is kept fixed at $100$. \label{cooray_ok_ellipse}}
\end{flushleft}
\end{figure}

The degeneracy between curvature and dark energy has been well-studied even for perfect measurements of any single observable such as $H(z)$ \cite{m1, CCB,de_degen,linder, huang}, implying that marginalising over the curvature is important when performing parameter estimation and forecasting of constraints on dark energy. The degree to which curvature affects dark energy constraints is shown here as a simple example of \name{}. Figure~(\ref{cooray_ok_ellipse}) shows Fisher error ellipses for the dark energy parameters $w_0, w_a$, after marginalising over curvature, as the  prior information on curvature is changed from Prior$(\Omega_k) = 10$ (weak) to Prior$(\Omega_k) = 10^6$ (strong) for the observables $H, d_A$ and $G$ considered separately and in combination. While uncertainty in the curvature of the universe (represented by a small prior value on $\ok$) degrades all ellipses, this is much less pronounced when the observables are considered in combination, showing the importance of combining multiple probes of dark energy.
\begin{figure}[htbp!]
\centering
$\begin{array}{l@{\hspace{-0.2in}}l@{\hspace{-0.2in}}l@{\hspace{-0.2in}}l}
	\includegraphics[width = 2.3in]{\fig_dir{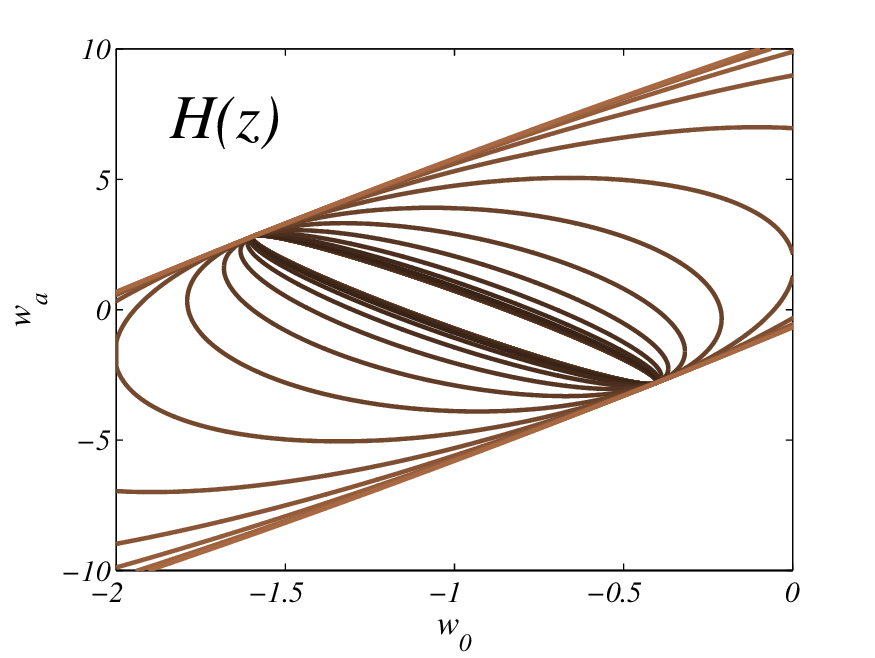}} &
	\includegraphics[width = 2.3in]{\fig_dir{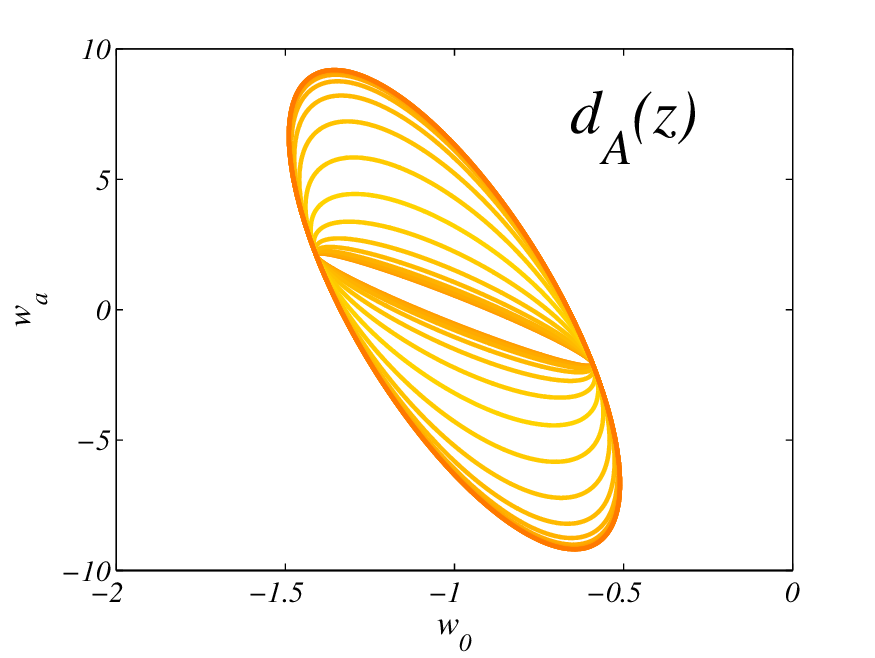}} &
	\includegraphics[width = 2.3in]{\fig_dir{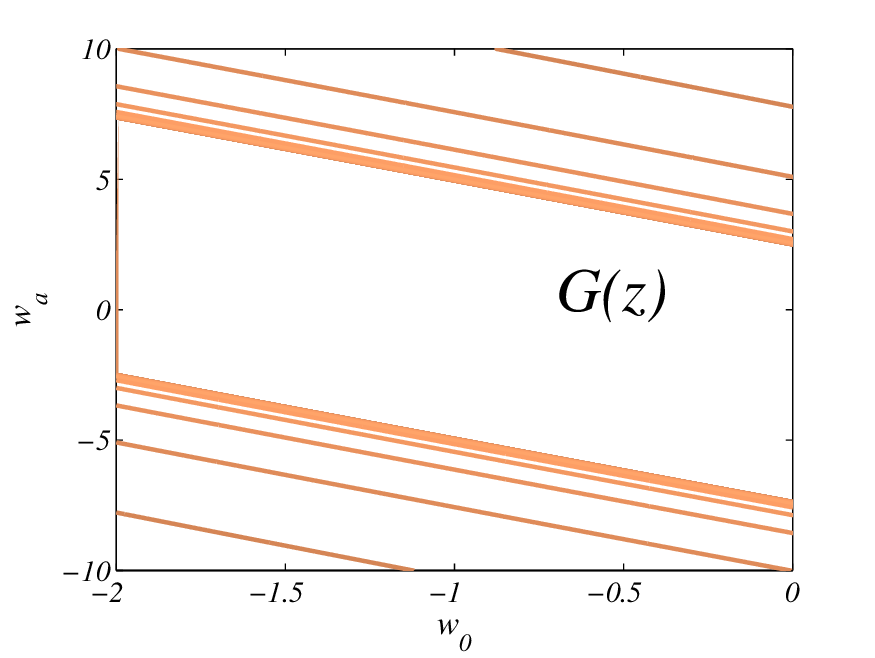}} \\ [0.0cm]
	\includegraphics[width = 2.3in]{\fig_dir{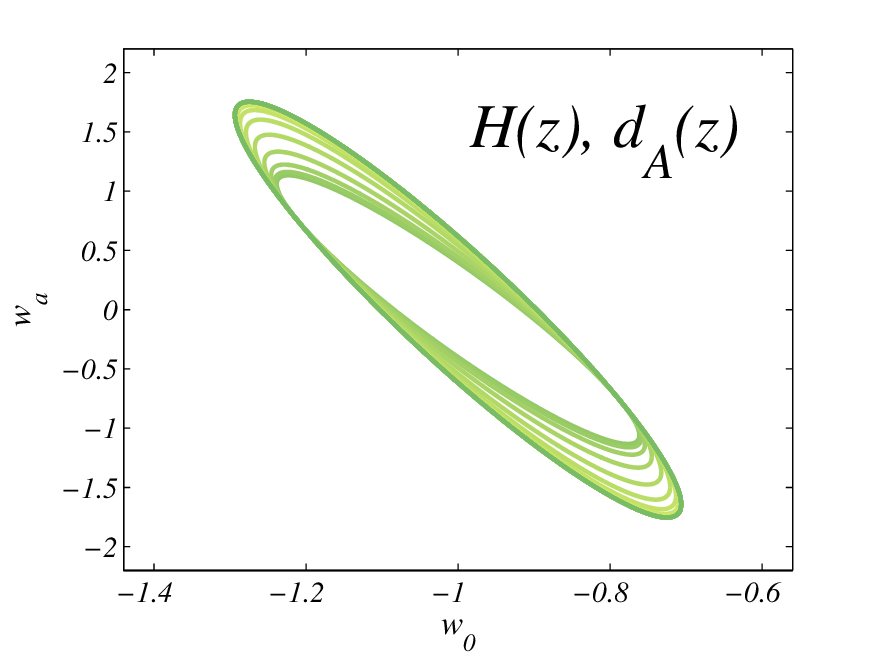}} &
	\includegraphics[width = 2.3in]{\fig_dir{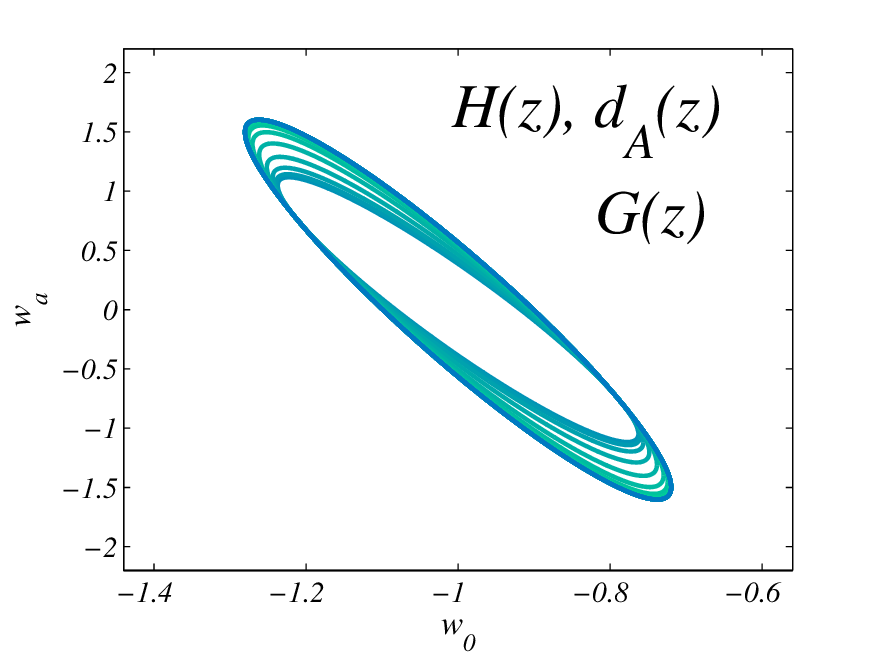}} &
	\includegraphics[width = 2.3in]{\fig_dir{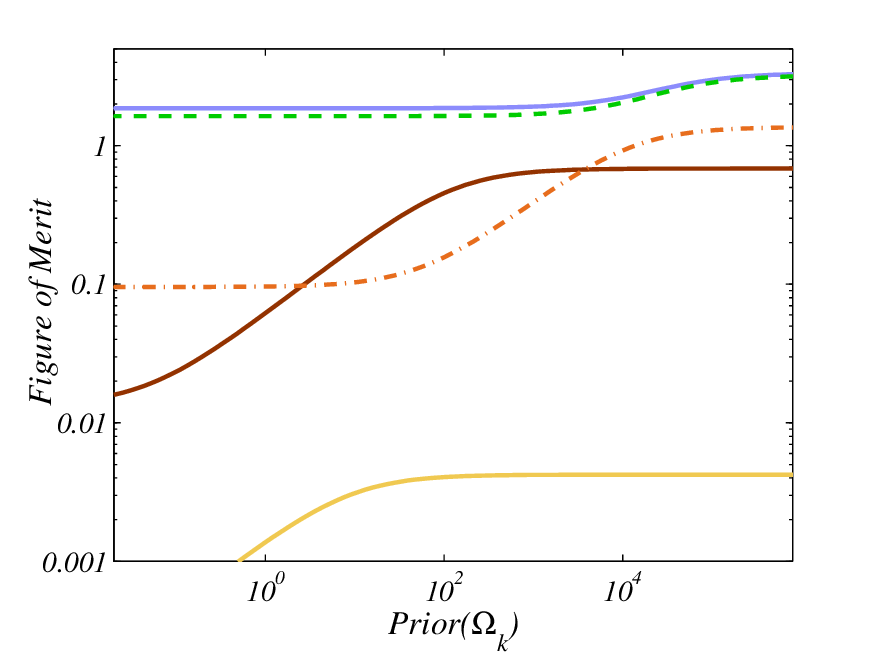}} \\ [0.0cm]
 \end{array}$
\caption{{\bf Curvature marginalisation degrades dark energy constraints} - using the same input survey data as Figure~(\ref{cooray_ok_ellipse}), the error ellipses in the dark energy parameters for prior values on $\Omega_k$ from $10^6$ (inner curves) to $10$ (outer curves) are shown. {The prior on $\Omega_k$ is decreases logarithmically from $10^6$ so that each values is 0.4 times the previous prior.} Ellipses are produced for measurements of (clockwise from top left) the Hubble parameter {\bf(dark brown curves)}, the angular diameter distance {(orange)}, the growth function {(light brown)}, a combination the Hubble parameter and the angular diameter distance {(green)} and a combination of all three observables {(blue)}. The bottom right panel shows the DETF Figure of Merit \cite{correct_detf} (the inverse of the area of the ellipse in the $w_0-w_a$ plane) for these ellipses as a function of the prior on $\Omega_k.$ The colours correspond to the ellipses: green (dashed) is $H$ and $d_A$ combined while orange (dot-dashed) corresponds to $d_A$ alone. In all cases the shape of the curves is roughly the same, with no change in the FoM beyond a certain value of the prior, both in the case of very large and very small priors. Interestingly the curves for the Hubble parameter (brown) and angular diameter distance (orange) cross - a survey consisting only of measurements of $H$ may seem to yield tighter (or weaker) constraints on dark energy than a survey only of distance measurements depending on the curvature prior assumed. \label{curvature_plot} }
\end{figure}
For each of the parameters, Figure~(\ref{curvature_plot}) shows that the constraints are eroded and the ellipse increases in size with the decrease in the prior - which expresses our confidence in the flatness of the universe. While the ellipses from the single observables such as $H, d_A$ show a much greater increase in size with decreasing curvature prior, using a combination of parameters is more robust - since combining multiple probes helps break the curvature-dark energy degeneracy. The change in the size of the ellipse directly relates to a change in the Figure of Merit (FoM), or power, of the particular combination of observables. Various FoMs are described in Section~\ref{formalism}. In the case of the FoM of the Dark Energy Task Force (DETF) \cite{correct_detf}, the area of the ellipse and the DETF FoM are inversely proportional. The right-hand panel of Figure~(\ref{curvature_plot}) shows this change in the FoM with changing curvature prior. Interestingly, the DETF FoM flattens out for both very large and very small values of the prior. This can be explained by the fact that for very large prior values, marginalising over the curvature has no effect, since our error on the curvature is minute, whereas for very small values of Prior$(\Omega_k)\sim 0$ the ellipse is completely degenerate with the curvature and hence reaches a maximum size.

\subsection{New Features in \name{}}\label{newfeat}
\name{}~made its public debut (Version 1.0) in May 2008, with significant updates and revisions every six months since then. This work coincides with the release of Version 2.2 of the code. The main new features of Versions 1.2 through to Version 2.2 are listed below; a more comprehensive description of the changes in the new code is contained in the {\bf Readme.txt} file, while the new features are described in the \name{}~Users' Manual \cite{usersmanual}, provided with the distribution of the code suite.
\begin{itemize}
\item{\em FoMSWG}\\
We include an extension based on the the original code suite produced by Dragan Huterer for the Joint Dark Energy Mission (JDEM) FoM Science Working\footnote{see http://wfirst.gsfc.nasa.gov/science/} (FoMSWG) \cite{albrecht_fomswg}. The code uses Principal Components \cite{huterer_pcs} (PCs) to probe the sensitivity of various surveys with respect to dark energy parameters. The equation of state $w(z)$ is parametrised by a function of 36 piecewise constant parameters. The Fisher Matrix is marginalised over all other cosmological parameters except those concerning $w(z)$. This marginalised matrix is then decomposed into eigenvalues and eigenvectors which represent the principal components. This can give insight to the redshift at which each experiment has the greatest power to constrain $w$. A full description of the method is found in \cite{albrecht_fomswg, huterer_pcs}.  The FoMSWG GUI creates an easy to use interface to the original code which allows one to experiment with  combinations of different surveys and the associated impact to the dark energy parameters. A range of Fisher Matrices for surveys are calculated and included as default, including the Planck satellite\footnote{http://www.rssd.esa.int/index.php?project=planck}, the weak lensing component of the Dark Energy Survey\footnote{http://www.darkenergysurvey.org/}, the ESSENCE survey\footnote{http://www.ctio.noao.edu/essence/wresults/}, Supernova Legacy Survey (SNLS)\footnote{http://www.cfht.hawaii.edu/SNLS/}, Higher-Z HST supernova surveys; and the WiggleZ\footnote{http://wigglez.swin.edu.au/site/}, BOSS\footnote{http://www.sdss3.org/surveys/boss.php}, WFMOS (which has been cancelled as a Gemini project but may resurface in another form) and HETDEX\footnote{http://hetdex.org/} baryon acoustic oscillation surveys. For consistency and ease of comparison both the input and output are stored in the same format as the original FoMSWG code. This extension is also accessible via the command line.
 \item {\em Report Generating Features} \\
 Code is included in the current version for automated generation of both \LaTeX{}~and text reports containing information such as the input survey, fiducial cosmology, output Fisher matrices and Fisher ellipse figure.
\item {\em Fitting formulae for Baryon Acoustic Oscillation (BAO) Surveys} \\
Two extensions are provided to calculate the errors on the BAO oscillation scale according to the prescriptions of Blake {\em et al.} \cite{blake_etal_ff} and Seo and Eisenstein \cite{SE2007} (see the files in the directories called {\bf EXT\_FF\_Blake\_etal2005} and {\bf EXT\_FF\_SeoEisenstein2007}).  These are not available via the GUI currently, and must be run directly from the command line. However, these modules combined with the rest of \name{}~provide the capability of going directly from BAO survey specifications (volume, area, number density etc.) to dark energy constraints. For further description of the two methods, we refer the reader to the papers \cite{blake_etal_ff, SE2007}.
\item {\em ``Point-and-click'' ellipse plotting} \\
A new feature allows the user to click in the figure and have the ellipse automatically generated at that fiducial model. This is automatically linked to the GUI and is accessible via the ``F4C Extensions'' drop-down menu in the GUI.
\item {\em General growth function}\\
Growth constraints are computed exactly for any allowed $w(z)$ and any curvature $\Omega_k$.
\item {\em Improved numerical derivative treatment} \\
The numerical derivative routine was modified from a simple double-sided derivative, to one which uses the complex-step algorithm \cite{complex_step}, with improved numerical stability, especially in extreme cases.
\item {\em GUI button to include/exclude the prior information matrix}\\
A button has been added to the GUI to allow the user to either include or exclude the prior information matrix from the Fisher analysis. This makes it easy to compute and compare the impact of adding previous knowledge (e.g. Planck) on a given survey.
\end{itemize}
\subsection{User Extensions}
The general philosophy of \name{}~was to make it as easy as possible to mould and extend to the needs of a general user. The power of \name{}~does come at a price however and to modify the code requires the user to become familiar with the underlying structures used in the design and building of \name{}. The good news is that \name{}~was designed to be as logical, elegant and general as possible and quick but limiting fixes were avoided where possible in favour of flexibility. This means that once the user is familiar with the underlying code structure, \name{}~should be easy to extend in ways that were not even conceived of when it was written.

All \name{}~functions begin with the prefix {\bf FM} while extensions have the {\bf EXT} prefix (in addition the figure plotting code is prefixed by {\bf FIG}. To use \name{}~for a new application, the function (the ${\bf X}$ in the notation of Section~\ref{formalism}), should be coded as a Matlab function (or functions if you have multiple observables) and named appropriately: {\bf FM\_my\_function.m}. If no analytical derivatives (i.e. $\partial X/\partial \theta_A$, are supplied (and named appropriately, as e.g. {\bf FM\_my\_derivs.m}), the code will evaluate numerical derivatives of the specified observables. Care should be taken to ensure that all functions take in a data vector and base parameter vector (the fiducial model where the Fisher Matrix will be evaluated) as arguments and that they return a vector, as in the header documentation of {\bf FM\_function\_1.m} and {\bf FM\_analytic\_deriv\_1.m}. The input structure should be modified by replacing the entry {\bf `FM\_analytic\_deriv\_1'} in the field {\bf input.function\_names} with functions described above; {\bf `FM\_my\_function'}. Similarly, the fields specifying which derivatives should be used are replaced.

As an example consider alternative parameterisations for dark energy. Currently, the \name{}~GUI is hard-coded for three cosmological observables ($H$,$d_A$, and $G$), assuming the Chevallier-Polarski-Linder (CPL) parameterisation of dark energy with parameters ($w_0,w_a$), see Eq.~(\ref{wparam}). This is true of both the functions themselves, and the analytical derivatives included in the \name{}~suite. The general framework of \name{}, however, means that one is not restricted to this parameterisation. Given that the dark energy equation of state enters the cosmological observables through the evolution of the dark energy, (via $f(z)$, defined in Eq.~(\ref{feq})), for any given $w(z)$ one only needs to specify the names of the functions (in the {\bf input} structure) that will replace the current versions of {\bf FM\_function\_1.m} ($H(z)$), {\bf FM\_function\_2.m} ($d_A(z)$) and {\bf FM\_function\_3.m} ($G(z)$). The same is true for the derivatives - either they can be coded analytically for the particular parameterisation of dark energy, or the derivatives will be evaluated numerically from the functions specified in the {\bf input} structure.

As a caveat, the GUI can only be used if the new parameterisation of dark energy still contains only two coefficients. If this is not the case, \name{}~must be run from the command line.
\section{Conclusions \label{conc}}
The Fisher Matrix formalism is the standard forecasting method in cosmology and assuming one is familiar with it, allows rapid and widely understood results. In principle it is easy to learn and code for oneself. In practice there is a threshold below which it simply is not worthwhile to reinvent the wheel.

\name{}~works both as a didactic and a research tool, yielding production-quality Fisher
Matrix forecasts. Written in Matlab, the \name{}~Graphical User Interface (GUI) allows easy exploration of
cosmological constraints and has both an interactive `point-and-click' facility for automatic ellipse generation (when activated from the ``F4C Extensions" drop-down menu) and an automatic \LaTeX{}~summary and results file generator, making direct inclusion of any output into research documents straightforward.

In this paper we focused on illustrating novel uses of the \name{}~suite by exploring the landscape of Fisher Matrix cosmology, as illustrated by Figures~(\ref{fig:wa_w0_vary_ellipse})-(\ref{fig:slice}), as well as highlighting the effects of growth and curvature on Fisher Matrix forecasts of future cosmological surveys. These illustrate a limited set of applications of \name{}~which we hope stimulates members of the community to use and extend the code. To date, research that has thus far used \name{}, either for forecasts or for producing plots, includes \cite{rassat_bao}, \cite{parkinson09} and \cite{BAOChapter}. 

\section{Acknowledgements} We would particularly like to thank Dragan Huterer, Martin Kunz and Cristiano Sabiu for their significant contributions to \name{}~and would like to thank Kishore Ananda, David Bacon, Chris Blake, Chris Clarkson, George Ellis, Daniel Holz, Wayne Hu, Eiichiro Komatsu, Bob Nichol, Patrice Okouma, David Parkinson, Varun Sahni, Charles Shapiro, Mat Smith, David Spergel, Alexei Starobinsky, Roberto Trotta and Melvin Varughese for ideas, discussions and comments related to \name{}. RH would like to thank Jo Dunkley, Joe Zuntz and James Allison for useful discussions. The authors acknowledge funding from the NRF (BB, JK), SKA South Africa and the Rhodes Trust (RH), NASSP and SISSA (YF) and the Royal Society (JK). JK thanks the ICG, Portsmouth, for its wonderful hospitality and support during the completion of this project along with the Astrophysics Department at Oxford University who were very patient in allowing him to complete the development of additional extensions for F4C. 
\bibliography{arxiv_man}
\bibliographystyle{apsrev1}

\appendix


\section{Derivatives for $H(z),d_A(z),G(z)$ used in the Fisher Matrix\label{app_deriv}}
We present the analytical derivatives of the Hubble parameter, $H(z),$ given as Eq.~(\ref{eeq}) and the angular diameter distance, $d_A(z),$ given as Eq~(\ref{daeq}) with respect to the cosmological parameters $(H_0, \om, \ok, w_0, w_a)$, where $\w0,\wa$ are the CPL dark energy parameters, and assume the forms $f(z)$ and $E(z)$ as given in Eqs.~(\ref{feq}) and (\ref{eeq}) respectively. In all cases the derivatives are taken in a general Friedmann-Lema\^{i}tre-Robertson-Walker background without assuming flatness.

{\bf{ --- The Hubble parameter}}\\
As the Hubble constant, $H_0$ only appears as a multiplicative term in $H(z)$, the derivative of the Hubble parameter with respect to the Hubble constant $H_0$ is simply
\be
\frac{\p H}{\p H_0} = E(z)
\ee
Derivatives of the function $\mathscr{E}(z) \equiv H^2(z)/\h0^2 = E^2(z)$ are found in all derivatives of both $H$ and $d_A$ and are worth defining separately:
\bea
\frac{\p \mathscr{E}(z) }{\p \om} &=& \nonumber (1+z)^3 -f(z) \\
\frac{\p \mathscr{E}(z) }{\p \ok} &=& \nonumber (1+z)^2 -f(z)\\
\frac{\p \mathscr{E}(z) }{\p \w0}&=&  3(1-\om -\ok)f(z)\ln(1+z)\nonumber \\
\nonumber \\
\frac{\p \mathscr{E}(z) }{\p \wa}&=& 3(1-\om -\ok)f(z)\(\ln(1+z) - \frac{z}{1+z}\) \,. \\ \label{qderivs}
\eea
For all the cosmological parameters we consider other than the Hubble parameter $\h0$, the derivatives with respect to the Hubble parameter can then be expressed as
\be \frac{\p H(z)}{\p \theta_i} =  \frac{H_0}{2E} \frac{\p \mathscr{E}(z) }{\p \theta_i}, ~~~ \theta_i \in (\om, \ok, \w0,\wa)
\ee

{\bf{ --- Angular Diameter Distance}}\\
In a FLRW background, $d_A(z)$ is given by Eq~(\ref{daeq}). 
The Hubble parameter appears only in the pre-factor of the angular diameter distance, and hence $\p d_A(z)\p H_0 = \(-\frac{1}{H_0}\) d_A(z)$.
Whereas the other parameters such as $\Omega_m, w_0, w_a$ contribute solely to the comoving distance, or $\chi(z)$ term, and hence can be expressed using Eqs.~(\ref{chider}) and (\ref{qderivs}) as
\bea
\frac{\p d_A(z)}{\p \theta_i} &=& \frac{1}{1+z}\frac{c}{H_0} \cosh \( \sqrt{\ok} \chi(z) \) \frac{\p \chi(z)}{\p \theta_i},
\label{ddadom}
\eea
where 
\be
 \frac{\p \chi (z)}{\p \theta_i} =-\int_0^z \frac{1}{2E^3(z')} \frac{\p \mathscr{E}(z')}{\p \theta_i} dz', ~~~~ \theta_i \in (\om,\ok,\w0,\wa) \label{chider}.
\ee
The curvature parameter is found both in the pre-factor and the $\sinh$ term of the angular diameter distance, hence
  \bea
 \frac{\p d_A(z)}{\p \ok} &=& -\frac{1}{1+z}\frac{c}{H_0} \frac{1}{2\ok^{3/2}} \sinh\( \sqrt{\ok} \chi(z) \) + \frac{1}{1+z} \frac{c}{H_0} \frac{1}{\sqrt{\ok}} \cosh\( \sqrt{\ok} \chi(z) \) \left[ \frac{\chi(z)}{2\sqrt{\ok}}  + \sqrt{\ok}\frac{\p \chi(z)}{\p \ok} \right]\nonumber \\
 &=&-\frac{1}{2\ok}d_A(z) + \frac{1}{1+z}\frac{c}{H_0}\cosh \( \sqrt{\ok} \chi(z) \)\left[\frac{\chi(z)}{2\ok} + \frac{\p \chi(z)}{\p \ok}\right].\nonumber \\
 \label{ddadok}
 \eea
The Taylor series expansion of Eq~(\ref{ddadok}) is used $\Omega_k \rightarrow 0$, namely:
\bea
 \left.\frac{\p d_A(z)}{\p \ok}\right|_{\ok \rightarrow 0} &=& \frac{c}{H_0}\frac{1}{1+z}\left\{\frac{1}{6}\chi^3(z,0) + \frac{\p \chi(z,0)}{\p \ok}\right\} \label{taylorok}
 \eea
where again $X(z,0)\equiv \left.X(z)\right|_{\ok\rightarrow 0}$ are the functions (for example $E(z),~\chi(z)$) assuming flatness.
Using the definitions Eqs.~(\ref{chider}) and (\ref{qderivs}), the derivatives of the angular diameter distance with are expressed similarly for $\theta_i \in (\w0,\wa)$ as:
\be   \frac{\p d_A(z)}{\p \theta_i} = \frac{1}{1+z} \frac{c}{H_0} \cosh \( \sqrt{\ok} \chi(z) \) \frac{\p \chi(z)}{\p \theta_i}\,. \ee

{\bf{ --- Growth}}\\
All derivatives related to the growth are computed numerically (in the routine FM\_num\_deriv.m) by solving Eq.~(\ref{newgln}) and using either the complex step or central finite difference algorithm, depending on the choice of the user.

\cleardoublepage
\section{\name{}~Figure~Code\label{fig_code}}
\subsection{Code to produce Figure~(\ref{derivs}): plot of function values and derivatives}
\begin{verbatim}
% ------------------------------------------------------------------------
function FIG_function_derivative_plot(deriv_flag, function_flag)
% This function generates a plot of the derivatives of the specific
% observables included in Fisher4Cast, H(data), d_A(data) and G(data).
% It must be included in the directory in which Fisher4Cast is contained,
% or that directory must be added to the Matlab path.

% The flags deriv_flag and function_flag are set to 1 (0) if you do (don't)
% want to plot the function or derivatives. If no input is given these are
% both set to 1 and you get plots of all functions and derivatives.

% As a default example it calls the Seo_Eisenstein_2003 input structure,
% but then generates a redshift vector from 0.1:10;
% The colours for line plots must be specified as 1x3 RGB vectors,
% normalised to 1 (i.e. so each entry divided by 255). See the default
% colours as an example.
close all

% Flags to control what you want to plot, either derivatives only, of
% function only, or both
if nargin == 0
    deriv_flag = 1;
    function_flag = 1;
elseif nargin == 1;
    function_flag = 0;
end
%--------------------------------------------------------------------------
% Specify the colours of the derivatives
hcolour =[147 50 0]./255;
gcolour = [240 201 81]./255;
dacolour = [231 109 29]./255;
colourmat = [hcolour
            gcolour
            dacolour];
styles = {'-', '-.',  '--',':', '-.'};
legendmat{1} = {'dlnH/dH_0'  ' dlnH/dln\Omega_m'...
 'dlnH/d\Omega_k' 'dlnH/dw_0' 'dlnH/dw_a'};
legendmat{2} = {'dlnd_A/dH_0' ' dlnd_A/dln\Omega_m'...
 'dlnd_A/d\Omega_k' 'dlnd_A/dw_0'  'dlnd_A/dw_a'};
legendmat{3} = {'dlnG/dH_0' ' dlnG/dln\Omega_m'...
 'dlnG/d\Omega_k' 'dlnG/dw_0' 'dlnG/dw_a'};
%--------------------------------------------------------------------------
% Generate the Input data for the derivative plot
input = Seo_Eisenstein_2003; % initialise the input structure
data = 0.1:0.1:10; % the redshift range we want to consider
data = data(:);
input.growth_zn = 0;
input.growth_zn_flag = 1;
% make sure the Growth is normalised at data = 0;
input.observable_index = [1 2 3]; % Use all three observables
input.num_observables = length(input.observable_index);
% Re-assign the redshift vectors in the input structure
input.data{1} = data;
input.data{2} = data;
input.data{3} = data;

% Re-assign the errors vectors in the input structure
input.error{1} = 0.1.*ones(1,length(input.data{1}));
input.error{2} = 0.1.*ones(1,length(input.data{2}));
input.error{3} = 0.1.*ones(1,length(input.data{3}));

% Use the analytical formula for H, d_A and
%numerical derivatives for G
input.numderiv.flag{1} = 0;
input.numderiv.flag{2} = 0;
input.numderiv.flag{3} = 1;
%--------------------------------------------------------------------------
% Run Fisher to get the parameter values and derivatives
output = FM_run(input);
close(1); % Close the figure of the Fisher Ellipse

%--------------------------------------------------------------------------
% PLOT THE DERIVATIVES
% We will plot dlnX/dtheta_i = dX/Xdtheta_i
%--------------------------------------------------------------------------
if deriv_flag == 1

    for i = 1:input.num_observables % Loop over the observable functions

        x = input.observable_index(i);
        % Set the figure properties
        figure(x*100)
        axes( 'FontName', 'Times', 'FontAngle', 'italic',...
         'FontSize', 14 , 'XScale', 'log', 'XTickLabel', {'0.1';'1';'10'} )
        hold on
        box on
        xlabel('Redshift', 'FontName', 'Times',...
         'FontAngle', 'italic', 'FontSize', 16 )
        ylabel(['Fisher Derivatives for ', input.observable_names{x},...
         '(z)'], 'FontName', 'Times', 'FontAngle', 'italic', 'FontSize', 16)

        for j = 1:5
            if j==2
             % The Omega_m derivative, this is actually dlnH/dlnOm
                semilogx(data, input.base_parameters(j)...
                .*output.function_derivative{x}(:,j)./output.function_value{x},...
                    'LineStyle', styles{j},'LineWidth', 2,...
                      'Color', colourmat(x,:))
            else
                plot(data, (output.function_derivative{x}(:,j)...
                ./output.function_value{x}),...
                'LineStyle', styles{j},'LineWidth', 2,   'Color', colourmat(x,:))
            end % end the check to see if we are plotting Omega_m derivatives
        end % end the loop over the parameters

        legend(legendmat{x}, 'Location','NorthWest')
         % plot the legend for the function according to the observable
    end % end the loop over the observable functions

end % end the if loop for plotting of derivs

%--------------------------------------------------------------------------
% PLOT THE FUNCTIONS
%--------------------------------------------------------------------------

if function_flag ==1

    for i = 1:input.num_observables
        x = input.observable_index(i);
        % Set the figure properties
        figure(100*x +1)
        axes( 'FontName', 'Times', 'FontAngle', 'italic', ...
        'FontSize', 14 ,'XScale', 'log', 'XTickLabel', {'0.1';'1';'10'} )
        hold on
        box on
        xlabel('Redshift', 'FontName', 'Times', 'FontAngle',...
         'italic', 'FontSize', 16 )
        ylabel([input.observable_names{x}, '(z)'], 'FontName',...
         'Times', 'FontAngle', 'italic', 'FontSize', 16 )

        % Plot the data
        semilogx(data, output.function_value{x}, 'LineWidth',...
         2,  'Color', colourmat(x,:));

    end % end the loop over observables

end  % end the if loop for function plotting
\end{verbatim}
\subsection{Code to produce Figure~(\ref{fig:slice}): volume slice plots}
\begin{verbatim}
% ------------------------------------------------------------------------
%This function takes a three dimensional matrix, fom_vol_out, and plots a
%slice plot for this data.
%
%fom_vol_out can be generated by calling
%
%>>fom_vol_out = FIG_generate_fom_volume_data(vol_res)
%
%If no vol_res is passed a default value of 30 is assumed.
%
%Example of using this function:
%
%>>FIG_plot_slice_fom_volume(fom_vol_out)
%
%if no fom_vol_out is given then the code checks to see if there is a default
%.mat file, default_fom_vol_out.mat, to load the data from, else the
%function FIG_generate_fom_volume_data is called with a set of default values.
%-----------------------------------------------------------------------------
function FIG_plot_slice_fom_volume(fom_vol_out)

%check if a matrix of the volume space, fom_vol_out, is passed to the function
if nargin<1
    %see if the default .mat file exists and load the data
    if exist('default_fom_vol_out.mat')
        load default_fom_vol_out;
        fom_vol_out = default_fom_vol_out;
    else
        %if not then generate the fom_vol data
        fom_vol_out = FIG_generate_fom_volume_data(30);
    end
end

%set the colormap
colormap(jet);

%select the planes to intersect for the slice plot
%---------------------------------------------------------------------------
%this is an additional subsection of code to make the selection of redshift
%planes to plot more generic and easy to manage for a range of
%fom_vol_out's produced. Please note you must still manually specify the
%redshift range
[x_col y_col z_col] = size(fom_vol_out);
x_redshift_range = [0,5];
y_redshift_range = [0,5];
z_redshift_range = [0,5];
%calculate a relationship from column to redshift
column_to_redshift_ratio_x = x_col/x_redshift_range(end);
column_to_redshift_ratio_y = y_col/y_redshift_range(end);
column_to_redshift_ratio_z = z_col/z_redshift_range(end);
%specify the slices redshift to intersect the fom volume space
x_redshift_slice = [0.6667];
y_redshift_slice = [0.4167, 1.6667];
z_redshift_slice = [2.5];
%calculate the slices in column numbers
xslice = column_to_redshift_ratio_x.*x_redshift_slice;
yslice = column_to_redshift_ratio_y.*y_redshift_slice;
zslice = column_to_redshift_ratio_z.*z_redshift_slice;
%---------------------------------------------------------------------------

%plot using slice
s = slice(fom_vol_out,xslice,yslice,zslice);

%set the labels so they match the range of
% redshift as opposed to the column numbers
set(gca,'xtick',[0:column_to_redshift_ratio_x:x_col],...
'xticklabel',[0:x_redshift_range(end)],'ytick',...
[0:column_to_redshift_ratio_y:y_col],'yticklabel',...
[0:y_redshift_range(end)],'ztick',[0:column_to_redshift_ratio_z:z_col],...
'zticklabel',[0:z_redshift_range(end)]);

%set the x y and z labels
ylabel('H Redshift');
xlabel('d_A Redshift');
zlabel('G Redshift');

\end{verbatim}

\subsection{Code to produce Figure~(\ref{curvature_plot}): Fisher ellipses as a function of changing curvature prior}
\begin{verbatim}
% ------------------------------------------------------------------------
function FIG_vary_fom_curvature_prior
global input plot_spec axis_spec
% This function generates a plot of the Fisher ellipse as one changes the
%  prior value on curvature, and a corresponding plot of the Dark Energy
% Task Force Figure of Merit (FoM) as a function of the prior.
% See the User's Manual for definitions of the FoM.

% This code must be included in the directory in which
% Fisher4Cast is contained, or that directory must be
% added to the Matlab path.

% As a default example it calls the Seo_Eisenstein_2003 input structure,
% The Matlab 'colormap' command is used to generate the line colours,
% specific to each observable and the number of iterations is given by N.

% NOTE That this code uses getfigdata.m by M.A. Hopcroft,
% which is code from the Matlab File
% Exchange (http://www.mathworks.co.uk/matlabcentral/fileexchange/14081).
% It is included in this package.

close all
%--------------------------------------------------------------------------
% Choose your colour schemes for the various combinations
linecolor{1} = sort(colormap(copper), 'descend');
linecolor{2} = colormap(autumn);
linecolor{3} = colormap(copper);
linecolor{4} = colormap(winter);
linecolor{5} = colormap(summer);
input.fill_flag = 1;
valinit = 35; % the starting colour for the plots
num_obs = [1 2]; % The vector of combinations you want:
% 1 = Hubble
% 2 = Angular Diameter distance
% 3 = Growth Function
% 4 = Hubble parameter + Angular Diameter distance + Growth
% 5 = Hubble parameter + Angular Diameter distance

line_width = 2;
%--------------------------------------------------------------------------
% Set the Input structure for the survey you will use
input = Seo_Eisenstein_2003;
input.data{3} = input.data{1};
input.error{3} = 0.1.*ones(1,length(input.error{1}));
input.observable_index = [1 2 3]; % We will use all observables
input.fill_flag = 0;
input.numderiv.flag{3} = 1;
%--------------------------------------------------------------------------
% Set up the range you wish to consider
start_prior= 1e6;
range = 1e8; % No of orders of magnitude in the prior
N = 20; % Number of points
amp = (range)^(1/N);

% Initialise the priors
prior_orig = input.prior_matrix; % this will be the default value
input.prior_matrix(3,3) = start_prior;
input.prior_matrix(2,2) = 0;
% Initialise the prior on the matter density to zero
%--------------------------------------------------------------------------
% Initialise the global FoM plot
    figure(3000)
    axes( 'FontName', 'Times', 'FontAngle',...
     'italic', 'FontSize', 14,'XScale', 'log','YScale', 'log' );
    hold on
    box on
    xlabel('Prior(\Omega_k)', 'FontName', 'Times',...
     'FontAngle', 'italic', 'FontSize', 16 )
    ylabel('Figure of Merit', 'FontName', 'Times',...
     'FontAngle', 'italic', 'FontSize', 16 )
%--------------------------------------------------------------------------
% Call Fisher4Cast in a loop
for ni = num_obs(1):num_obs(end)
    input.prior_matrix(3,3) = start_prior;
    if ni ==4
    % Compute the Fisher Ellipse for combination of Hubble, d_A and G
        input.observable_index = [1 2 3];
    elseif ni ==5
    % Compute the Fisher Ellipse for combination of Hubble and d_A  only
        input.observable_index = [1 2 ];
    else
        input.observable_index = ni; % use index value as specified
    end

    for i=1:N
        figure(1)
        hold on
        input.prior_matrix(3,3) = input.prior_matrix(3,3)./amp;
        % modify the Prior
        output = FM_run(input); % Call Fisher4Cast
        val(i) = input.prior_matrix(3,3);
        % save the value of the prior for plotting
        outv(i,:) = output.fom;
         % Save the full FoM vector
        out(i) = outv(i,1); % Save the DETF FoM
         h = getfigdata(1);
         % call getfigdata.m to rip off the ellipse
         x{i} = h{1}.x;
         y{i} = h{1}.y;
         close(1) % close the figure
    end
    %--------------------------------------------------------------------------
    % Initialise the plotting figures with the axis specs etc
    figure(1000+ni)
    axes( 'FontName', 'Times', 'FontAngle', 'italic', 'FontSize', 14 )
    hold on
    box on
    xlabel('w_0', 'FontName', 'Times', 'FontAngle',...
     'italic', 'FontSize', 16 )
    ylabel('w_a', 'FontName', 'Times', 'FontAngle',...
     'italic', 'FontSize', 16 )
    axis([-3 1 -10 10 ])
    count = 0;

%--------------------------------------------------------------------------
    % Plot the resulting ellipses
    for i = 1:N
        figure(1000+ni)
        hold on
        % Use increasing or decreasing colour to get a gradient
        if i < ceil(N/2)
            count = count +1;
        else
            count = count -1;
        end
        cval = valinit+ 2*count;
        plot(x{i}, y{i}, 'Color', linecolor{ni}(cval,:),...
         'LineWidth',line_width)
    end

    figure(3000)
    hold on
    plot(val,out, 'Color', linecolor{ni}(20,:),...
      'LineWidth', line_width)
end
%--------------------------------------------------------------------------

\end{verbatim}
\section{Quick Start Guide to \name{}\label{quickstart}}

\subsection{Hardware and software requirements}

This software is written to be run in Matlab (Linux, Windows and under Mac OSX, although this has not been extensively tested). The user needs Matlab installed (Tested on Version 7) to be able to run this code. Free disk space of approximately 2MB and the minimum recommended processor and memory specifications required by the Matlab version you are using is suggested.

\subsection{Downloading Fisher4Cast}
Currently the code is available for download at one of the following websites
\cite{cosmo_org, mathworks}. Save this .zip file into the directory you want to run the Fisher4Cast suite from.
\subsection{Getting started}
The code can be run from the command line or the Graphical User Interface (GUI). We describe the command line below, and mention how to get the GUI started. More information on the GUI can be found in the Users' Manual \cite{usersmanual}.
\subsection{The Graphical User Interface}
\begin{itemize}
\item{ Running the GUI}

The GUI can be started from the Matlab editor. The file {\bf FM\_GUI.m} must be opened from the directory, and once the file is opened (click on the file icon from within the Command-line
interface to open it with an editor) press F5 to run the code. This will open up the GUI screen.

You can also launch the GUI from the command line by typing: \begin{verbatim} >>FM_GUI\end{verbatim}
This then functions in the same way as using FM\_run in the command line (as explained in the following section).

For more information on the technicalities of the GUI, see the full manual.
\item{GUI Screenshots}

We include some screenshots of the Graphical User Interface.

\begin{figure}[htb!]
\begin{center}
\includegraphics[width=0.5\textwidth]{\fig_dir{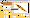}}
\caption{{\bf Plotting multiple ellipses on one axis} -  using the `Hold on' multiple error ellipses can be overlaid on one axis. The `Area Fill' command allows you to choose the colours for the error ellipses. Also shown is the `Running' window which indicates the code is running to calculate the Fisher ellipses. \label{qs_screen1}}
\end{center}

\end{figure}
\begin{figure}[htb!]
\begin{center}
\includegraphics[width=0.5\textwidth]{\fig_dir{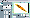}}
\caption{{\bf Different background images and colour schemes} - the background images and colour schemes (skins) allow for a fully customisable Graphical User Interface. \label{qs_screen2}}
\end{center}
\end{figure}
\begin{figure}[htb!]
\begin{center}
\includegraphics[width=0.5\textwidth]{\fig_dir{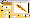}}
\caption{{\bf Various Figures of Merit can be plotted} - the drop-down list allows for a choice between various Figure~of Merit options. \label{qs_screen3}}
\end{center}
\end{figure}

\end{itemize}

\newpage
\subsection{ The Command Line}
\begin{itemize}
\item{ Running the code}

Open your version of Matlab and change the working directory to be the same as where you saved \name{}~in. To run the code from the command line with one of the standard test input structures supplied, type:\begin{verbatim} >>output = FM_run(Cooray_et_al_2004)\end{verbatim}

This will call the code using the pre-supplied test input data (Cooray\_et\_al\_2004) and then generate an error ellipse plot for the parameters and observables supplied in the chosen input. All the relevant generated output is written to the output structure. You can see the range of outputs to access by typing:\begin{verbatim} >>output \end{verbatim}
and then examine each output individually by specifying it exactly.
For example: \begin{verbatim}>>output.marginalised_matrix\end{verbatim} will access the marginalised Fisher Matrix from the output structure.

You can use the supplied input files as a template for generating new input files with your own customised parameters and values. All fields shown in the example structures must be filled in any user-defined structure.

The code can also be run from the Matlab editor. Once the code is opened (open it from inside the Matlab window), you can press
F5 to run the code. Note that if the code is run from the Editor it will call the default input structure, which is the
{\bf Cooray\_et\_al\_2004.m} file. This is an example file containing input data from the paper by Cooray {\em et al.} \cite{cooray2004}. This output can be directly compared to that of Figure~1 of that paper. If your output compares correctly, you have a working installation of the code. Another input available is {\bf Seo\_Eisenstein\_2003.m} \cite{seo2003}.
\end{itemize}

\end{document}